\documentclass[12pt, modern]{aastex61}

\shorttitle{Kepler AutoRegressive Planet Search}
\shortauthors{Caceres et al.}

\begin{document}

\title{AutoRegressive Planet Search: Application to the Kepler Mission}
\author{Gabriel A. Caceres}
\affiliation{Department of Astronomy \& Astrophysics, Pennsylvania State University, University Park, PA 16802}

\author{Eric D. Feigelson}
\altaffiliation{Center for Astrostatistics, Pennsylvania State University, University Park PA 16802}
\altaffiliation{Center for Exoplanets and Habitable Worlds, Pennsylvania State University, 
University Park PA 16802}
\affiliation{Department of Astronomy \& Astrophysics, Pennsylvania State University, University Park, PA 16802}

\author{G. Jogesh Babu}
\altaffiliation{Center for Astrostatistics, Pennsylvania State University, University Park PA 16802}
\affiliation{Department of Statistics, Pennsylvania State University,  University Park, PA 16802}

\author{Natalia Bahamonde and Alejandra Christen}
\affiliation{Instituto de Estad\'{i}stica, Pontificia Universidad Cat\'{o}lica de Valpara\'{i}so, Valpara\'{i}so, Chile}

\author{Karine Bertin and Cristian Meza}
\affiliation{Centre for Research and Modeling of Random Phenomena, CIMFAV, Universidad de Valpara\'{i}so, Valpara\'{i}so, Chile}

\author{Michel Cur\'{e}}
\affiliation{Instituto de F\'{i}sica y Astronom\'{i}a, Universidad de Valpara\'{i}so, Valpara\'{i}so, Chile}

\begin{abstract}
The 4-year light curves of 156,717 stars observed with NASA's Kepler mission are analyzed using the AutoRegressive Planet Search (ARPS) methodology described by Caceres et al. (2019).  The three stages of processing are: maximum likelihood ARIMA modeling of the light curves to reduce stellar brightness variations; constructing the Transit Comb Filter periodogram to identify transit-like periodic dips in the ARIMA residuals; Random Forest classification trained on Kepler Team confirmed planets using several dozen features from the analysis.  Orbital periods between 0.2 and 100 days are examined.  The result is a recovery of 76\% of confirmed planets, 97\% when period and transit depth constraints are added.  The classifier is then applied to the full Kepler dataset; 1,004 previously noticed and 97 new stars have light curve criteria consistent with the confirmed planets, after subjective vetting removes clear False Alarms and False Positive cases.  The 97 Kepler ARPS Candidate Transits mostly have periods $P<10$ days; many are UltraShort Period hot planets with radii $<1$\% of the host star.  Extensive tabular and graphical output from the ARPS time series analysis is provided to assist in other research relating to the Kepler sample.  
\end{abstract}


\accepted{for publication in the Astronomical Journal}

\section{Introduction}  \label{intro.sec}

NASA's Kepler satellite, during its 4 year photometric observations of nearly 200,000 stars, has been superbly successful in detecting over 2500 confirmed exoplanets.  But the mission encountered difficulties in efficiently detecting smaller planets; it has detected only a handful of Earth-like planets in their host star's Habitable Zones, far below the estimate of several hundred made before the mission launch \citep{Borucki97}.  The instrument was designed to achieve a differential photometric precision of $\leq 20$ parts per million (ppm) for 6.5 hr transit around a  $V=12$ star, including a guessed 10 ppm contribution from stellar variability based on solar observations.  However, both the instrumental noise and the intrinsic stellar variability were greater than expected, giving a median around 30 ppm for $V \simeq 12$ stars \citep{Batalha14}.  Over half of stars have intrinsic stellar variations $1-3$ times solar levels and an additional $\sim 20\%$ show intrinsic noise $3-7$ times above solar levels, even after removal of large-scale trends and outliers \citep{Gilliland11}.  

To reduce these stellar variations that impede small planet transit detection, the Kepler team developed a complex pipeline of statistical procedures that involve wavelet transforms, autoregressive gap filling, local polynomial trend removal, signal whitening, and periodic transit template matching with transit durations ranging from 1.5 to 15 hours.  Stars satisfying intermediate criteria are collected into Multiple Event Statistics and Threshold Crossing Events categories which are then subject to four False Alarm veto steps.  At several stages in the processing, critical threshold levels are subjectively chosen to reduce false alarms and concentrate promising transit cases.  Stars that pass the various criteria are called Kepler Objects of Interest (KOIs) and are subject to further vetting to reduce astronomical False Positives and isolate exoplanet Candidates. The Kepler Transit Search Pipeline is outlined by \citet{Seager15} and described in detail by \citet{Jenkins17}.  

From a statistical viewpoint, the methodology underlying the Kepler pipeline's effort to reduce intrinsic stellar variability involves mostly nonparametric methods, such as wavelet transforms and local regressions, where `nonparametric' here means that no low-dimensional mathematical model for the global variability behavior is assumed. Other researchers use nonparametric procedures like autocorrelation functions \citep{McQuillan14}, and Gaussian Processes (GP) regression \citep[][and many others]{Gibson12, Petigura13, Haywood14} to assist with transit detection.  The GP approach involves fitting a statistical model to the data, but the model is high-dimensional with $O(N)$ parameters for $N$ photometric observations. 

Several research groups have searched for planets in the full Kepler 4-year photometric dataset independently of the Kepler Team's pipeline methodology. 
\begin{enumerate}
\item  A non-algorithmic search by thousands of Citizen Scientists has resulted in two new candidates from Q2 data that survive False Positive tests  \citep{Lintott13}.  

\item \citet{Ofir13} apply a procedure called `Simultaneous Additive and Relative SysRem' (SARS) that reduces stellar variability with median and Savitzky-Golay filters and collective instrumental effects from a subset of intrinsically constant stars.   The Box Least-Square periodogram \citep[BLS,][]{Kovacs02} is then applied with an arbitrary threshold, followed by a variety of tests and manual inspection.  Eighty-four new transit signals are identified with depths ranging from $\sim 100-1000$ ppm. 

\item \citet{Huang13} detrend with median and discrete cosine filters, Trend Filtering Algorithm to remove instrumental effects on spatially proximate stars, the BLS  to find periodic transit-like patterns with an arbitrary peak threshold criterion, followed by manual inspection to reduce False Positives. They recover $\sim 82$\% of KOI planet candidates from Q1-Q6 data, and report 150 new candidates with depths similar to \citet{Ofir13}. 

\item \citet{Jackson13} scanned the Kepler dataset for planets with orbital periods $<12$~hr.  They apply a boxcar filter to reduce stellar variability and apply a multiscale binning procedure to test for Gaussianity in the residuals.  Peaks in the BLS periodogram are identified with arbitrary thresholds and a truncation to remove long-duration periodicities.  Candidate transits are identified after manual inspection and various astronomical tests (even-odd transit depth variation, astrometric shifts).  Four candidates with periods $0.18-0.45$~days, transit depths $\sim 10$~ppm, and planet radii $0.6-3$~R$_\earth$.  

\item \citet{SanchisOjeda14} use the Fourier periodogram to identify Ultra-Short-Period (USP) planets with orbital periods $P<1$~day because it suppresses subharmonics compared to the BLS periodogram.  They remove outliers, apply a medial filter with arbitrary width, and apply an arbitrary threshold to the Fourier power, applied a variety of filters and False Positive tests accompanied by manual vetting of the spectrum and folded light curve.  They identify 18 new, and catalog a total of 106, USP planets smaller than $\sim 2$~R$_\earth$, 

\end{enumerate}

\citet[][henceforth Paper I]{Caceres19} present a different statistical approach that has been widely used in time series analysis, signal processing and econometrics since the 1970s.  Known as `autoregressive modeling' or the `Box-Jenkins' method, the central procedure involves regression for low-dimensional  models related to ARIMA, the `autoregressive integrated moving average' model.  This is a flexible approach that combines application of a differencing operator to reduce trends (the `I' component in ARIMA) with regression to model  deterministic and stochastic dependencies on recent past values (AR component) and past changes (MA component) in the light curve.  Long-memory power law `red noise' is treated in ARFIMA models (F component).  The model is fit to regularly spaced time series with maximum likelihood methods without free parameters or thresholds, except for a limit on model complexity.  The ARIMA approach to time series analysis is described in textbooks such as \citet{Chatfield04}, \citet{Hyndman14}, and \citet{Box15}.   \citet{Feigelson18} discuss ARIMA-type models and its continuous-time variants (CARMA, CARFIMA) in an astronomical context.

Paper I describes a multi-stage AutoRegressive Planet Search (ARPS) procedure summarized in Figure~\ref{ARPS_flowdiag.fig} here: fitting ARIMA and ARFIMA models, calculating a newly developed Transit Comb Filter (TCF) periodogram to detect transit-shaped periodicities in the model, and applying a Random Forest (RF) classifier trained on confirmed planets to reduce False Alarms and False Positives.  The RF classifier learns from scalar `features' from various stages of the ARPS analysis, trained on the Confirmed Candidates sample derived from the Kepler Team pipeline.  We do not adopt the more computational intensive approach of Deep Learning that classifies directly from the light curves \citep{Pearson18, Shallue18, Zucker18}. A single threshold is applied to the classifier probabilities based on ROC curves, followed by a subjective vetting procedure to produce a sample of candidate new planets. 

This study applies the ARPS procedure to the 4-year Kepler mission data.  The principal goal to characterize the light curves using different mathematical approaches that used previously and to discover thereby new transit candidates. Sections~\ref{data.sec}-\ref{methods.sec} briefly review the Kepler data and the ARPS methodology.  Sections~\ref{preprocessing.sec}-\ref{dataprod.sec}  presents intermediate stages of the analysis: ARIMA and ARFIMA modeling, TCF periodograms, and RF classification.  Results are presented in sections~\ref{results.sec}-\ref{KACT.sec} and discussed in section~\ref{Disc.sec}.  The main scientific result is the identification of 97 candidate small-diameter transiting exoplanets absent from the Kepler Team Candidate Planet sample, many of which are UltraShort Period (USP) planets with periods $P \leq 1$ day (Table~\ref{KACT.tbl} and Figure Set~\ref{KACT_pngs.fig}).  Results in this study do not address long-period planets relevant to the populations of Earth-like planets in Habitable Zones.  A variety of tabular and graphical data products are provided to assist researchers interested in different criteria for planet detection or different aspects of Kepler star variability (\S\ref{dataprod.sec}).

\section{Data, Methods, and Software} \label{sec2.sec}

\subsection{Kepler 4-Year Light Curves} \label{data.sec}

The ARPS methodology is applied to long-cadence photometry from the prime phase of NASA's Kepler mission \citep{Borucki10}.  The dataset used in the ARPS analysis here is long-cadence light curve data from Data Release 25 (DR25) for Quarters 1 through 17 obtained in April 2016 from the Kepler Data Products residing at NASA's Mikulski Archive for Space Telescopes (MAST).  The light curves are available in files entitled $kepler\_id.[stop\_times]\_llc.fits$; the stop times refer to the 17 3-month quarters of observation. Each light curve records  $\sim 70,000$ measurements of stellar flux in a regularly spaced cadence of 29.4~minutes, measured in instrumental units of electrons per second continuously for about 4 years.  Many stars are not observed for the full 17 3-month quarters and thereby exhibit long gaps in the datastream.  All stars show gaps for a variety of instrumental causes totaling $15\%$ or more of the data stream.   

Our effort here is restricted to the limited problem of extracting periodic behaviors resembling planetary transits after `Pre-search Data Conditioning' (PDC) by the Kepler Science Operations Pipeline.  No other Kepler satellite raw data are considered here; for example, pixel-level fluxes are not examined, and barycentric corrections were not applied to the time stamps.  The PDC processing corrects light curves for instrumental effects such as pointing offsets, thermal transients, focus changes due to reaction wheel heater cycling, random flux discontinuities from cosmic rays or other causes, and sky crowding. Outliers are removed and data gaps are filled.  Details of PDC processing are provided by \citet{Twicken10}, \citet{Stumpe12}, and \citet{Jenkins17}.  

The following supplementary datasets were collected for use at various stages of ARPS analysis: 

{\it Kepler Input Catalog (KIC) stars} ~~ Stellar properties and KIC designations of 197,096 stars were acquired from NASA's Mikulski Archive for Space Telescopes (MAST) Web page entitled {\it Kepler Q1-Q17 DR25 Stellar Parameters} \citep{KSPWG16}.  Note that these have not been corrected for improved distances obtained by the ESA Gaia satellite. 

{\it Kepler Objects of Interest (KOIs)} ~~ The official Kepler exoplanet catalog for DR25 was downloaded from the NASA Exoplanet Archive at Caltech's Infrared Processing and Analysis Center on April 1, 2018 \citep{Thompson18}. It contains results for 8,054 KOIs (from 6,923 unique stars), of which 4,034 are candidates and 4,020 are considered false positives. We also used a list with 1,510 additional entries from previous data releases which were not part of DR25. 

{\it Kepler Certified False Positive Table}  ~~ The Kepler Certified False Positive Working Group reexamined many KOIs in detail to provide a high-reliability sample. Many KOIs from all data releases were categorized as `Certified False Positives'. This table was  downloaded from the MAST archive on April 1, 2018. 

{\it Kepler Threshold Crossing Events (TCEs)} ~~ TCEs are potential transit signals flagged in the initial stage of the Kepler analysis \citep{Twicken16}. A list of 34,032 TCEs originating from 17,230 unique stellar targets was downloaded. 

{\it Kepler Eclipsing Binary Catalog}  This catalog, produced by the research group at Villanova University, lists all known eclipsing binary systems in the Kepler field \citep{Kirk16}. We downloaded the catalog updated on April 27, 2017 containing 2,909 entries. 

\subsection{Methods and Software Overview} \label{methods.sec}

\begin{figure}
\centering
\includegraphics[width = 0.7\textwidth]{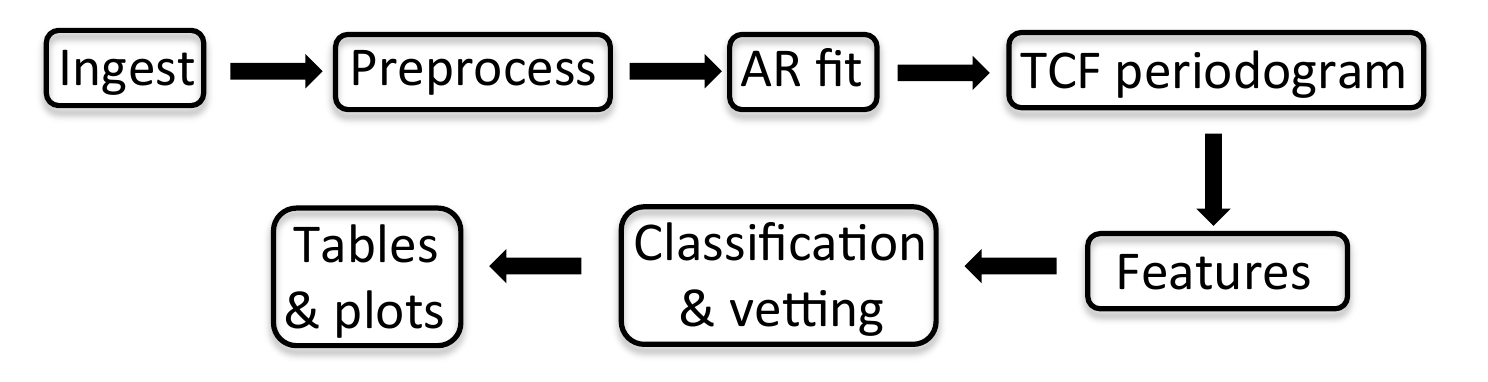}
\caption{Flow diagram summarizing the steps involved in the ARPS analysis pipeline.} 
\label{ARPS_flowdiag.fig}
\end{figure}

The mathematical foundations and algorithms for ARPS are presented and discussed in Paper I.  A flow diagram summarizing the analysis procedures is shown in Figure~\ref{ARPS_flowdiag.fig}. It shows the principal steps in ARPS analysis: autoregressive modeling with ARIMA and ARFIMA for detrending and whitening the time series; matched filtering with the TCF algorithm to search for planets in the ARIMA/ARFIMA residuals; automated identification and selection of promising exoplanet candidates using a Random Forest classifier.   The mathematical and algorithmic foundations for the methods are described in Paper I.  Implementation issues for the Kepler application are discussed in the following subsections. 

Our analysis is implemented in the public domain R statistical software environment \citep[version 3.2.0,][]{RCoreTeam15} with the computationally intensive TCF algorithm coded in Fortran. Most of the ARPS pipeline is written in base R with crucial autoregressive functions {\it auto.arima} and {\it arfima} provided by the {\it forecast} CRAN package \citep{Hyndman08, Hyndman18}.  Other specialized functions, such as the tests for normality, autocorrelation and stationarity are obtained from CRAN packages {\it tseries} \citep{Trapletti16}, {\it lmtest} \citep{Zeileis02} and {\it nortest} \citep{Gross15}. The Fortran code for the TCF filter was written by G. Caceres and incorporated into the R pipeline script using the standard {\it .Fortran} function. For the classification stage of the analysis, we use the {\it randomForest} function from CRAN package {\it randomForest} \citep{Liaw02} and the {\it roc} function from CRAN package {\it pROC} \citep{Robin11}. 

The computational time for the ARPS pipeline for each Kepler star is $\sim 30$ CPU-minutes on an Intel XEON E5-2680 processor.  A full ARPS analysis requires over 100,000 CPU-hours.  Processing was performed using the Advanced CyberInfrastructure high-performance computing infrastructure provided by The Institute for CyberScience at The Pennsylvania State University. The execution of this pipeline for a multitude of Kepler stars is an `embarrassingly parallel' computational problem due to the lack of interdependence between different stars.  The GNU {\it parallel} utility \citep{Tange11} is used to distribute each star as an individual job for cores in the cluster.

\subsection{Light Curve Preprocessing} \label{preprocessing.sec}

ARPS begins with the long-cadence PDC Kepler light curves detrended from experimental systematics to allow the analysis to focus on modeling the intrinsic stellar variability and uncovering planetary transits.  The datasets provided by the MAST archive are in the Flexible Image Transport System (FITS) format which are ingested into R with function {\it readFITS} from the {\it FITSio} CRAN package \citep{Harris16}.  The input light curve is the column `PDCSAP\_FLUX' in units of electrons per second.  We ignore data associated with the preliminary Quarter 0; Quarter 1 starts at cadence 1105 (131.5126 days) and Quarter 17 ends at cadence 71427 (1591.001 days). 

\begin{figure}
\centering
\includegraphics[width=\textwidth]{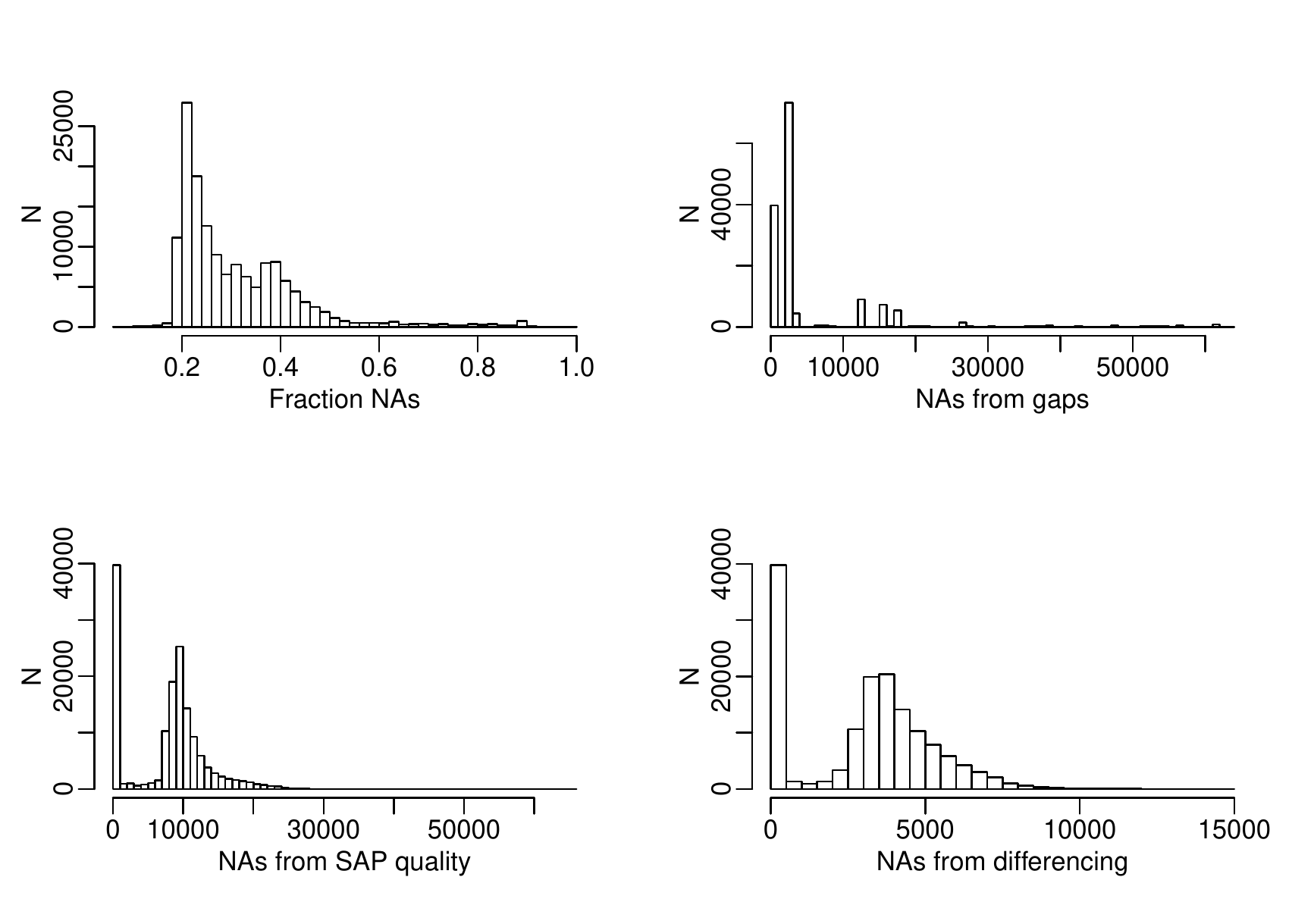}
\caption{ Frequency histograms for the  dataset showing: the fraction of all missing values (top left); number of missing values due to inter-quarter gaps (top right); number of missing values due to the Kepler data quality flags (bottom left); and number of missing values induced by differencing (bottom right). The ordinates give the number of Kepler stars with the chosen characteristics. } 
\label{preprocessing_NAs.fig}
\end{figure}

A significant fraction of cadence values do not have reliable flux measurements and are filled with R's {\it NA} (Not Available) logical variable. NA values are used for inter-quarter gaps and missing quarters, and for points marked with non-zero data quality flags by the Kepler team. These data quality flags, that include conditions such as satellite attitude problems and cosmic ray hits, are described in Table 2-3 of the {\it Kepler Archive Manual} \citep{Thompson16}. We choose not to apply additional outlier detection and removal beyond the Kepler data quality events because it is difficult to discriminate between instrumental problems and stellar variability such as flares or deep planetary transits.  The fractions of NAs produced in the ensemble of Kepler light curves analyzed with ARPS due to these causes are shown in Figure~\ref{preprocessing_NAs.fig}. 

With this preprocessing, the flux measurements for each star correspond to a 71427-element vector with missing values where appropriate. Since the ARIMA models discussed here require equally spaced data, we work with cadence number and not time (e.g. barycentric corrections are ignored). 

\begin{figure}
\centering
\includegraphics[width=\textwidth]{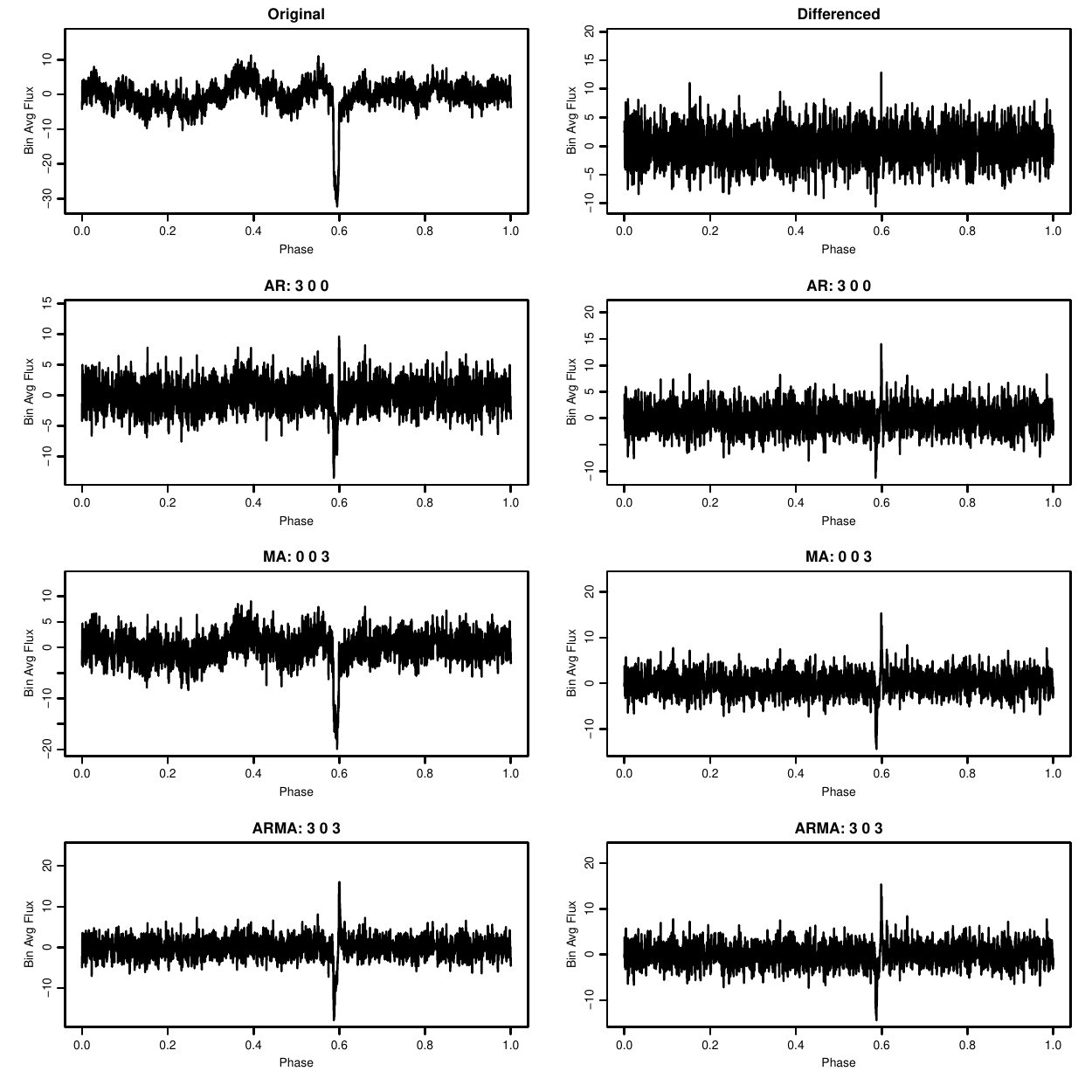}
\caption{Effects of the differencing operator and ARMA model orders on the shape of a transit. The top row shows the original light curve, and the following rows show the effects of selected low-dimensional AR, MA and ARMA models. The left column shows application to the original light curve and the right columns shows application to the differenced light curve. The ordinates give the PDC flux in electrons sec$^{-1}$ with median values removed. }
\label{Effects_ARMA_order.fig}
\end{figure}

Fluxes during the 4-year light curve are subject to shifts in the measured stellar flux from quarter to quarter due to variations in the hardware (e.g. instrument temperature).  The ARPS preprocessing module `stitches' them together first by subtracting the mean flux within each quarter, and then by applying the differencing operator to the full 4-year light curve.  The differencing operation is commonly used to make time series (approximately) stationary (Paper I, \S2.3).  While its main purpose in ARPS is to remove trends due to stellar variability, it has the additional beneficial effect of removing the inter-quarter flux offsets. Time series analysts often omit an explicit differencing step and let it be incorporated into the ARIMA fitting as required to maximize the likelihood via the Akaike Information Criterion. But our application of differencing during the preprocessing step guarantees that any box shape transit has been transformed into a double spike for all stars in the sample.  Without the explicit differencing step, the shape of the transit in the residual light curve can differ depending on the exact ARMA order used. Figure~\ref{Effects_ARMA_order.fig} shows examples of this effect, where the transit shape is less stable when autoregressive models are applied without differencing (left columns) than when applied to the differenced light curve (right columns). 

Since differencing reduces the series to the point-to-point variations, a data point is lost after this operation. We therefore add a leading zero as the initial point of each light curve so that time-related estimates (e.g., period, phase, duration) remain correctly aligned. Differencing also creates additional missing values in the data since the difference between a known value and a missing one is undefined, the corresponding point in the differenced light curve will be NA. The last panel of Figure~\ref{preprocessing_NAs.fig} shows this effect. 

It is useful to have a reference noise level to quantify any improvements achieved by ARIMA modeling. The original un-stitched light curve has a high and uninformative variance, so we will use zero-mean centered light curves (hereafter referred to as the `stitched light curves') as a point of comparison to evaluate how our approach improves the background noise.  We measure noise level with the InterQuartile Range (IQR), the range between the 25\%-th and 75\%-th quantiles of the flux distributions.  The IQR is a robust measure of spread that is insensitive to non-Gaussianity and outliers.  

\subsection{Autoregressive Models} \label{armod.sec}

The R statistical environment provides many tools for the development and application of the autoregressive models presented in Paper I (\S2), as well as other more complex models. We apply the {\it auto.arima} and {\it arfima} functions provided by the widely used {\it forecast} package \citep{Hyndman18} for our AR(F)IMA modeling of the differenced light curves.  The {\it arfima} function provides equivalent functionality with the addition of allowing for fractional differencing \citep{Palma07}. While all light curves were differenced in advanced as part of our preprocessing, the algorithms allow for automatic selection of differencing order and may apply additional differencing. 

The {\it auto.arima} function provides automated autoregressive order selection and fitting, as described by \citet{Hyndman08}. We use their step-wise algorithm for efficient order selection, although this can result in a non-optimal model selection.  The code also provides a more computationally intensive procedure to exhaustively test all permutations of autoregressive and moving average orders, up to a chosen maximum order. Each model is fitted by maximum likelihood estimation and the multiple fits are compared through the Akaike Information Criterion (AIC) to select the optimal parsimonious model. The AIC is a penalized likelihood model selection measure widely used in parametric time series analysis \citep{Sakamoto86}. 

Maximum likelihood estimation allows for the presence of missing values, so the {\it auto.arima} estimation is not hindered by NA's in the series. Unfortunately, fractional differencing does not permit missing values,  so missing values must be imputed in some way for the {\it arfima} code.  We implement the simple approach of replacing NA's in the differenced series with a value of zero.  This has a minimal effect on model parameter estimates. After ARFIMA modeling, NAs are placed back into the residuals at the locations where values were previously missing. 

The success of ARIMA and ARFIMA modeling can be evaluated with time series diagnostics that compare the residuals to Gaussian white noise.  These are commonly in econometrics \citep[e.g.][]{Enders14}; references can be found in Paper I (\S2.2).  The Durbin-Watson test evaluates serial (lag=1) autocorrelation, the Breusch-Godfrey test evaluates lagged autocorrelation (we choose lag=5), the augmented Dickey-Fuller and KPSS tests evaluate stationarity.

\begin{figure}
 \centering
  \hspace*{-0.5in} \includegraphics[width=0.9\linewidth,angle=-90]{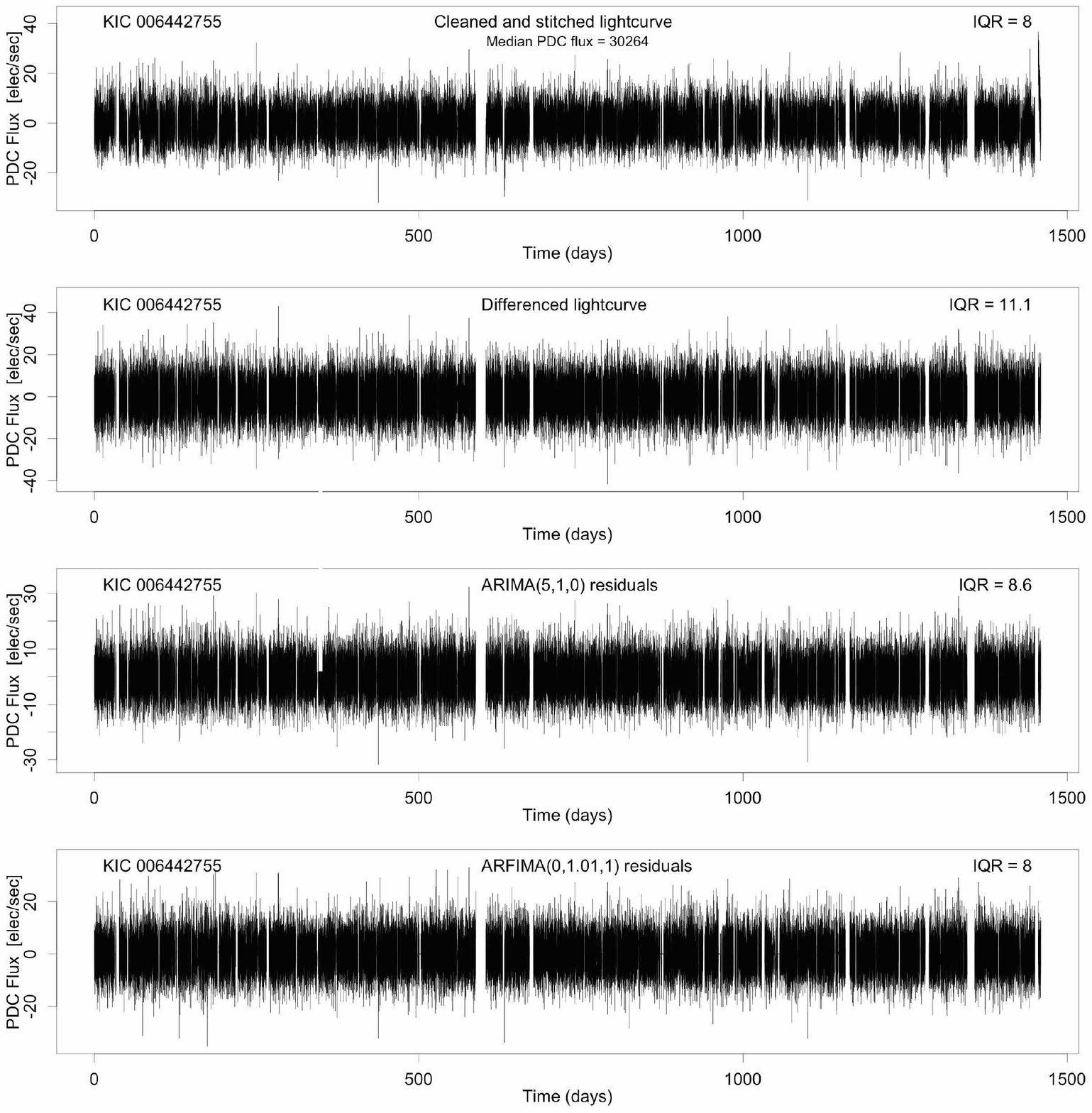}\vspace{-0.8in}
  \includegraphics[width=0.7\linewidth,angle=-90]{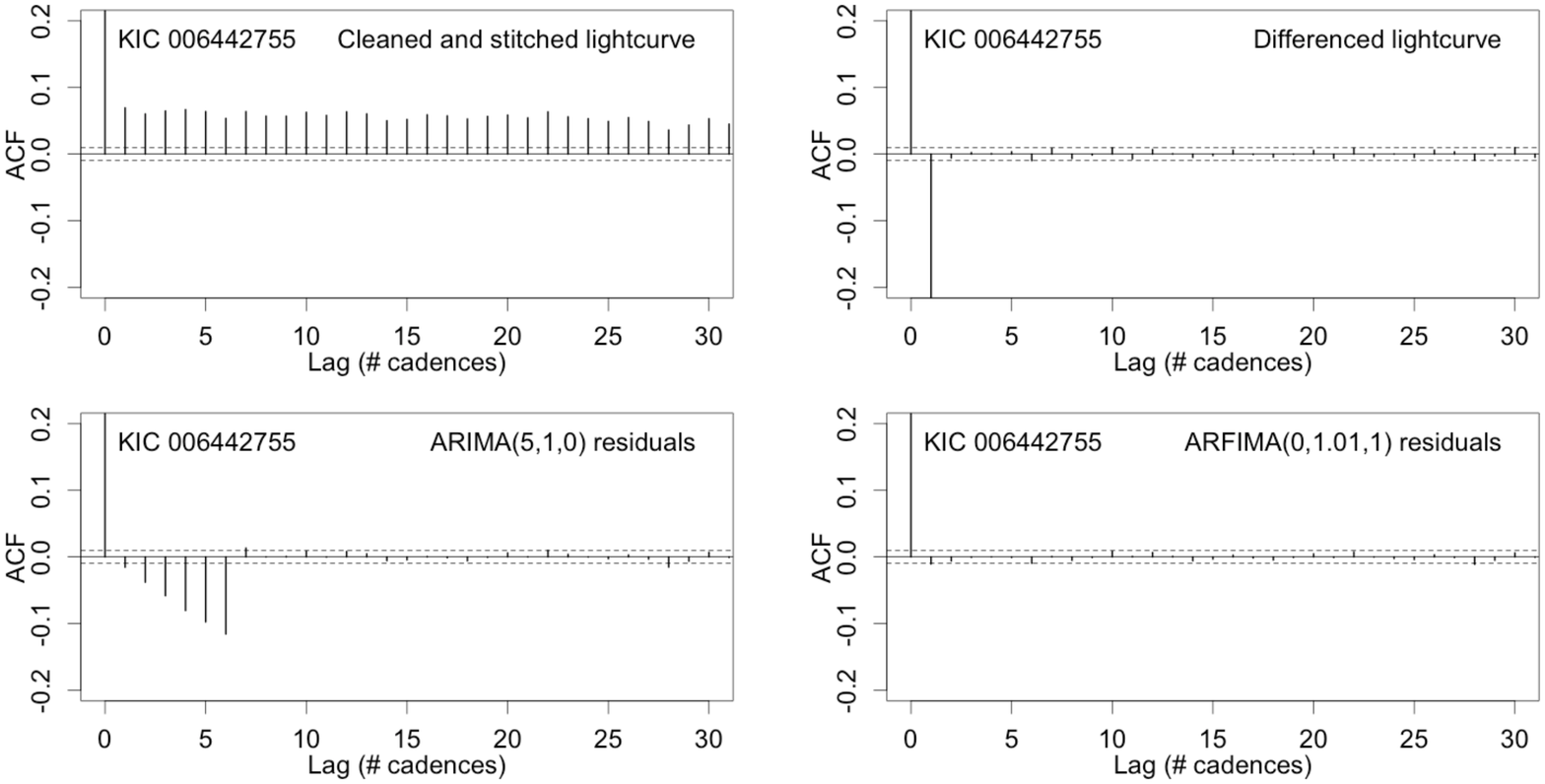}\vspace{-0.6in}
 \caption{Quiet star (KIC 6442755) example of ARIMA and ARFIMA performance.  The top four panels show the 4-year light curves at four stages of analysis: stitched and cleaned light curve; differenced light curve; ARIMA residuals; and ARFIMA residuals. IQR gives the interquartile range at each stage.  Units are in electrons/second.  The bottom panels show the autocorrelation function at each stage.  The ARIMA residual IQR of 8.6~elec~s$^{-1}$ corresponds to 280 ppm.   }
 \label{ARmod1.fig}
 \end{figure}

\begin{figure}
 \centering
  \hspace*{-0.5in}   \includegraphics[width=0.9\linewidth,angle=-90]{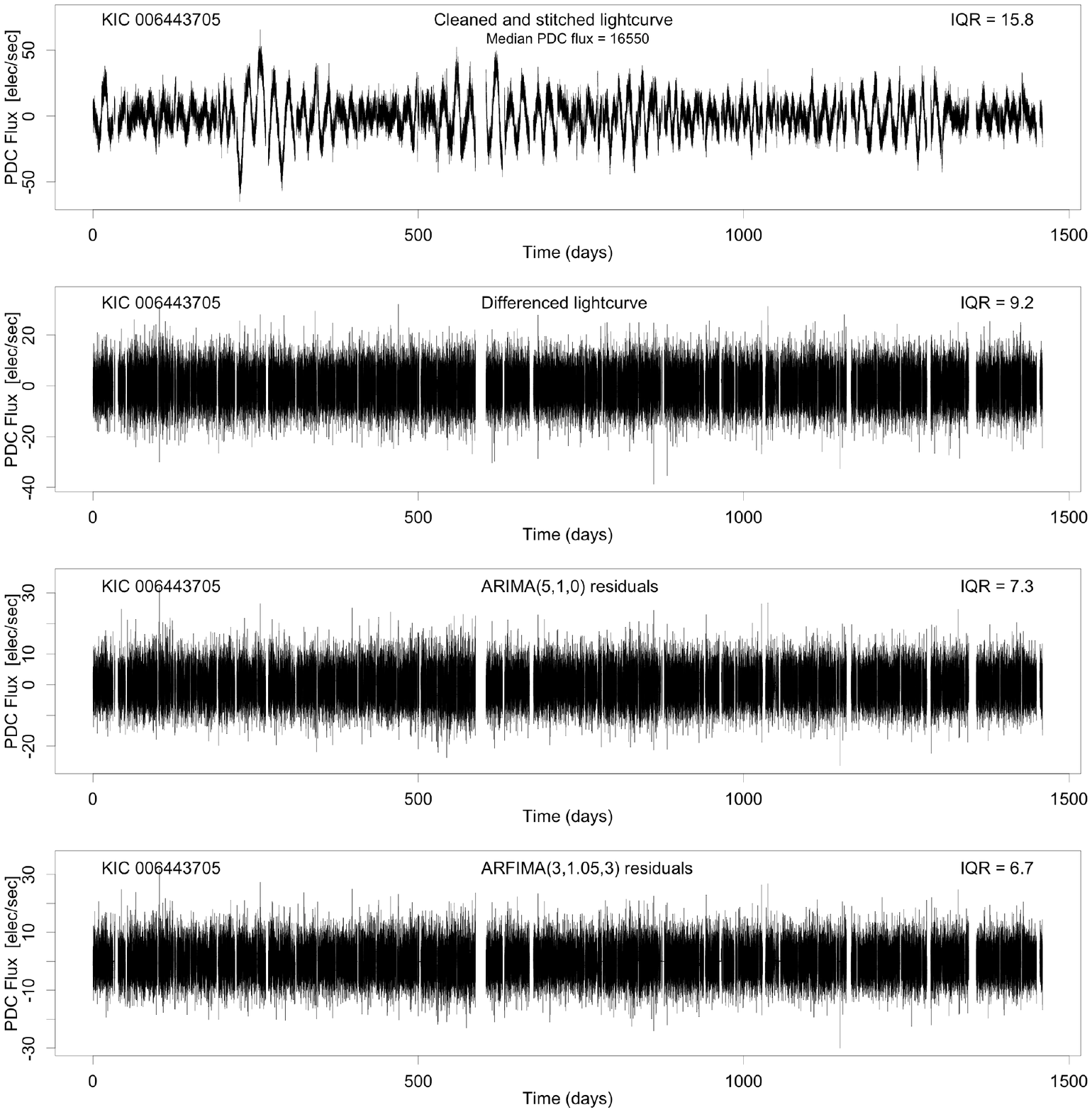}\vspace{-0.8in}
  \includegraphics[width=0.7\linewidth,angle=-90]{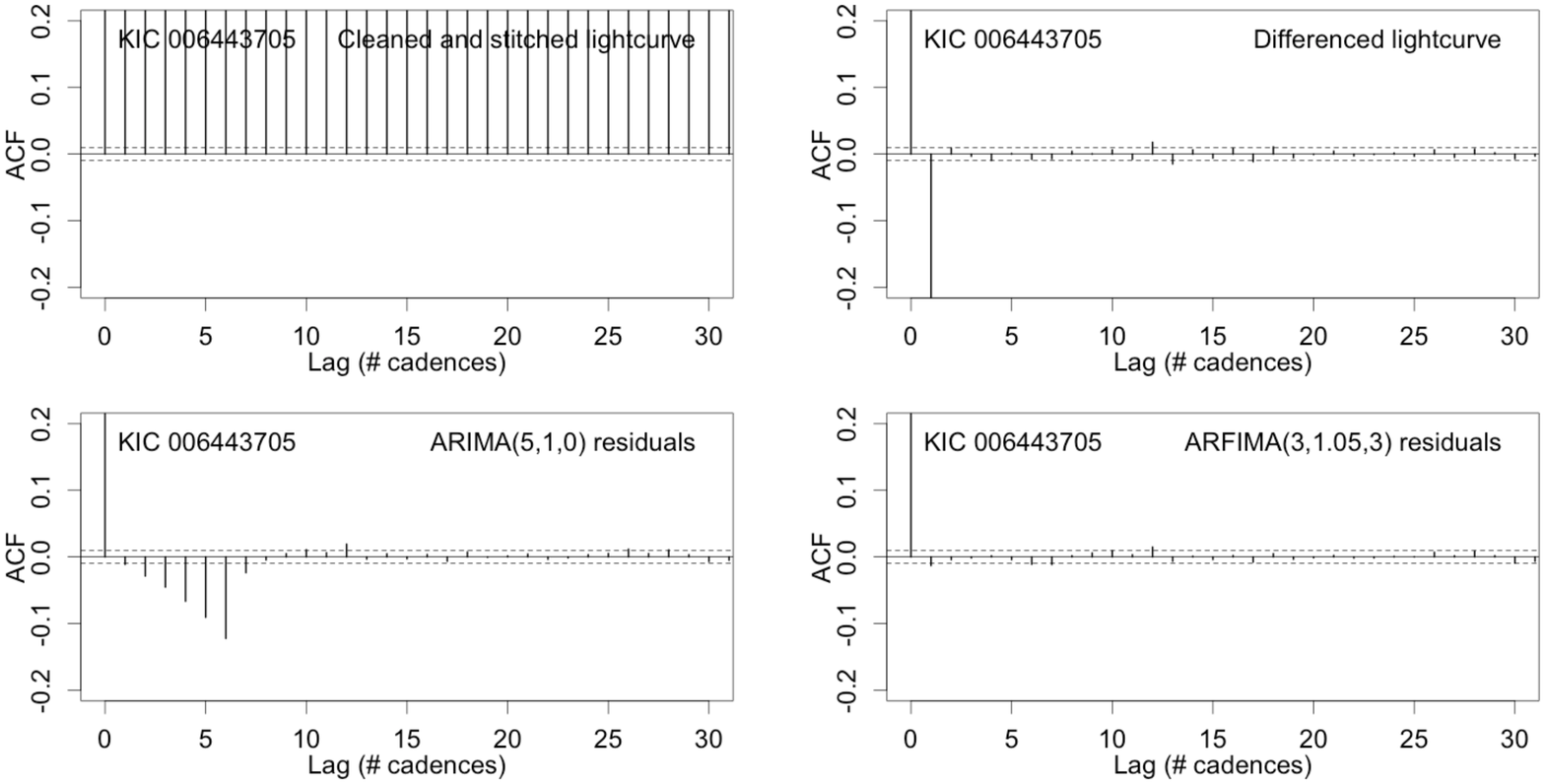}\vspace{-0.5in}
 \caption{Same as Figure~\ref{ARmod1.fig} for a spotted star, KIC 6443705. The ARIMA residual IQR of 6.7~elec~s$^{-1}$ corresponds to 440 ppm.}
 \label{ARmod2.fig}
 \end{figure}

\begin{figure}
 \centering
  \hspace*{-0.5in}   \includegraphics[width=0.9\linewidth,angle=-90]{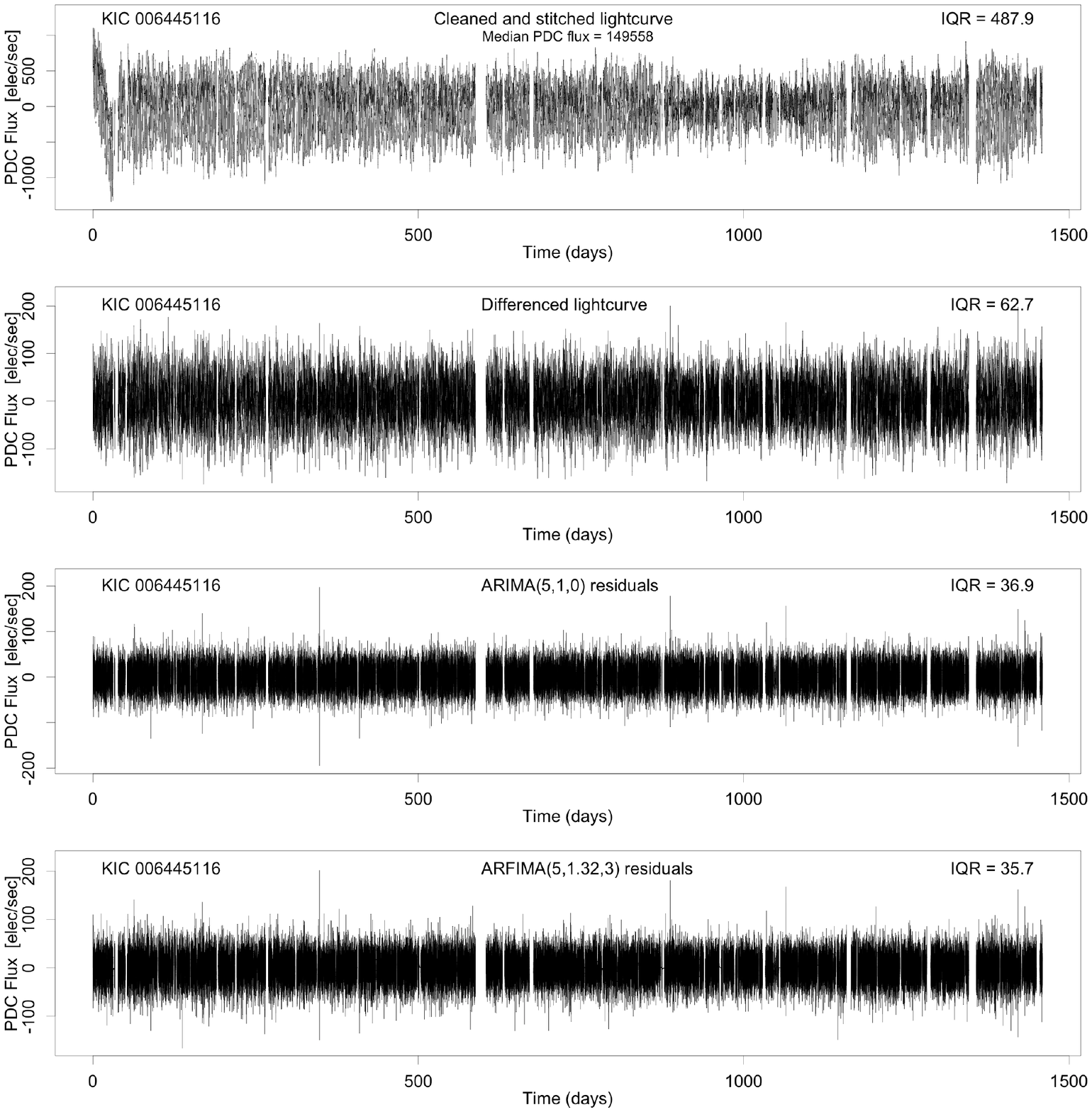}\vspace{-0.8in}
  \includegraphics[width=0.7\linewidth,angle=-90]{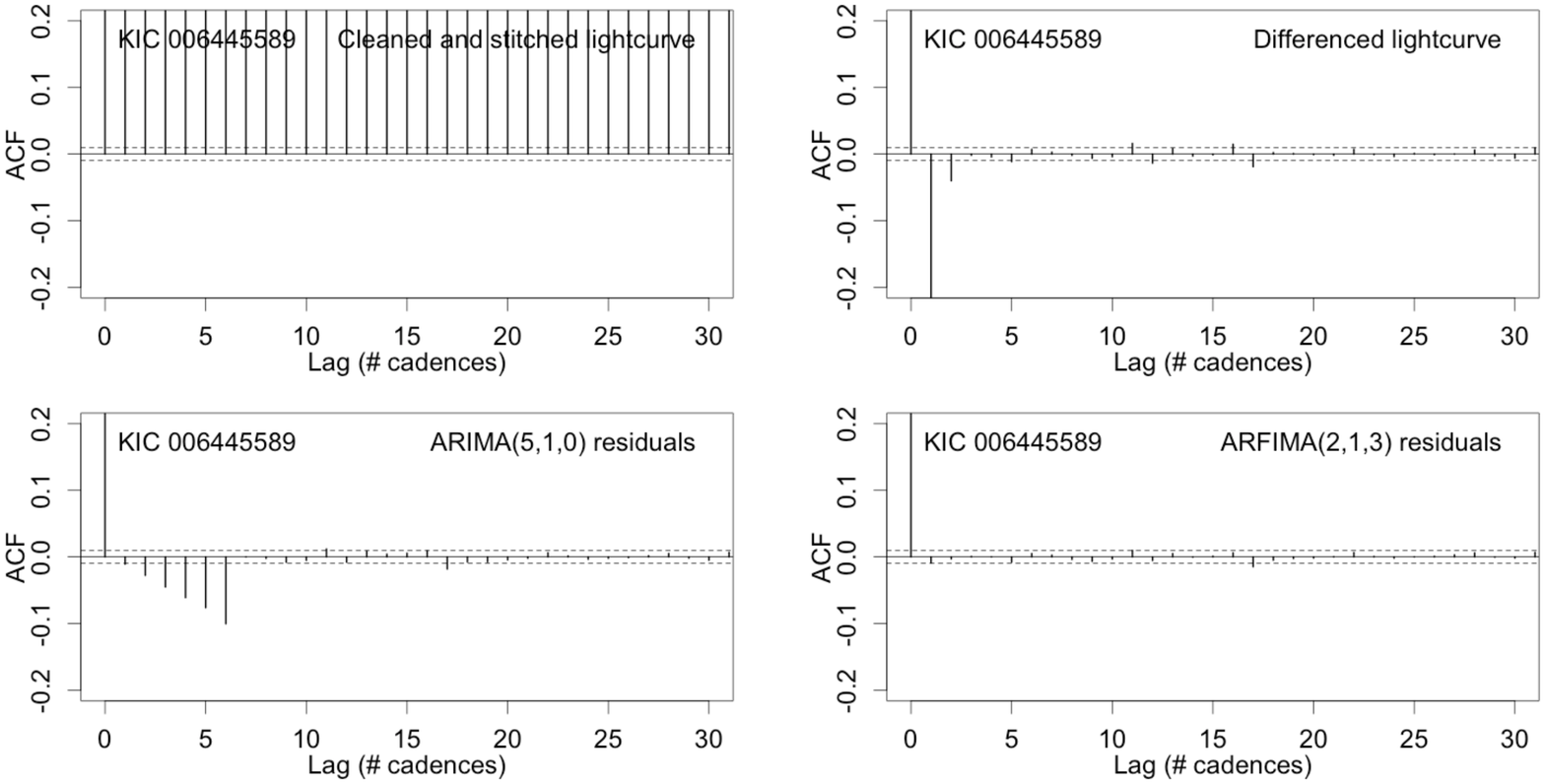}\vspace{-0.5in}
 \caption{Same as Figure~\ref{ARmod1.fig} for a red giant star, KIC 6445116. The ARIMA residual IQR of 35.7~elec~s$^{-1}$ corresponds to 250 ppm.}
 \label{ARmod3.fig}
 \end{figure}

\begin{figure}
 \centering
  \hspace*{-0.5in}   \includegraphics[width=0.9\linewidth,angle=-90]{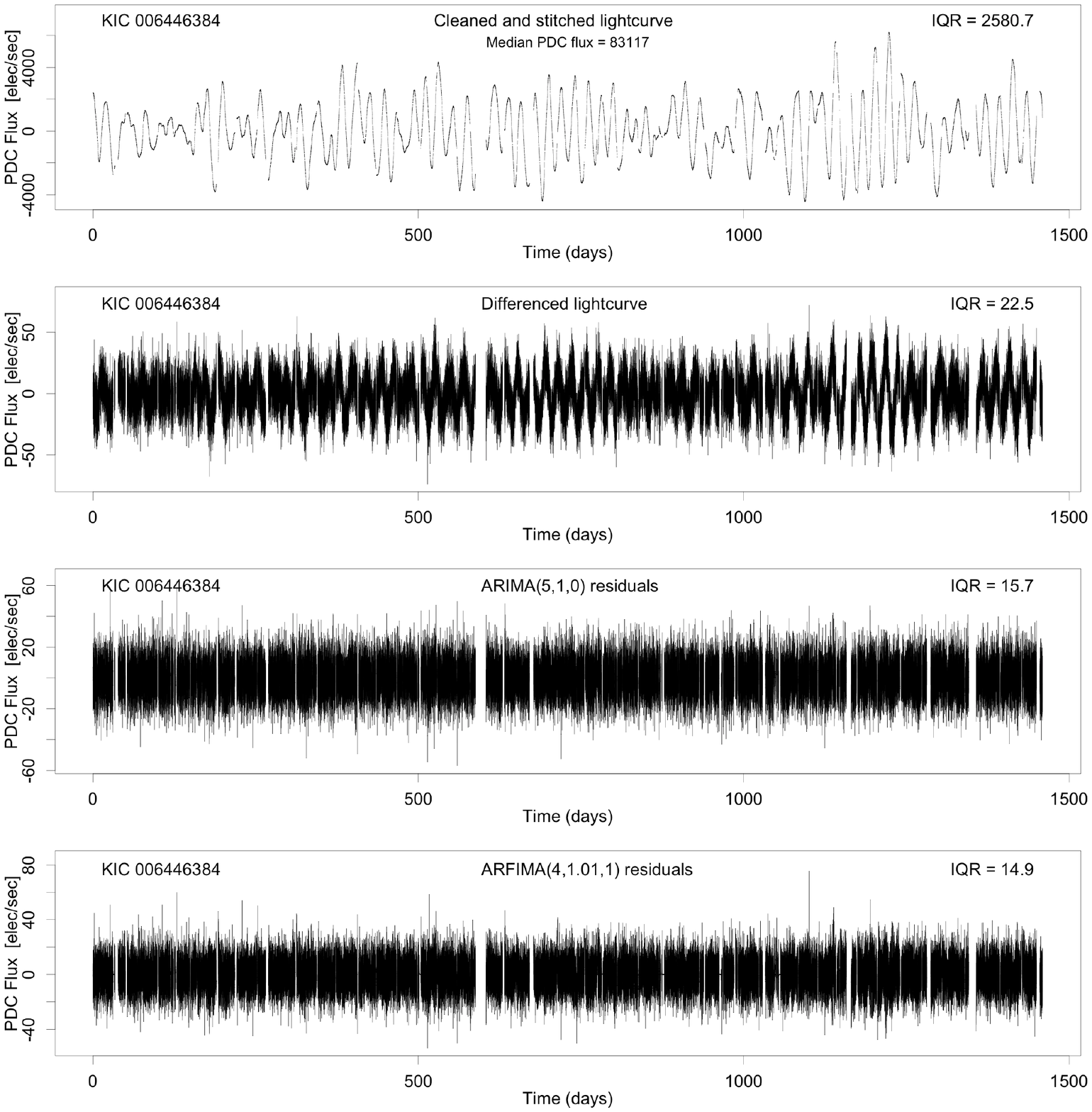}\vspace{-0.8in}
  \includegraphics[width=0.7\linewidth,angle=-90]{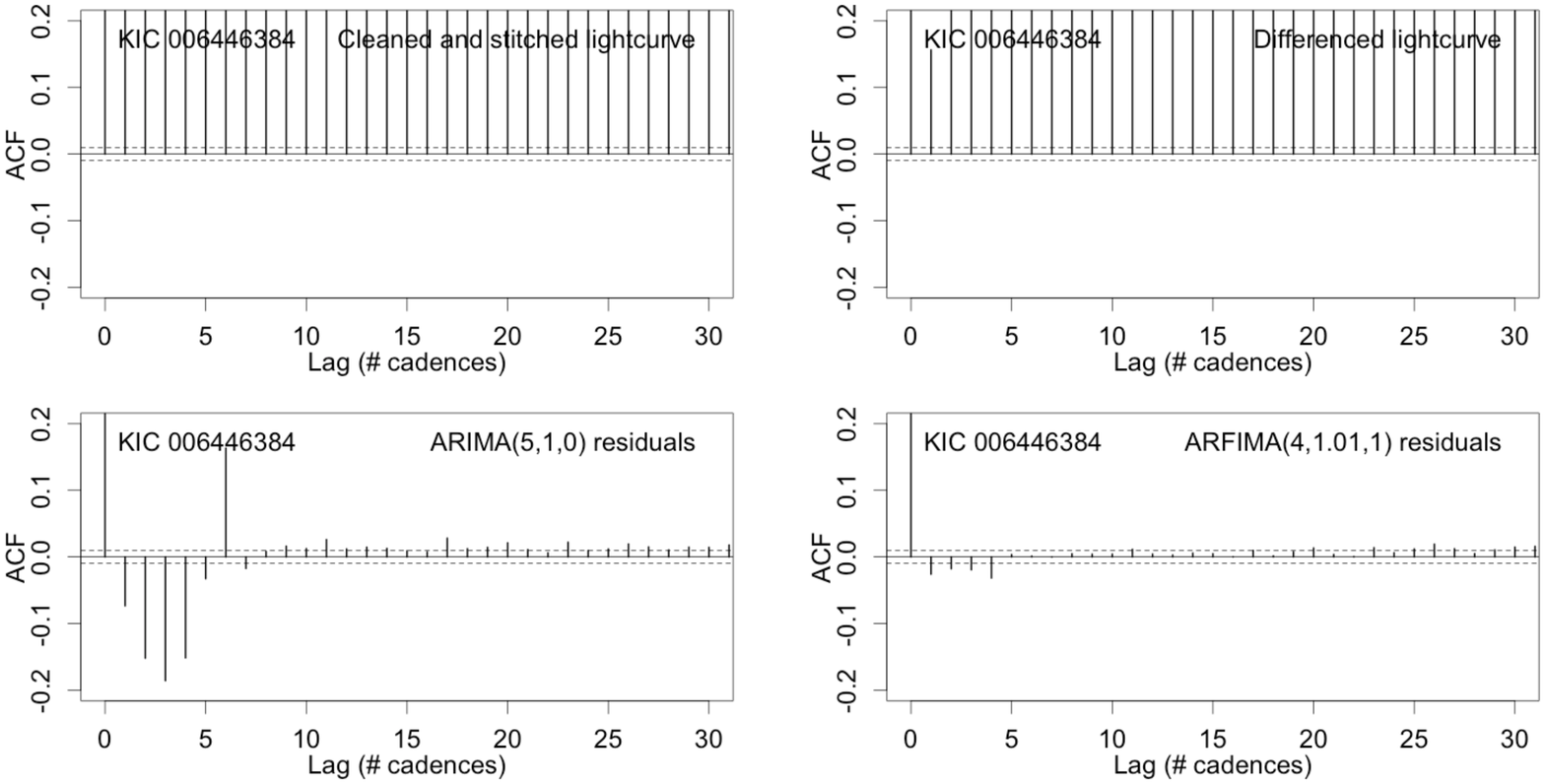}\vspace{-0.5in}
 \caption{Same as Figure~\ref{ARmod1.fig} for a high-amplitude variable star, KIC 6446384. The ARIMA residual IQR of 15.7~elec~s$^{-1}$ corresponds to 190 ppm.}
 \label{ARmod4.fig}
 \end{figure}

We find that autoregressive modeling can effectively reduce a wide variety of stellar behaviors displayed by Kepler light curves. Figures~\ref{ARmod1.fig}-\ref{ARmod4.fig} provide detailed views of light curves and autocorrelation functions (ACFs) for three typical cases: a quiet star, a magnetically acctive star with rotationally modulated starspots, a red giant, and a high-amplitude variable star. In each figure, we can see how ARIMA and ARFIMA do an excellent job at reducing both the variability (quantified with the IQR statistic) and the autocorrelation of each of these light curves.  Even light curves that appear by casual visual inspection to be white noise can be autocorrelated, as shown in Figure~\ref{ARmod1.fig}.

The ARIMA residuals' ACF often display some degree of autocorrelation, as see in Figures~\ref{ARmod1.fig}-\ref{ARmod4.fig}. We also find that for a large fraction of stars an ARIMA(5,1,0) model is selected as the best fit. This indicates to us that an optimal model might need higher (or different) orders than what was allowed by the current fitting approach. This problem did not occur for the ARFIMA modeling.  The cause of the non-optimal ARIMA model selection is our use of default settings of {\it auto.arima} that restrict the autoregressive and the moving average components to a maximum of five lags each, and that reduce the computational load with a step-wise algorithm \citep{Hyndman08}. For at least some of these objects, it may have been preferable to try all possible orders, and allowing for longer lags, rather than the subset tested by the step-wise algorithm.  We emphasize that this is a problem model selection, not model estimation. The coefficients for the selected model still correspond to a maximum-likelihood solution. 

In all four cases of Figures~\ref{ARmod1.fig}-\ref{ARmod4.fig}, and a large fraction of the full dataset, both the IQR noise level and the ACF values are lower for ARFIMA residuals than ARIMA residuals.  This is expected, as ARFIMA is a larger mathematical family that incorporates ARIMA within it.  Therefore, ARFIMA is preferred for scientific purposes relating to modeling the stellar variability.  However, in the majority of Kepler light curves we have explored, the TCF periodograms exhibits better performance when applied to ARIMA residuals rather than ARFIMA residuals. Figure~\ref{AR_ARF_peak.fig} discussed below shows how the TCF periodogram signal-to-noise of confirmed planetary transits is higher using ARIMA residuals rather than ARFIMA residuals. 

Two factors reduce the effectiveness of ARFIMA for planet discovery compared to ARIMA.   First, while it often outperforms ARIMA at modeling light curve variability, this enhancement can remove part of the transit signal; that is, it `overfits' the light curve for the goal of transit detection.  Second, due to the fractional differencing, the transit shape is distorted in a more complex way that TCF designed for periodic double-spike detection is no longer an optimal matched filter\footnote{We are grateful to Prof. Soumendra Lahiri (Statistics, NCSU) for raising this last point.}.  Therefore, although ARFIMA is most promising for modeling stellar variability, our analysis will mainly focus on ARIMA residuals since they appear better suited for our scientific goal of exoplanet discovery.

\subsection{Transit Comb Filter} \label{tcf.sec}

After Kepler light curves have been processed through ARIMA and ARFIMA modeling, we apply our TCF to the model residuals to search for signatures of a periodic exoplanetary transit signal. Since removing the stellar variability component affects the shape of the transit signal, the TCF algorithm presented in Paper I (\S3) matches a template of down-and-up spikes corresponding to the transit ingress and egress to the autoregressive residuals.  From geometric considerations alone, only a few percent of stars are expected to have transiting planets.  But other types of periodic variability can produce peaks in the TCF periodogram, most notably eclipsing binaries.  These are astronomical False Positives that we will seek to remove in the classification stage (\S\ref{RF.sec}). 

The TCF algorithm produces a periodogram of TCF power corresponding to how well the input signal (ARIMA/ARFIMA residuals) matches the template signal (periodic double-spike shape) at each of the tested periods.  The selection of periods where TCF power is calculated is discussed in Paper I (\S3.2).  We select here a minimum period $P_0 = 10$ cadences, corresponding to 294 minutes = 4.9 hours = 0.2 days for the Kepler 29.4 minute cadence,  and a maximum period around 500 days, corresponding to a minimum of 3 cycles during the 4 year span of the observations.  Specifically, we calculate the TCF periodogram at 558,854 periods with $P_{max} = 25,000.33$ cadences corresponding to $P_{max} = 511$ days.  As described in Paper I (\S3.4), the TCF procedure has lower sensitivity than the commonly used Box Least Squares algorithm as period increases (assuming circular orbits).

Two other quantities are required for the TCF periodogram calculation.  First, the width of the `teeth' of the TCF is set to range from 1 cadence at short periods to 5 cadences at long periods, accounting for the finite duration and misalignment of ingresses and egresses with respect to the cadence bins.  Second, durations tested at each period are set to range from a minimum of 15\% of the period to a maximum of 50 cadences ($\simeq 25$ hours).   

At each period, the algorithm gives the depth, duration, and phase that maximizes the TCF power, as well as a measure of the transit depth significance with respect to the overall variability of the folded and binned light curve.  As explained in Paper I (\S3.2), the transit parameters derived from the TCF calculation may only be rough estimates of the true properties of a physical planetary transit.

\subsubsection{Sample Periodograms} \label{periodograms.sec}

\begin{figure}
\centering
  \includegraphics[width=\textwidth]{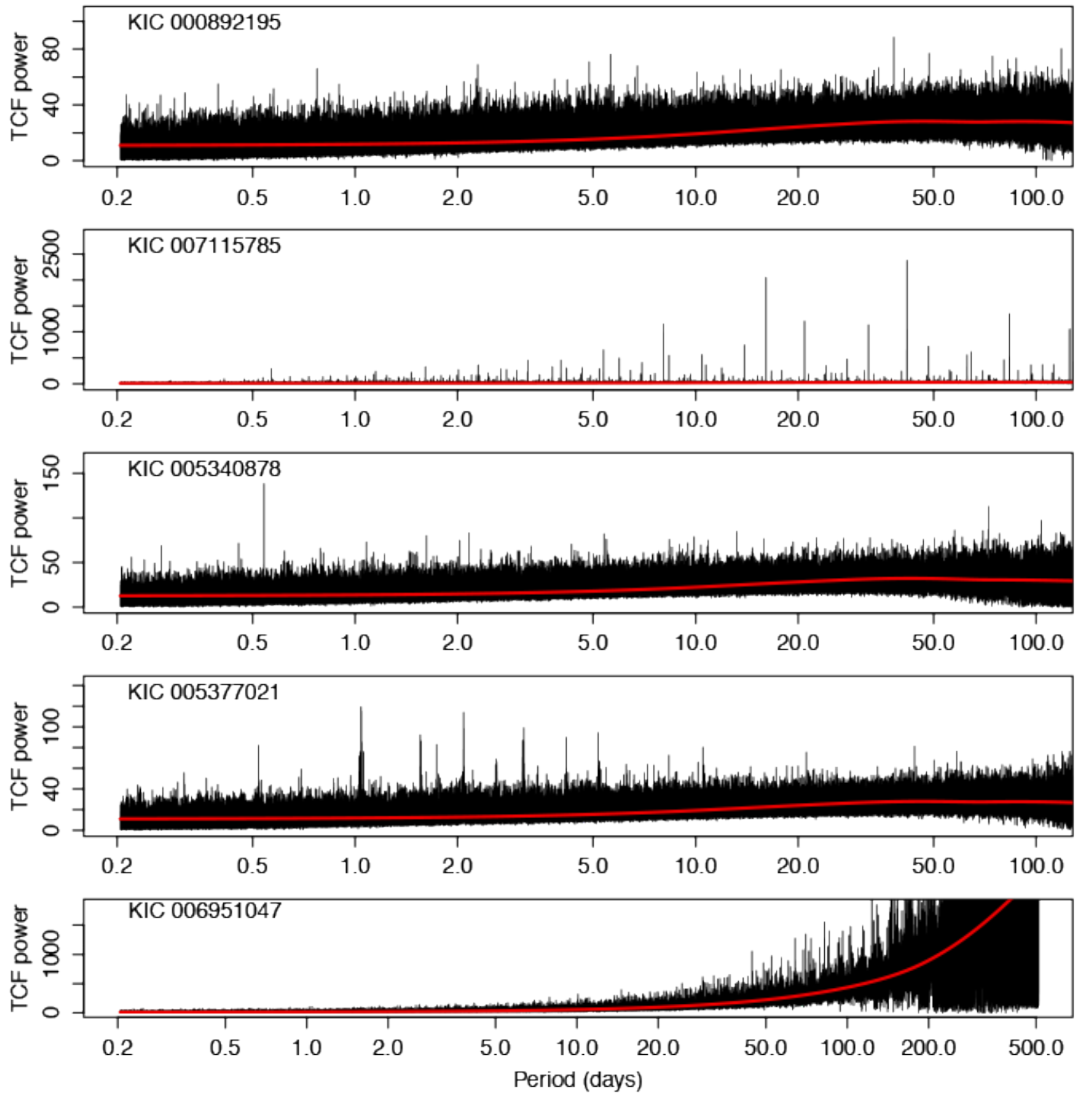}
\caption{Examples of TCF periodogram behaviors applied to ARIMA residuals. From top to bottom: clean periodogram with no clear signal of any type (KIC~005111742); strong KOI with many harmonics from a 2-planet system (KIC~007115785 = KOI~672 = Kepler~209); weak KOI with no discernible harmonics (KIC~005340878 = KOI~4199 = Kepler~1566); rapidly rotating spotted star (KIC~005377021); no significant transit signal but displaying high noise at long periods (KIC~006951047.  Note the extended range of the horizontal and vertical axes in the last panel. The red line shows a LOESS curve tracing the median of periodogram power values; this is used to remove large-scale trends in the periodogram.} \label{TCF_sample.fig}
\end{figure}

Figure~\ref{TCF_sample.fig} shows examples of the variety of periodogram behaviors observed in the Kepler stars. 

\begin{enumerate}

\item The great majority of stars observed by Kepler do not display significant periodic variability in the ARIMA residuals. A typical TCF periodogram of this type is shown in the top panel of Figure~\ref{TCF_sample.fig} showing only noise.  

\item The second panel exhibits the strong signal and harmonic structure created by an easily noticeable planetary transits. This is a confirmed 2-planet system with $P=16.1$ day (depth = 612 ppm) and 41.7 day (depth=1055 ppm) transits \citep{Rowe14}. Both of these transits produce prominent spikes in the periodogram, along with a complement of harmonics at shorter and longer periods.  

\item A more subtle planetary transit signal is displayed in the third panel for a much weaker KOI. \citet{SanchisOjeda14} identified this ultrashort period planet with $P=0.53$ days (depth = 96 ppm) with an estimated planetary radius of 0.7~$R_\earth$. 

\item The fourth panel displays another type of periodogram peaks, but of noticeably different structure. These correspond to quasi-periodicities from spots on a rapidly rotating star with a period of 1.02 days \citep{McQuillan14}.  Much of the this variability has been removed with the differencing operation within the ARIMA fitting procedure, but some signal remains in the ARIMA residuals. 

\item The last panel exhibits strongly increased periodogram noise at long periods. This excess noise is due alignments of stellar flares superposed on quasi-periodic variations on a strongly spotted star with rotational $P \simeq 9$ days \citep{Reinhold13}.  The TCF is particularly sensitive to chance alignments of sudden outliers of stellar or instrumental origin.  
\end{enumerate}

Many periodograms exhibit some amount of trend as period increases; \citet{Ofir14} describes this effect in BLS periodograms.  Therefore the highest power peak in a Kepler TCF periodogram is often a noise peak at long periods ($100 \lesssim P \lesssim 500$ day).  To more effectively capture astrophysically interesting periodicities, we reduce this effect by estimating the power of each periodogram value with respect to the local median of power values.  Local medians are estimated with a robust semi-parametric LOESS regression curve \citep{Cleveland94}, calculated using R's {\it loess} function, shown as red curves in Figure~\ref{TCF_sample.fig}.  The SNR of each TCF power value is calculated by subtracting the LOESS curve value at that period and dividing by the  median absolute deviation (MAD) within a window of 20,000 points around that period.  This window includes about 5\% of the periodogram periods. 

After determining SNR values for each of the 558,854 periods of the periodogram, the highest SNR peak is extracted as the selected value for possible exoplanet identification.  Although possible in future work, the current study does not perform iterative filtering to identify multi-planet systems.  Multiple planets can readily be seen by visual inspection in some systems, such as the second panel of Figure~\ref{TCF_sample.fig}. However, we collect the 250-strongest peaks to search if harmonics of the strongest peak are present among them. 

Finally, in the present study we restrict interest to TCF spectral peaks with $P<100$ day.  First, this reduces the effects of increased periodogram noise at long periods mentioned above.  Second, it recognizes the reduced sensitivity of TCF signals compared to BLS periods at longer transit durations, which typically is associated with longer periods.  This effect is described in the TCF $vs.$ BLS comparison in Paper I (\S3.4).  It is also seen in the right panel of Figure~\ref{ARPS_vs_missed_KOI.fig} discussed below, showing that the TCF best peak is more likely to miss the confirmed KOI period if $P \gtrsim 100$ day than for shorter periods. 

Therefore, after considerable experimentation, we have chosen to define the `best' period from a TCF periodogram to be the period with the highest TCF SNR after subtracting the LOESS median curve restricted to periods to $P< 100$.  

\subsubsection{Evaluating TCF Results with Kepler Objects of Interest} \label{Eval_KOI.sec}

To gauge the performance of the autoregressive modeling and TCF periodogram, we compare ARPS results with the list of DR25 Kepler Objects of Interest that have `Candidate' planet disposition \citep[KOIs][]{Thompson18}.  Of the 156,717 objects that completed ARPS processing, 2,529 are KOI Candidates. For this sample, we assess recovery if the ARPS TCF `best peak' (restricted to $P<100$ days) matches the Kepler team period (or the 2- or 1/2-times harmonic) to within 1\%.   More details on the comparison of ARPS and Kepler team findings for KOIs appear in the Appendix.

These results are plotted in Figure~\ref{ARPS_vs_KOI.fig}. When a star has multiple planet candidates, the TCF period was compared with the periods of all planets and at least one match was required.  Considering the full sample of 2,529 KOI candidates, 1,910 (76\%) have a matching period in the ARPS analysis.  For the periods that match, transit depths are well-correlated although the TCF estimate is typically $\sim$25\% lower than the KOI value.   

\begin{figure}
\centering
  \includegraphics[width=\textwidth]{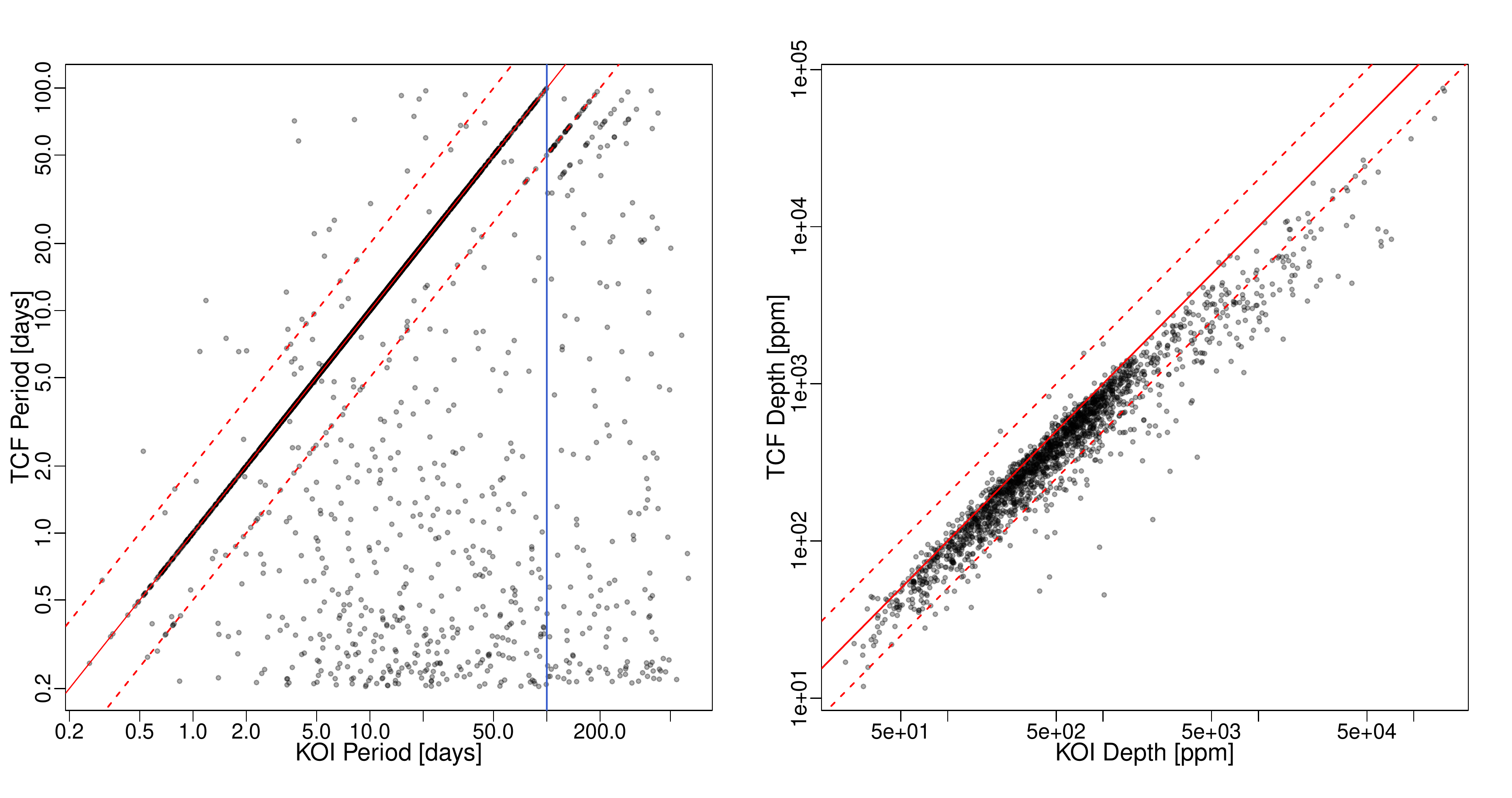}
\caption{Comparison of ARPS and Kepler pipeline periods KOIs. {\it Left:} TCF period compared with KOI periods for all KOIs.  The solid red line shows one-to-one correspondence, the two dotted red lines show the 2- and 1/2-times harmonics, and the vertical blue line shows 100 day where ARPS periods are truncated.  {\it Right:} Comparison between TCF depth and KOI depth for the matching periods.}
\label{ARPS_vs_KOI.fig}
\end{figure}

\begin{figure}
\centering
  \includegraphics[width=\textwidth]{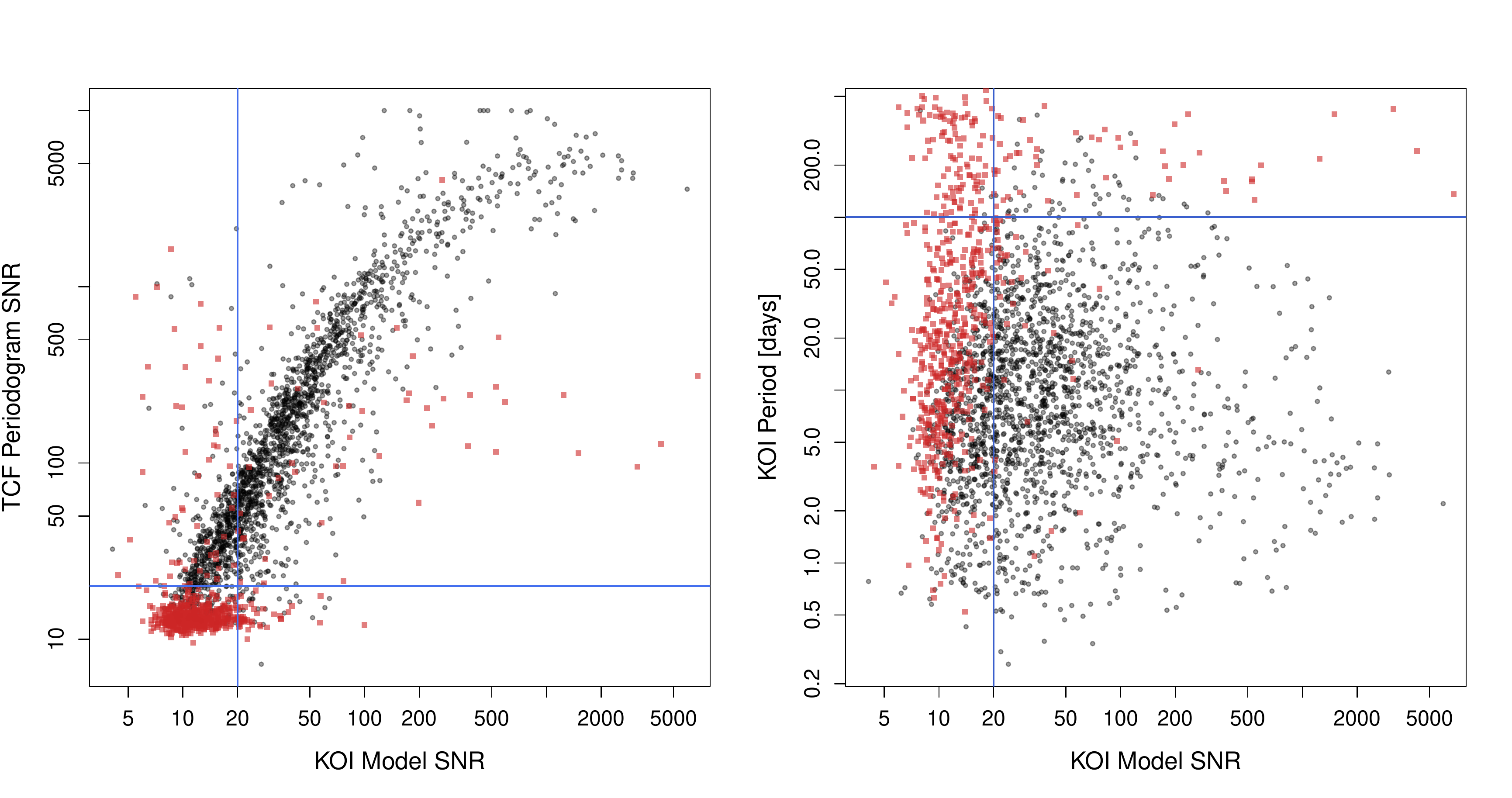}
\caption{Exploration of matching and missed KOIs.  Black circles correspond to ARPS and Kepler matching periods and red squares to missed periods. {\it Left:} Comparison of TCF peak SNR and KOI ModelSNR. {\it Right:} KOI period and KOI SNR.}
\label{ARPS_vs_missed_KOI.fig}
\end{figure}

Figure~\ref{ARPS_vs_missed_KOI.fig} shows that nearly all of the KOIs with periods missed by ARPS have weak planetary signals.   Here we use the Kepler team's measure of transit significance measure called `Model FIT SNR' that is one of the top three attributes in their Random Forest classifier \citep{McCauliff15}.  Examination of their RF classifier outcomes (Figure 5 in McCaulliff et al.) suggests that $ModelSNR<20$ stars produce decision tree splits with relatively weak discriminatory power between true transits and False Positive populations.

When considering only KOIs with $ModelSNR> 20$, 94\% of the 1,454 stars are recovered with ARPS.  If we add the criterion that the period period is less than 100 days, then 97\% of 1,349 KOI candidates are recovered.   

We conclude that the ARPS procedure captures nearly all of the Kepler Transit Search Pipeline confirmed planetary candidates providing the transit is not too weak and the period is not too long. This success is gratifying as the mathematical foundations and detailed procedures for identifying planets in ARPS and the Kepler pipeline are very different.  It is not clear which  failures of ARPS to confirm planets with KOI Model SNR$<$20 are due to False Alarms in the Kepler pipeline analysis, and which failures arise for true planets where the Kepler pipeline is simply more sensitive to these planets than ARPS. We suspect that both cases are present.  In any case, an imperfect overlap between the two methods for smaller planets is desirable, as the ARPS method must differ from the Kepler pipeline if it is to discover new planets.

\begin{figure}
\centering
  \includegraphics[width=0.7\textwidth]{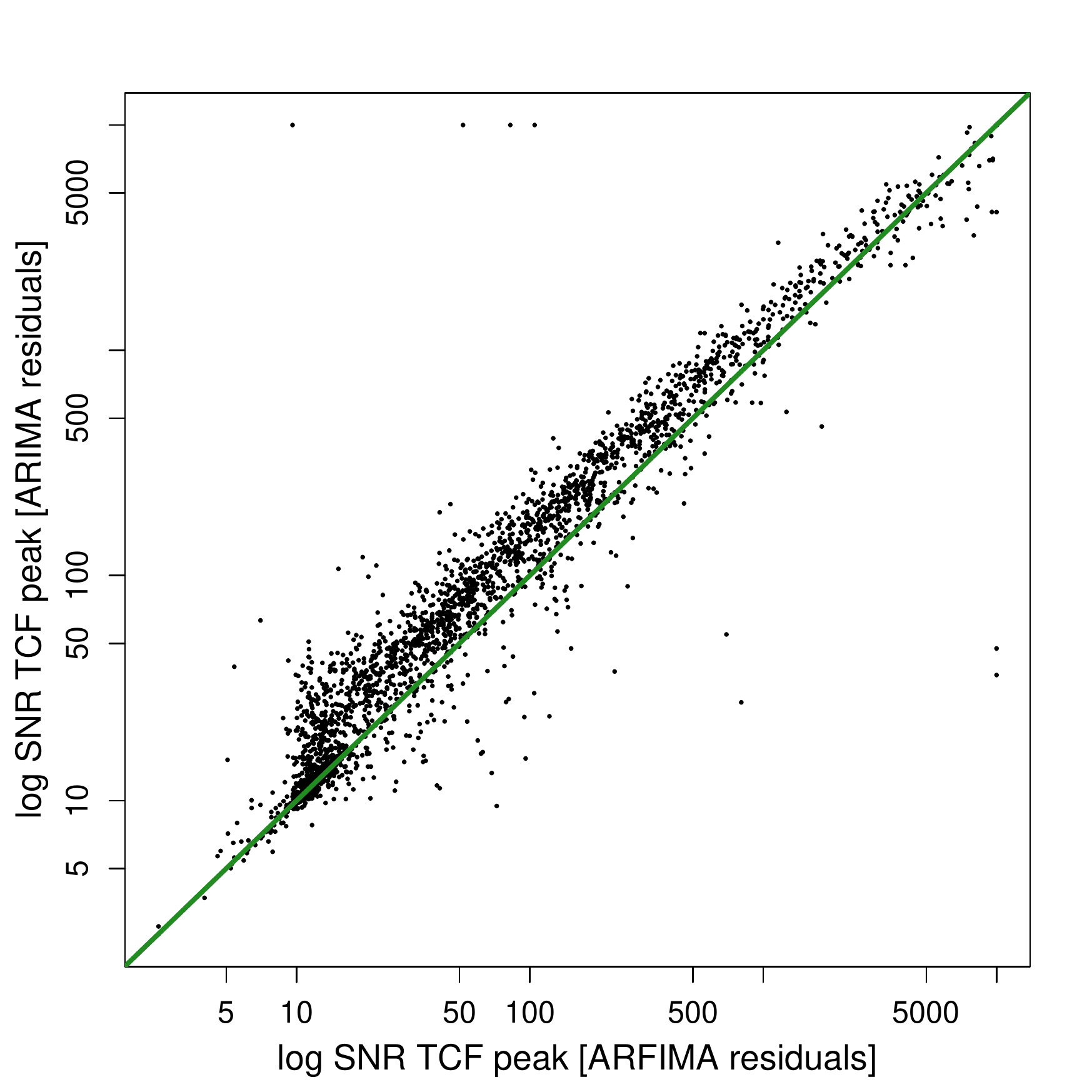}
\caption{Comparison of TCF periodogram power of ARIMA and ARFIMA residuals for known KOI candidates. The green line shows equal signal-to-noise values.}
\label{AR_ARF_peak.fig}
\end{figure}

A distinctive feature in the right panel of Figure~\ref{ARPS_vs_KOI.fig} and left panel of Figure~\ref{ARPS_vs_missed_KOI.fig} is a curvature at high depth and SNR: TCF does not capture the entire signal for for the strongest planetary transits.  Figure~\ref{AR_ARF_peak.fig} shows that the best-period SNR in TCF periodograms found from ARFIMA residuals found to be consistently stronger that those found from ARIMA residuals, except for the strongest planetary signals. We believe this is an indication that the more elaborate nonlinear ARFIMA model `overfits' the light curves when the transits are extremely strong, partially incorporating the planetary signal into the autoregressive model for the stellar variations and thereby weakening their signal in the model residuals (\S\ref{armod.sec}).  The more flexible family of ARFIMA model is more effective than the ARIMA models in absorbing some of the planetary signal.  If we restrict consideration to ARIMA residual, this behavior affects only very large planets and has little importance for our science goals of discovering new weak-transit planets.

\subsection{ARIMAX with Box Model} \label{ARIMAX.sec}

As described in Paper I (\S3.3), we now revisit to autoregressive modeling after identification of a promising transit-like periodicity as the strongest peak TCF SNR after TCF trend removal.  Here a simplified box-shaped signal is added as an `exogeneous' variable to an ARIMA model; this is known as an ARIMAX model \citep{Hyndman14}. After transit period, duration and phase are specified from the strongest peak in the TCF periodogram, the transit depth becomes the only new parameter added to the autoregressive model.  The maximum likelihood estimation then jointly models the transit depth and the autocorrelated noise, giving a best-fit value and confidence band for the depth; an estimate of the depth SNR thereby emerges.  This contrasts with the earlier ARPS procedure where the aperiodic (ARIMA) and  periodic (TCF) component  were estimated sequentially rather than jointly.  As with TCF, this step it is not expected to be a perfect representation of the transit, as it uses a simple box model rather than an astrophysically motivated model \citep{Mandel02}.   

The ARIMAX analysis is easily implemented using the {\it auto.arima} function in the {\it forecast} CRAN package by including the {\it xreg} argument corresponding to `external regressors' with an indicator vector consisting of zeros (out of transit) and ones (in transit) for the full light curve. 

\begin{figure}
\centering
  \includegraphics[width=\textwidth]{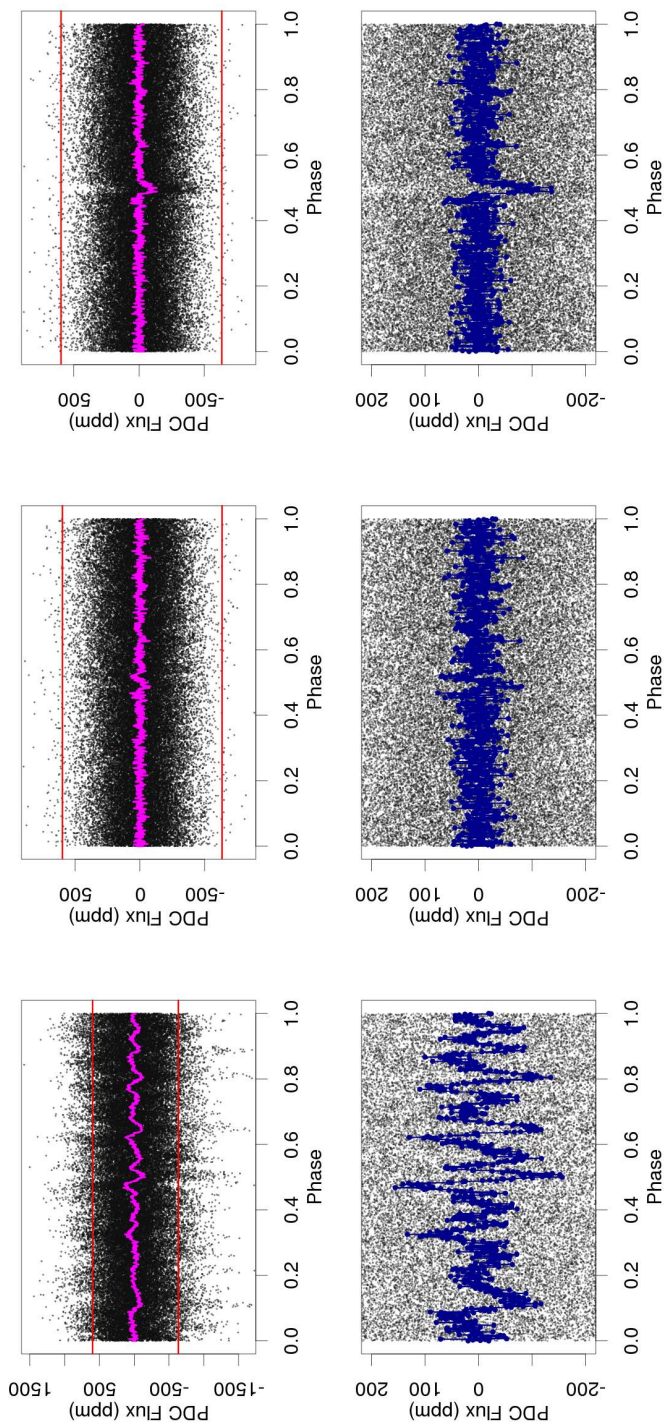}
\caption{ Example of ARIMAX modeling for KIC 008573332 (= KACT 69 in Table~\ref{KACT.tbl} below), one of our reported new planetary candidates. Red lines label the same flux values in all plots to guide the eye.  In the top row, the panels show: the original stitched light curve folded with the best TCF period (left); residuals from ARIMA modeling showing a double-spike transit (center); and residual from ARIMAX modeling showing a box-shaped transit (right). The magenta line shows the smoothed data after averaging points into one-cadence bins.  The red lines are the same in all panels to guide the eye.  The bottom row is the same as the top row with an expanded scale for flux. }
\label{arimax.fig}
\end{figure}

An additional advantage of this step is that the residuals of the ARIMAX model permit a useful visualization of a box-shaped (rather double-spike) transit on a folded light curve with autoregressive noise removed.  Having two separate components for the model, we can simply subtract the autoregressive contribution to create a `cleaned' light curve and superpose the exogeneous box contribution.  This is illustrated in Figure~\ref{arimax.fig}; the inclusion of a periodic box as an external variable (right panel) maintains the advantages of ARIMA modeling (middle panel) while also preserving the expected transit shape. 

The signal-to-noise ratio derived using the ARIMAX box depth and error is distinct and complementary to the TCF periodogram SNR, although they are typically highly correlated when a true transit is present.  The depth SNR shows the detectability of the in-transit points relative to the out-of-transit photometry, while the periodogram SNR helps address the uniqueness of a particular signal compared to other periodogram values. Taking both of these values into account can help differentiate interesting signals from background sources.   For light curves without transits, the ARIMAX box depth is highly variable and uncorrelated with TCF properties.  In such cases, it can have the incorrect sign; that is, a box-shaped rise rather than dip is obtained.  In the classification stage (\S\ref{RF.sec}), we find that the ARIMAX box depth SNR is the single best variable for discriminating true transits from false alarms and false positives.  

\subsection{Random Forest Classification} \label{RF.sec}

The development of a Random Forest classifier for planet detection with False Positive reduction is outlined in Paper I (\S4).  We perform the classification with the {\it randomForest} function from the CRAN package {\it randomForest}  \citep{Liaw02}. ROC curves of RF results are plotted and evaluated with the {\it roc} function provided in CRAN package {\it pROC} \citep{Robin11}.  After experimentation, we chose to grow a relatively large number of trees, $ntree = 2, 500$ (where the code default is 500 trees), and use $mtry = 7$ features at each split, slightly larger than the code default of the square-root of the total number of features. 

\subsubsection{Feature Selection} 
From the ARIMA modeling, TCF periodograms, and ARIMAX modeling, we can generate a suite of properties of each Kepler light curve that focuses on selecting stars with exoplanetary transits (True Positives, TPs) and rejects other stars (False Positives, FPs) as well as spurious indications of periodicity (False Alarms, FAs).   Statistically, this is a binary classification problem.  As discussed in Paper I (\S4.2), Random Forest (RF) is an extremely effective and widely used machine learning algorithm  for a wide variety of multivariate regression and classification problems \citep{Breiman01, Cutler12}. Receiver Operating Characteristic  (ROC) curves and diagnostic tests on the confusion matrix are then used to compare and validate trial RF classifiers with different input variables and training sets.  Scientific goals guide the final decision criteria for classification.  A major methodological difficulty is the extreme `imbalance' in this problem; the FPs outnumber the TPs by a factor of $50-100$.   This issue is discussed in Paper I (\S4.3).

We performed a number of experiments to develop a strong list of features for the RF classification, particularly to add features to help remove different types of FPs from the classifier.  Inclusion of many features should help us reject stars with many different characteristics, including several types of non-planetary periodic behaviors such as pulsating stars, multiply periodic red giant stars, rotationally modulated starspots, mutually illuminated binaries, flare stars, and blended eclipsing binaries.  We purposefully include characteristics from different stages of the analysis (host star, stitched light curve, ARIMA modeling, TCF periodogram, `best' period folded light curve, and ARIMAX modeling), and features specifically designed to capture classes of FPs such as eclipsing binaries and flare stars.  The RF algorithm benefits from the inclusion of a wide selection of features that may contribute even slightly towards improving the classification.   Decision trees can struggle to capture relationships involving combinations of features \citep{Cutler12}; this is ameliorated by including quantities such as the ratios of features already present in the classifier.  Finally, we include characteristics of the noise, as well as the putative transit signal, in the full light curve, transit dips, and the periodogram.  

\begin{deluxetable}{ll} 
\tabletypesize{\footnotesize}
\tablewidth{0pt}
\centering
\tablecaption{Features for the Random Forest Classifier}
\startdata 
& \\ [-0.05in]
\multicolumn{2}{c}{\bf Stellar properties} \\ [-0.05in] 
ST\_Kepmag & Kepler magnitude \\ [-0.05in]
ST\_Teff & Effective temperature ($^\circ$K)\\ [-0.05in]
ST\_logg & Surface gravity (log cgs)\\ [-0.05in]
ST\_rad & Radius (R$_\odot$) \\ [-0.05in]
ST\_EB & Indicator for star in Eclipsing Binary Catalog \\ \hline
\multicolumn{2}{c}{\bf Light curve properties} \\ [-0.05in] 
LC\_Npts    & Number of points omitting NAs \\ [-0.05in]
LC\_med   & Median of unstitched PDC light curve (elec s$^{-1}$) \\ [-0.05in]
LC\_IQR  & InterQuartile Range of stitched light curve \\ [-0.05in]
LC\_POM & Positive Outlier Measure of stitched light curve \\  [-0.05in]
LC\_impr & IQR improvement between stitched light curve to ARIMA residuals \\ \hline
\multicolumn{2}{c}{\bf ARIMA residuals properties} \\  [-0.05in] 
AR\_DW	& Probability of Durbin-Watson test for autocorrelation at lag=1 \\ [-0.05in]
AR\_BG	 & Probability of Breusch-€"Godfrey test for autocorrelation at lag=5 \\ [-0.05in]
AR\_IQR	& InterQuartile Range \\ \hline
\multicolumn{2}{c}{\bf Transit Comb Filter periodogram: Best peak properties\tablenotemark{a}} \\ [-0.05in]
TCF\_SNR    & Best peak SNR  \\ [-0.05in]
TCF\_shp	     & Best peak shape statistic   \\ [-0.05in]
TCF\_dep	     & Best peak depth \\ [-0.05in]
TCF\_har      & Best peak harmonics indicator   \\ [-0.05in]
TCF\_lsratio  & Ratio of best period in `full' ($0.2<P<511$ day) vs. `short' \\ [-0.05in]
                      & ($0.2<P<100$ day) periodograms\\ [-0.05in]
TCF\_ARF     & Ratio of best period in ARIMA vs. ARFIMA residuals  \\ [-0.05in]
TCF\_dupkoi  & Indicator for duplicate periods in KOI list \\ \hline
\multicolumn{2}{c}{\bf Transit Comb Filter periodogram: 250 peaks properties\tablenotemark{a}\tablenotemark{b}} \\  [-0.05in] 
GPS\_mnsnr & Mean peak SNRs  \\ [-0.05in]
GPS\_sdsnr  & Standard deviation of peak SNRs \\ [-0.05in]
GPS\_mnpow & Mean power  \\ [-0.05in]
GPS\_sdpow & Standard deviation of power \\ [-0.05in]
GPS\_sdper & Standard deviation of periods \\ [-0.05in]
GPS\_dupct   & Number of duplicate periods \\ [-0.05in]
GPF\_mnsnr & Mean peak SNRs  \\ [-0.05in]
GPF\_sdsnr  & Standard deviation of peak SNRs \\ [-0.05in]
GPF\_mnpow & Mean power  \\ [-0.05in]
GPF\_sdpow & Standard deviation of power \\ [-0.05in]
GPF\_sdper & Standard deviation of periods \\ \hline 
\multicolumn{2}{c}{\bf Folded light curve properties} \\  [-0.05in] 
ARbox\_SNR      & Box depth SNR from ARIMAX model, $P<100$ day \\ [-0.05in]
Fold\_EvSNR     & Mean SNR of even transit depths \\  [-0.05in]
Fold\_OddSNR  & Mean SNR of odd transit depths \\  [-0.05in]
Fold\_EvOddt     & $t$-test p-value for even vs. odd transit depth SNRs \\ [-0.05in]
Fold\_TransSNR & Mean SNR of all transits \\ [-0.05in]
Fold\_Transt       & $t$-test p-value for all in-transit vs. out-of-transit depths \\
\enddata
\tablenotetext{a}{LOESS trend removed }
\tablenotetext{b}{GPS refers to the `short' periodogram with $0.2 < P < 100$ days.  GPF refers to the `full' periodogram with  $0.2<P<511$ days.}
\label{RFfeatures.tbl}
\end{deluxetable}

Our final selection of features is listed in Table~\ref{RFfeatures.tbl} and are grouped as follows:
\begin{description}

\item[Stellar properties] Basic properties of the Kepler stars are provided by \citet{KSPWG16}.  We include the effective temperature, surface gravity, radius, and Kepler magnitude. Stellar mass is omitted because it is not available for $\sim 3,000$ stars.  A binary flag indicates whether a star appears in the Eclipsing Binary catalog of \citep{Kirk16}.

\item[Stitched light curve] Three statistics capture some of the information contained in the original light curve: IQR, median PDC flux (in electrons s$^{-1}$ obtained prior to stitching), and the cadence count of non-missing values.  Our Positive Outlier Measure seeks to identify the small fraction of stars with unusually strong flares; these produce spurious peaks in the TCF periodogram when then align by chance at long periods.  Finally we select a measure of how much the noise improved before and after ARIMA modeling using the ratio of IQR values.  A large ratio indicates that the autoregressive modeling was highly effective in reducing stellar variability.  

\item[ARIMA modeling]  This includes the noise level of the ARIMA residuals, and two measures of residual autocorrelation: the log of the probabilities of the Durbin-Watson and Breusch-Godfrey (lag=5) tests.  Small probabilities indicate that autocorrelation is still present in the ARIMA residuals. 

\item[TCF periodogram] A considerable number of features are drawn from the TCF periodogram.  Several refer to the `best' spectral peak defined as the period with the highest signal-to-noise ratio with respect to local periodogram noise after subtraction of the LOESS trend curve. These include the dimensionless SNR, transit depth (in ppm), a shape statistic of the folded light curve (defined as the depth divided by a binned standard deviation), and a 4-valued indicator of the presence of harmonics to the best peak among the 250 strongest peaks.  We avoid including period and duration because our experiments show this biases predictions against very short periods due to the paucity of ultrashort candidates available in the DR25 Gold training set (\S\ref{training.sec}).  To avoid problems with rising periodogram noise at long periods, all of these values are obtained from the `short' periodogram truncated to periods under 100 days. 

Three additional features refer to the `best' TCF peak.  One is the ratio of best period in the full ($0.2<P<511$ day) vs. short ($0.2<P<100$ day) periodograms; if this ratio is unity, then there is increased confidence a true periodicity is present.  Another is the ratio of best period obtained from the TCF of the ARIMA vs. ARFIMA residuals ($0.2<P<100$ day).  Again a value of unity lends supports that the peak represents a true periodicity.  Finally, an indicator tells whether this period is found repeatedly in the Kepler Team KOI lists, suggesting it is a spurious effect from overlapping signals of different stars in the detector \citep{Coughlin14}.  

Ten additional features are included to characterize the ensemble of the strongest (in SNR) 250 peaks of the ARIMA-residual TCF periodogram, not just the `best' peak.  These are the standard deviation of the periods, mean power, standard deviation of the power, mean SNR, and standard deviation of the SNR.  These features are provided both for the full periodogram with $0.2<P<511$ days and the periodogram truncated at $P<100$ days.  This is helpful in promoting periodograms with strong peaks and harmonics, and demoting light curves with outliers that produce exceedingly noisy periodograms at long periods.  Finally, two features flag cases where multiple stars near to each other in the sky have identifical periods.  These spurious cases, known as `ephemeris matches', arise from overlapping point spread functions in the Kepler detector \citep{Coughlin14}. 

\item[Folded light curve]  Six features associated with the light curve folded at the `best' period are developed to help reduce FAs and FPs.  The most important is the SNR of the transit depth in the ARIMAX model fit to the best period; this emerged as the most important feature in the Random Forest classifier.  Three features measure differences between alternate even-odd transits.  This is a well-known technique of reducing FPs from eclipsing binaries where the primary and secondary eclipses often have different depths.  Finally, two measures are provided comparing the fluxes of points inside and outside the transits, as defined by the ARIMAX box fits. This helps to remove cases where extremely non-Gaussian noise in the light curve produces a false peak in the periodogram.  

\end{description}

\begin{figure}
\centering
  \includegraphics[width=0.75\textwidth]{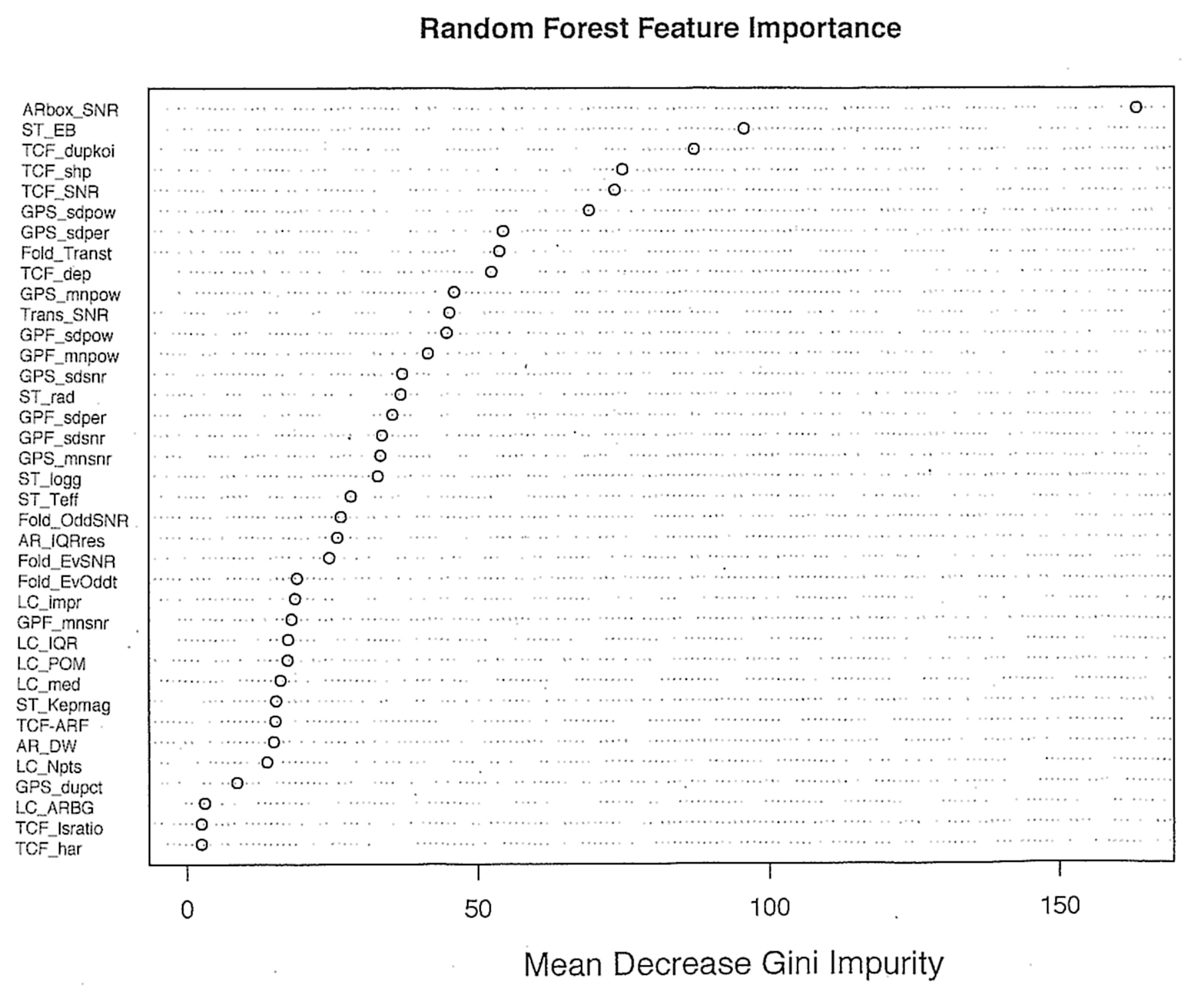} 
\caption{Sequence of feature importance based on the mean reduction in Gini impurity for the Random Forest classifier applied to the Kepler training sets.} 
\label{training.fig}
\end{figure}

A comparison of feature importance resulting from the `final' RF classifier applied to the `final' training sets (\S\ref{training.sec}) is shown in Figure~\ref{training.fig}.   The quantity plotted is the mean decrease in Gini impurity contributed by each feature in the model \citep{Breiman01}.  The traditional univariate measure to find periodicities $-$ the periodogram peak SNR (feature labeled {\it TCF\_SNR}) $-$ is only the fifth most important variable.  More important features are: the SNR of the box depth fitted in the ARIMAX model ({\it ARbox\_SNR}); the binary flag that points away from the Eclipsing Binary Catalog ({\it is\_EB}); the shape statistic of the folded light curve measuring the duty cycle of the periodic variability({\it TCF\_shp}); and the flag that points away from  duplicated periods due to overlapping point spread functions in the Kepler detector({\it dup\_is\_koi}).  Two characteristics of the top-250 peaks in the periodogram ({\it GPS\_sdpow} and {\it GPS\_sdper}) and a measure of noise within transit flux values ({\it Fold\_Transt}) are next in importance.   It is nontrivial to distinguish individual effects from interactions between features, and features important for predictive success do not necessarily indicate importance for physical interpretation \citep{Genuer10}. 

\subsubsection{Training Set Construction} \label{training.sec} 

Given the complexity of this classification problem, strong training sets are needed.  For the TP  training set, we start with the DR25 KOI stars with disposition 'Planet Candidate' provided by the Kepler team \citep{Thompson18}.  This is based on a complicated collection of tests and criteria (nicknamed `Robovetter') to identify candidate planets. To assess the confidence of the Robovetter's decision, inputs are randomly varied and then a `disposition score' is assigned corresponding to the fraction of times the Robovetter assigned the disposition `Planet Candidate'.  Following the suggestion of Thompson et al., we consider KOI Planet Candidates that have a disposition score of at least 0.7.  We restrict the TP training set in two additional ways: only 1,910 of KOI candidates which had a matching TCF period (\S\ref{Eval_KOI.sec}) are considered, and objects in the Certified False Positive Table are omitted.  The latter include FAs without discernible periodic signal and FP periodic signals such as eclipsing binaries. 

Altogether, 11,669 stars were removed as possible members of training sets to avoid biasing the classifier for or against being a transit. These include:
\begin{enumerate}
\item TCEs which were not promoted to KOIs
\item KOIs from previous data releases with were not included in the DR25 list 
\item DR25 Candidates with a Robovetter score less than 0.7
\item DR25 Candidates where the TCF period does not match the KOI period
\item DR25 FPs which were not on the Certified False Positive list
\end{enumerate}

For the True Negative training set, we start with the 156,717 Kepler stars that are successfully processed through the ARPS pipeline.  All stars with possible periodicities from any cause are omitted.  This is the union of $\sim 7,000$ stars in the DR25 KOI catalog, the $\sim 17,000$ stars in the TCE listing, and $\sim 3,000$ stars in the eclipsing binary catalog (\S\ref{data.sec})\footnote{Noting that stars in the Eclipsing Binary catalog are omitted from both the True Positive and True Negative training sets, we experimented with RFs treating eclipsing binaries as a third training set. No improvement in classifier performance was found.}. 

The RF algorithm uses the `out-of-bag' (OOB) sample for predictions where each tree only makes predictions on samples that were not used for training. All 2,500 trees will make a prediction for stars omitted from the training sets, and different trees make predictions for different stars included in the training sets.  The OOB objects in each tree permit a global validation of the RF classification without segregating a validation subset from the training process \citep[][\S8.2]{James13}.

When a strong imbalance is present in the sizes of the training sets, biases often emerge in the learning process and prediction from machine learning algorithms \citep{Chawla02, Liu09}. Common techniques to overcome these problems involve giving a higher weight to rare samples, over-sampling the minority class, and under-sampling the dominant class \citep{Weiss04}. A key reason for our choice to use Random Forests for classification is how easily imbalanced classes can be taken into account. The approach of RF is to under-sample the larger True Negative class in the construction of individual trees with balanced training sets, but cover the full range of parameters in the ensemble of trees constituting the forest.  \citet{Chen04} discuss the application of Random Forest to imbalanced data. 

\subsubsection{Classifier Evaluation and Comparison} \label{ROC.sec}

The merits of a probabilistic (`soft') classifier like RF with two input classes is evaluated in two ways:  the ROC curve (plot of TP rate $vs.$ FP rate for a range of classifier probability thresholds), and scalar statistics based on the $2 \times 2$ `confusion matrix' of TP $vs.$ FP once a critical threshold is chosen.   As described in \S\ref{training.sec}, the training set's OOB estimates can be used to assess model performance without the need for additional cross-validation or a holdout set. The ROC curves and AUC values presented here are calculated using these out-of-sample predicted probabilities. 

We have examined the merits of RF classifiers with: different choices of sample sizes for the exoplanet (TP) and non-exoplanet (TN) training sets to overcome dataset imbalances; different treatments of Certified False Positives in the training sets (mostly eclipsing binaries); and different choices of features used for classification.  Some of the results are summarized in Table~\ref{RF_options.tbl}.  The Area Under Curve (AUC) using all the training data always exceeds 99.6\% and varies little between the various classifier runs; this is due to the ease of classifying light curves without periodic structure and with strong transit signals.  More contrast in classifier performance is seen when the TN sample contains only Certified False Positives, many of which are eclipsing binaries.  While most choices are good, we have chosen as the `final' RF classifier (first line of Table~\ref{RF_options.tbl}) one that uses 37 features (Table~\ref{RFfeatures.tbl}) and balanced training sets with 1400 TP DR25 Gold Planet Candidates, and 1400 TNs of which half are Certified False Positives.  

\begin{deluxetable}{rrrrrll}
\tablecaption{Training Set and Feature Trials for the Random Forest Classifier} 
\tabletypesize{\footnotesize}
\centering
\tablehead{  
\colhead{Model} &  \colhead{N$_{pos}$} & \colhead{N$_{neg}$} & \colhead{$N_{FP}$} & \colhead{N$_{feat}$} & 
\colhead{AUC$_{all}$} & \colhead{AUC$_{CFP}$}
}
\startdata
  Final & 1,400 & 1,400 & 700 & 37~~ & 0.9979 & 0.9229 \\   \hline
  1 & 1,000 & 1,000  & 0 & 71~~ & 0.9969 & 0.8542 \\ 
  2  & 1,000 & 2,000 & 0 & 71~~ & 0.9969 & 0.8628 \\ 
  3  & 1,000 & 5,000  & 0 & 71~~ & 0.9972 & 0.8753 \\   \hline
  4 & 1,000 & 1,000 & 500 & 71~~ & 0.9964 & 0.9198 \\  
  5 & 1,000 & 2,000 & 1,000 & 71~~ & 0.9966 & 0.9229 \\  
  6 & 1,000 & 5,000 & 1,000 & 71~~ & 0.9972 & 0.9169 \\    \hline
  7 & 1,000 & 2,000 & 1,000 & 33~~ & 0.9960 & 0.9226 \\ 
  8 & 1,000 & 2,000 & 1,000 & 43~~ & 0.9975 & 0.9244 \\ 
  9 & 1,000 & 2,000 & 1,000 & 71~~ & 0.9966 & 0.9229 \\    \hline      
  10 & 1,800 & 3,600 & 1,800 & 43~~ & 0.9978 & 0.9262 \\ 
  11 & 1,800 & 3,600 & 1,800 & 40~~ & 0.9978 & 0.9217 \\
  12 & 1,400 & 1,400 & 0 & 37~~ & 0.9979 & 0.8769 \\           \hline
\enddata
\tablecomments{Column Definitions: \\
Model: Arbitrary label for trials \\
N$_{pos}$: Number of ``positive'' labels sampled for each tree \\
N$_{neg}$: Number of ``negative'' labels sampled for each tree \\
N$_{FP}$: Number of Certified False Positives within N$_{neg}$ \\
N$_{feat}$: Number of features used in model \\
AUC$_{all}$: Area Under the Curve considering all samples in training set\\
AUC$_{CFP}$:  Area Under the Curve considering only Certified False Positives as ``negative'' labels.
} 
\label{RF_options.tbl}
\end{deluxetable}

\begin{figure}
\centering
  \includegraphics[width=\textwidth]{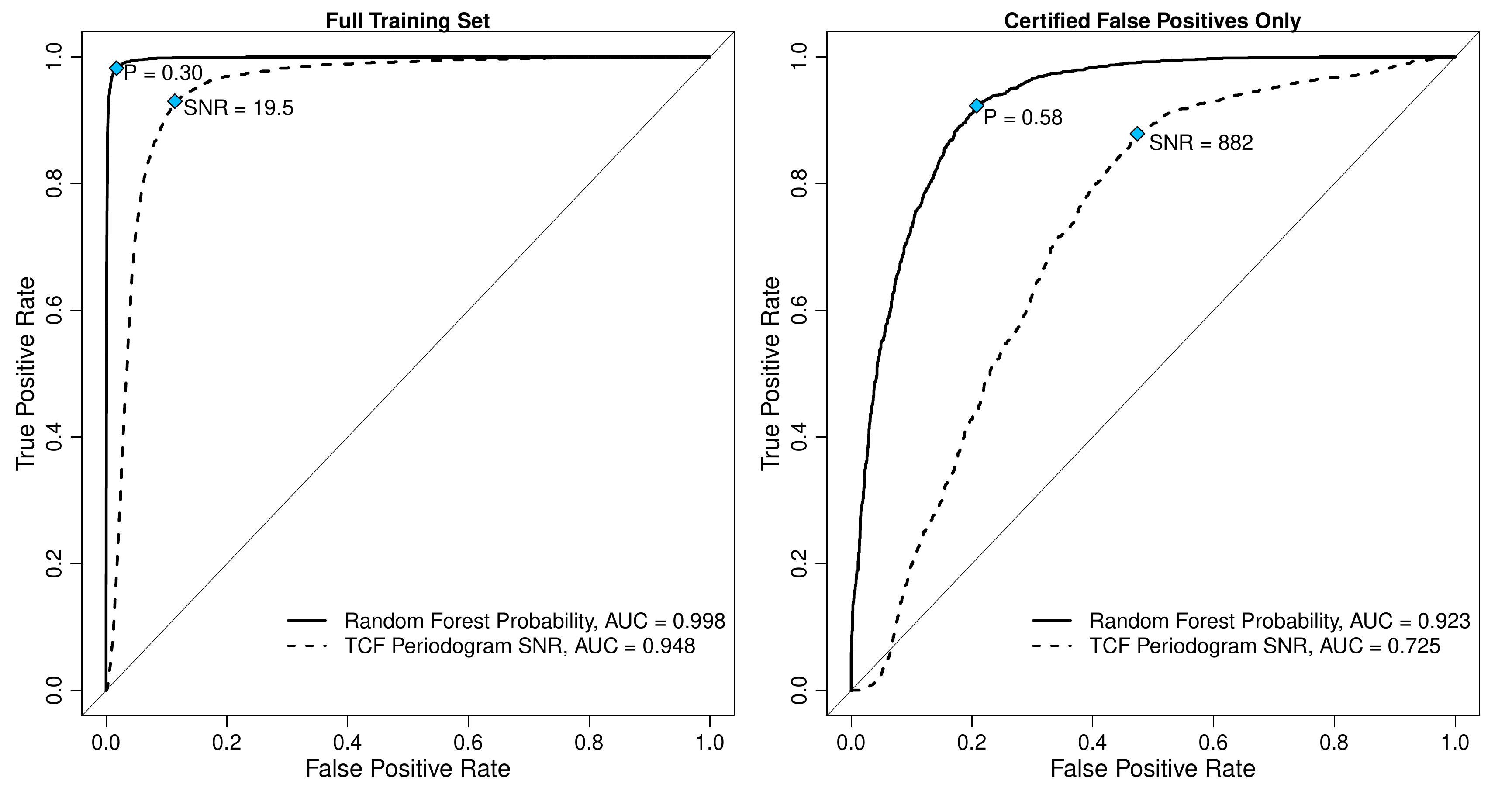} 
\caption{Receiver Operating Characteristic (ROC) curves for the class probability estimated by the Random Forest model (solid line) and only the Transit Comb Filter periodogram signal-to-noise (dashed line). Left panel shows ROC curve using all training data; right panel uses only Kepler Certified False Positives for the negative labels. Blue diamonds indicate optimal curve values using Youden's $J$ statistic.}
\label{ROC.fig}
\end{figure}

The ROC curve for the final RF classifier is shown as the solid curve in the left panel of Figure~\ref{ROC.fig} where the RF probability runs from $P_{RF}=1.00$ at the lower left to $P_{RF}=0.00$ at the upper right. There is no consensus on a single measure to define a `best' threshold value for a ROC curve; see the discussion by \citet{Powers11}.  An intuitive and commonly used measure is Youden's $J$ statistic which maximizes the vertical distance between the ROC curve and the diagonal line on the ROC diagram representing random class assignment \citep{Youden50}. These points are plotted as blue diamonds in Figure~\ref{ROC.fig}. Youden's $J$ statistic for the final RF classifier is optimized at $P_{RF} = 0.30$ where a high TP rate (around 98\%) is achieved simultaneously with a low FP rate (around 2\%).  But, due to the strongly imbalanced class sizes, a 2\% FP rate is still unpleasantly high as it implies that $\sim 3000$ False Positives would be mixed with the True Positives. Therefore, considering scientific reasons to reduce FPs with classifier performance, we will choose a higher $P_{RF}$ threshold than recommended by Youden's $J$.  

The right panel of Figure~\ref{ROC.fig}  shows that discrimination between TPs and Certified FPs is more difficult; 20\% or more of the eclipsing binaries will satisfy the classifier criteria at the Youden's $J$ threshold.  However, as there are many fewer eclipsing binaries than nonperiodic stars, this leakage can be tolerated providing a separate vetting procedure attempts to exclude them.   

Figure~\ref{ROC.fig} also compares the final RF classifier with that obtain using a traditional procedure where classification is based on a threshold for the TCF peak SNR as a single variable.  The univariate optimal threshold of $SNR=19.5$ give a much weaker performance than the multivariate RF classifier, with $\sim10$\% FP rate compared to 2\% for the multivariate classifier.  Every model presented in Table~\ref{RF_options.tbl} outperforms the univariate SNR decision criteria over the full range of values plotted in the ROC curve. This result emphasizes the effectiveness of including many features to identify transits.

\section{ARPS data products} \label{dataprod.sec}

The ARPS analysis outlined in Figure~\ref{ARPS_flowdiag.fig} and discussed above is applied to 156,717 light curves from the 4-year Kepler mission.  Quantitative results are provided in three forms and are available as High Level Science Products from NASA's Mikulski Archive for Space Telescopes (MAST)  :
\begin{description}

\item[Machine Readable Table]  Table~\ref{ARPSfull.tbl} provides 28 scalar or string quantities for each star.  Results for 5 stars are shown here to illustrate the structure of the table, while the full table is provided in ASCII format in the electronic version of this paper.  

The first two columns gives the Kepler Input Catalog number and stellar magnitude.  The precise location and other properties of the star can be obtained from NASA's {\it Kepler Data Search \& Retrieval} Web page at the Mikulski Archive for Space Telescopes (MAST)\footnote{\url{https://archive.stsci.edu/kepler/data_search/search.php}}.  The next 7 columns give properties from the MAST archive for 6252 Kepler Objects of Interest\footnote{\url{https://archive.stsci.edu/kepler/koi/search.php}}.  Orbital periods for entries in the Kepler Eclipsing Binary Catalog of \citet{Kirk16} appear in the next column.  

The remainder of the table gives scalar measures from the ARPS analysis: four quantities derived from the light curve, five quantities from the autoregressive modeling, eight quantities from the Transit Comb Filter periodogram, and one quantity from the Random Forest classifier.  

\item[Graphical output]  Four pages of plots with 18 panels are provided for each star.  One example is shown in Figure~\ref{KACT_pngs.fig} for a new Kepler ARPS Candidate Transit (KACT)  with similar pages for all KACTs Figure Set in the electronic version of this paper (\S\ref{KACT.sec}).  A collection of $\sim 0.6$ million plot pages for the full Kepler sample will be publicly available at NASA's MAST online repository as a `High Level Science Product'.  

\item[Binary output]  A binary FITS table for each star in the full Kepler sample will be available at MAST.  It contains vector values for the light curves and TCF periodograms at different stages of ARPS analysis, a large collection of scalar quantities (including those in the Machine Readable Table above), and intermediate results from the autoregressive, TCF and Random Forest analysis stages. 

\end{description}

\pagebreak

\vspace*{0.1in}
 \begin{deluxetable}{lrcccccrr}[h]
\tablecaption{ARPS Results for the Full Kepler Dataset}
\tabletypesize{\footnotesize}
\tablenum{3}
\tablewidth{0pt}
\centering
\tablehead{
\multicolumn{9}{c}{Kepler Team information}  \\ \cline{1-9}
\colhead{KIC0} & \colhead{Kepmag} & \colhead{Quarters} & \colhead{KOI.Name} &  \colhead{KOI.Kep} & \colhead{KOI.Disp} & 
\colhead{KOI.Per} & \colhead{KOI.Dep} & \colhead{KOI.SNR}  
}
\startdata
  000757137 & 9.196 &11111111111111111 &     \nodata &     \nodata &  \nodata  &   \nodata  &   \nodata  &   \nodata \\
  000757450 & 15.264 & 11111111111111111 & K00889.01 &   Kepler-75 b & CAND   &   8.884923  & 17670.4 &  505.3 \\
  000892010 & 11.666 & 11100000000000000   &  \nodata  &    \nodata  &  \nodata  &   \nodata  &   \nodata &   \nodata \\
  000892772 & 15.162 & 00011111111111111 & K01009.01   &  \nodata &           FP &     5.092442  &   283.3 &   16.2  \\
  001026032 & 14.813 & 11111111111111111 & K06252.01   &   \nodata &        FP   &   8.460438  & 79047.6  & 1882.9  \\
  \enddata
  \tablecomments{
Five entries are shown to illustrate the structure of the table.  The full table with 200,039 rows and 28 columns is given in the electronic version of this paper. \\
Column definitions:  KIC0 = Kepler Input Catalog identifier; Kepmag = Kepler magnitude; Quarters = 3-month quarters observed ($1-17$); KOI.Name = Kepler Object of Interest (KOI) designation (e.g. K00123.01); KOI.Kep = additional KOI name (e.g. Kepler-123b); KOI.Disp = KOI disposition (CAND = exoplanet candidates, FP = False Positive); KOI.Per = KOI period (days); KOI.Dep = KOI transit depth (ppm); KOI.SNR = KOI model signal-to-noise ratio.  
 }
\label{ARPSfull.tbl}
  
\end{deluxetable}

\vspace*{1in}
\nopagebreak 

\begin{deluxetable}{ccrrrccrcrr}[h]
\tabletypesize{\footnotesize}
\tablewidth{0pt}
\centering
\tablehead{
\colhead{KEBC.per} &
\multicolumn{4}{c}{Light curve properties} && \multicolumn{5}{c}{Autoregressive model} \\ \cline{2-5} \cline{7-11}
& \colhead{LC.NA} & \colhead{LC.med} & \colhead{LC.IQR} & \colhead{LC.dIQR} &&  \colhead{AR.mod} & \colhead{AR.IQR} & 
\colhead{ARF.mod} & \colhead{ARF.d} & \colhead{ARF.IQR} 
}
\startdata
\nodata &  0.40  & 2955743  &  2460.8  &  577.8~~ && 300  &  310.8 & 215 & 1.00  &  214.3 \\
\nodata  & 0.36  &   11430  &    71.9  &   11.5~~ &&100  &   11.4 & 414 & 1.16  &   5.5 \\
\nodata  &  0.90  &  305298  &   397.0  &   66.9~~ && 500   &  51.3 & 512 & 1.00  &   0.0  \\
\nodata  &  0.32   &  12192  &     8.4  &   11.1~~ && 500  &    8.5 & 211 & 1.01  &    5.3 \\
8.4604400 &  0.30  &   16800  &    24.3  &   13.4~~ && 500  &   20.9 & 515 & 1.00 &   12.7 \\
\enddata
\tablecomments{Column definitions: KEBC.per = Period (days) if listed in the Kepler Eclipsing Binary Catalog \citep{Kirk16}; LC.NA = fraction of 4-year time slots with `Not Available' entries; LC.med = median PDC flux (elec~s$^{-1}$); LC.IQR= InterQuartile Range of stitched light curve; LC.dIQR = InterQuartile Range of differenced light curve; AR.mod = ARIMA(p,d,q) model order; AR.IQR = InterQuartile Range of ARIMA residuals; ARF.mod = ARFIMA(p,d,q) model order; ARF.d = value of `d'; ARF.IQR = InterQuartile range of ARFIMA residuals.}
\tablecomments{The AR.mod and ARF.mod values represent the order of the autoregressive models.  A value like  '400' represents (p,d,q)=(4,1,0) where d is increased by 1 due to the earlier differencing operation. The value ARF.d for the ARFIMA model quantifies $1/f^\alpha$ long-memory red noise: $d=1$ implies $\alpha=0$.}
\end{deluxetable}
 
 \begin{deluxetable}{crrrrrrcrc}[h]
\tabletypesize{\footnotesize}
\tablewidth{0pt}
\centering
\tablehead{\multicolumn{8}{c}{Transit Comb Filter periodogram peak} && \colhead{RF.Prob} \\ \cline{1-8} 
\colhead{TCF.per} & \colhead{TCF.pow} & \colhead{TCF.SNR} & \colhead{TCF.shp} &  \colhead{TCF.dep} & \colhead{TCF.pha} & 
\colhead{TCF.dur} & \colhead{TCF.har} &&   
}
\startdata
 8.193797    &       419.0 &   8.0 &    40.4  &    732.0 & 3510~~  & 1~~~~~ & -1 && 0.030  \\
 8.884898    &   38689.5 & 3623.5  &  191.0  &   9595.3  &143~~  & 2~~~~~ &  2 && 0.785  \\
 2.577965    &      84.6 &  17.6  &    5.2 &      69.6  & 64~~ & 14~~~~~ &  2 && 0.062  \\
 5.092458    &     257.1  & 54.4  &   10.2  &    262.4 &  93~~  & 6~~~~~  & 2 && 0.940  \\
 8.460500    &   36532.8 & 3382.7  &  145.9 &    8939.0 & 107~~ &  3~~~~~ &  2 && 0.015  \\
\enddata
\tablecomments{
Column definitions: TCF.pdf = peak period (days); TCF.pow= peak power; TCF.SNR = peak signal-to-noise ratio after subtraction of LOESS trend curve; TCF.shp = folded light curve shape statistic (lhigh values indicate low-duty cycle periodic behaviors); TCF.dep = transit depth from TCF calculation (elec s$^{-1}$); TCF.pha = transit phase (29.4 min cadence \#); TCF.dur = transit duration (29.4 min cadence \#); TCF.har = harmonics indicator; RF.Prob = uncalibrated probability from Random Forest classifier. }
\tablecomments{The TCF harmonics indicator {\it TCF.har} is coded as follows: 0 = No harmonic, -1 = Half-period harmonic, 1 = Twice-period harmonic, 2 =  Both half- and twice-period harmonics}

\end{deluxetable}

\vfill
\pagebreak

\section{ARPS Results for the Full Kepler Sample} \label{results.sec}

\subsection{Autoregressive modeling} \label{resultsAR.sec}

\begin{figure}[ht]
\centering
  \includegraphics[width=0.65\textwidth]{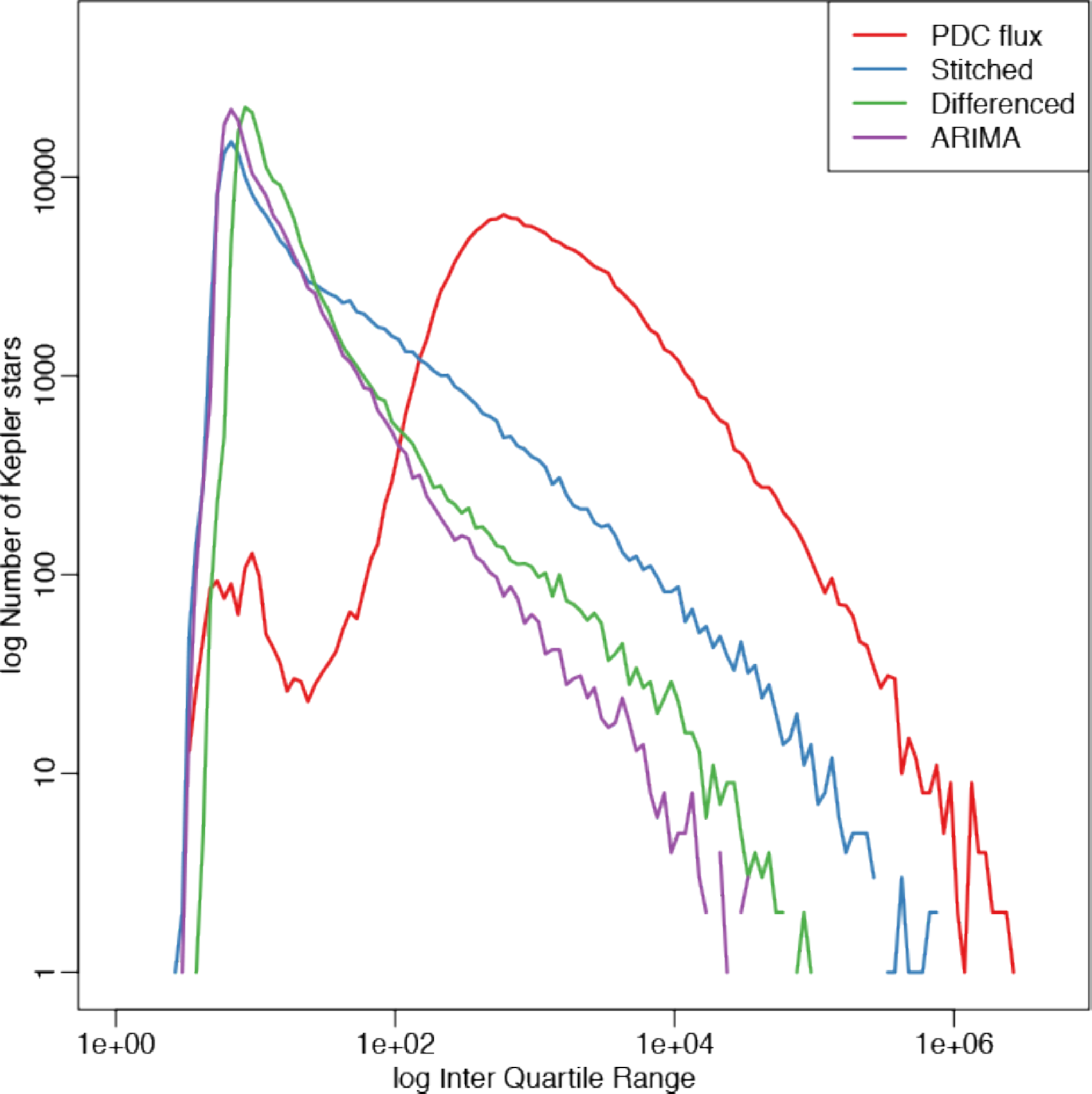}
\caption{Number of Kepler stellar light curves showing values of InterQuartile Range (IQR, in units of elec s$^{-1}$) at different stages of KARPS analysis: Original PDC light curve before stitching of quarters (red); light curve after stitching of quarters and removal of flagged values (blue); light curve after difference operator (green); ARIMA model residuals (purple). }\label{IQR1.fig}
\end{figure}

\begin{figure}[hb]
\centering
  \includegraphics[width=\textwidth]{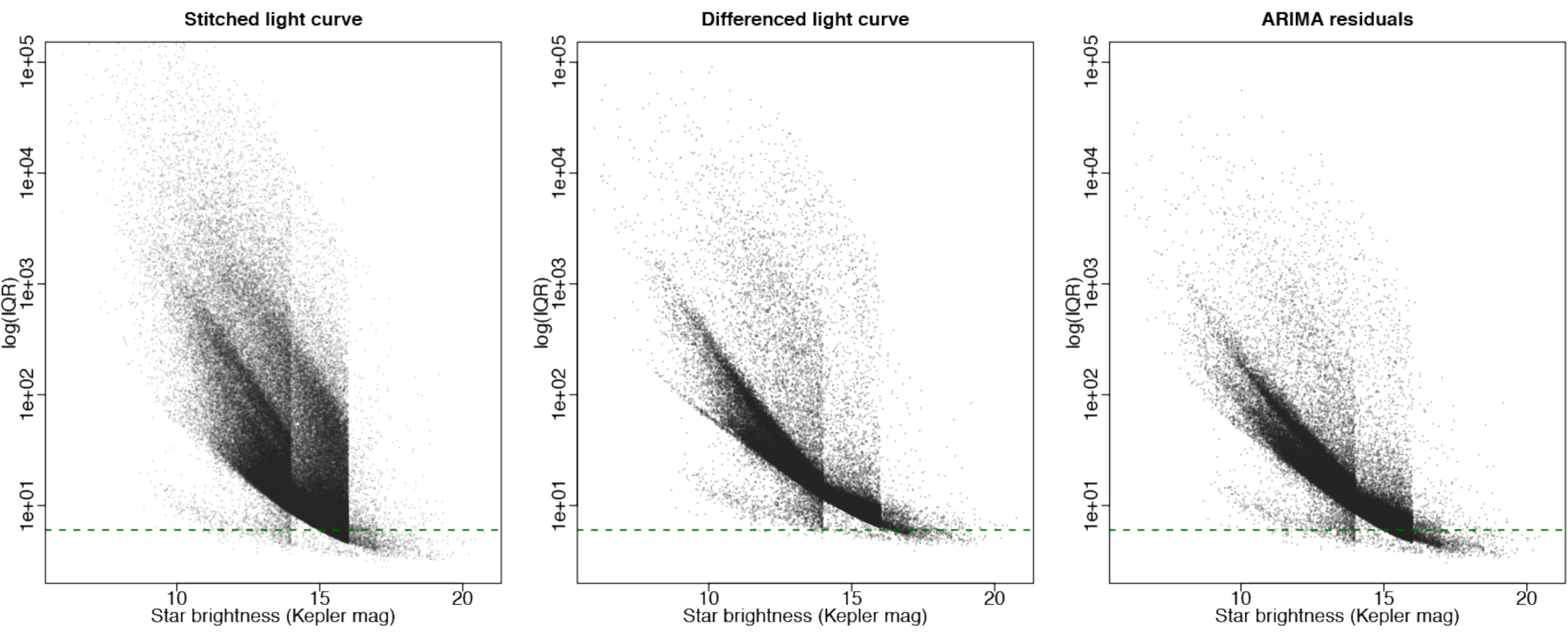}
  \includegraphics[width=\textwidth]{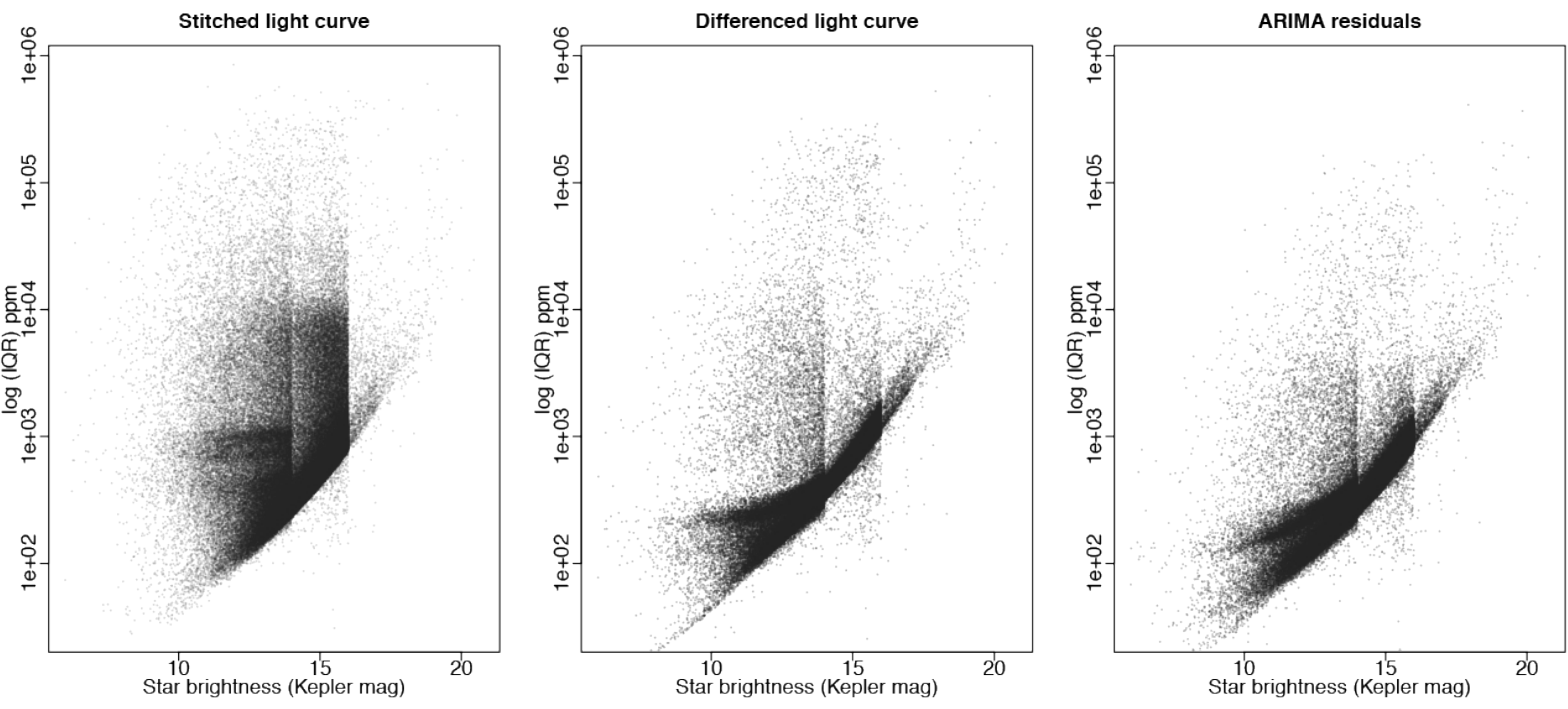}
\caption{Light curve InterQuartile Range (IQR) at three stages of KARPS analysis plotted against the Kepler magnitude for the full Kepler sample: stitched light curve (left); differenced light curve (center); ARIMA model residuals (right).  The top panels give IQR in units of elec~s$^{-1}$, and the bottom panels give IQR in scale-free units of parts-per-million (ppm).  In the top panels, the dashed line at IQR=6, a typical value for a faint quiet star, is plotted to aid the eye.}
\label{IQR2.fig}
\end{figure}

\begin{figure}[t]
\centering
  \includegraphics[width=\textwidth]{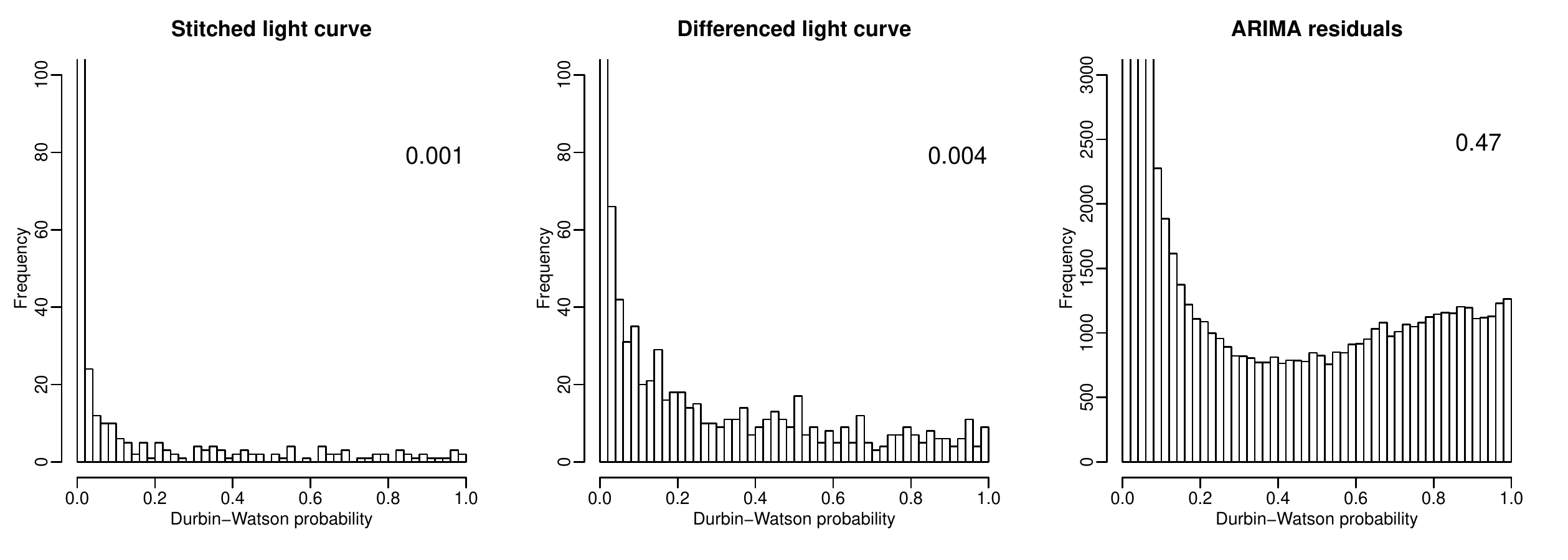}
\caption{Distribution of probabilities for the Durbin-Watson measure of serial autocorrelation for 156,717 Kepler stars at different stages of KARPS analysis: (a) Original light curve after stitching of quarters and removal of flagged data points; (b) Light curve after differencing operator; (c) ARIMA residuals. The annotation gives the fraction of stars for which the probability is greater than 1\%; this is the fraction of stars for which the ACF(lag = 1 cadence) value is consistent with uncorrelated white noise.}
\label{DW.fig}
\end{figure}

The InterQuartile Range of a light curve, combined with the Durbin-Watson statistic of serial (lag=1 cadence interval) autocorrelation, provide straightforward measures how the variations in stellar brightness are reduced at different stages of ARPS analysis.  Figures~\ref{IQR1.fig}-\ref{IQR2.fig} show univariate and bivariate distributions of IQR improvement, and Figure~ \ref{DW.fig} shows the Durbin-Watson (DW) probability improvement.  Small DW probability (e.g. below 1\%) indicate temporal structure inconsistent with Gaussian white noise.  Selected quantitative results are presented in Table~\ref{lc_improv.tbl}.  The DW test measures autocorrelation inconsistent with Gaussian white noise for 0.5 hr lag, while the Breusch-Godfrey test measures the same with 2.5 hr lag.  The Augmented Dickey-Fuller and KPSS tests measure deviations from stationarity (i.e., the presence of trends).   

In Figure~\ref{IQR1.fig}, the high noise level in the red curve reflects the instrumental calibration problems in the PDC flux measurements that are easily removed by preprocessing (`stitching', \S\ref{preprocessing.sec}).  More importantly, a marked reduction in variability after differencing  and ARIMA modeling is seen for many stars.  While ARIMA modeling produced a slight deterioration in noise for  $\sim 30$\% of the sample, half of the Kepler stars have noise levels reduced by over 10\% and about 10\% of stars show reductions by more than an order of magnitude. For the large number of stars with quiescent fluxes, the differencing operation (green curve) shifts the IQR distribution slightly higher than the stitched light curve distribution (blue curve).  This is the well-known effect known as `over-differencing' where the operation introduces spurious structure in time series that are close to Gaussian white noise (Paper I, \S2.3). But the original low noise values are recovered by the ARIMA fitting (purple curve).  

\begin{deluxetable}{llllll}[t]
\tablecaption{Light Curve Improvements with Stage of ARPS analysis} 
\tabletypesize{\small}
\centering
\tablehead{
\colhead{Property} & \colhead{Statistic} && \multicolumn{3}{c}{Fraction\tablenotemark{a}}  \\ \cline{3-6}
 & && \colhead{Stitched LC} & \colhead{Diff LC} & \colhead{ARIMA }
}
\startdata
Spread  					&  IQR  $>$5000 ppm      		&&  ~~~0.074  &  ~~~0.012  	&  ~~0.006 \\
        						&  IQR  $>$500 ppm        		&&  ~~~0.62  	&  ~~~0.55 	&  ~~0.43 \\
Serial autocorrelation 		& Durbin-Watson              		&& ~~~0.001  	& ~~~0.004 	&  ~~0.47 \\
Lagged autocorrelation		& Breusch-Godfrey            	&& ~~~0.0007 & $<$0.0001 	&  ~~0.11 \\
Stationarity 				& Augmented Dickey-Fuller       && ~~~0.09 	& ~~~0.997	&  ~~1.000 \\
                   				& KPSS                                      && ~~~0.005 	& ~~~0.996 	&  ~~0.998 \\
\enddata
\tablenotetext{a} {For IQR, this gives the fraction of stars with spread greater than the stated level. For the other measures, this gives the fraction of Kepler stars consistent with the null hypothesis (e.g., no autocorrelation, no nonstationarity) at a $p>0.01$ significance level.  The Breusch-Godfrey statistic is evaluated at lag = 5 cadence steps, or about 2.5 hours.  KPSS is the acronym for the Kwiatkowski-Phillips-Schmidt-Shin test for stationarity against the hypothesis of a unit root. See \citet{Enders14} or other texts in econometrics for background on these tests.}
\label{lc_improv.tbl}
\end{deluxetable}
   
More details are seen in the scatter plot of IQR against Kepler magnitude (Figure~\ref{IQR2.fig}).  Thousands of variable stars in the stitched light curves converge to the band where quiet stars reside after ARIMA modeling. Most of the effect is due to the differencing step, but the ARIMA modeling both reduces some additional structure and removes the spurious noise produced by over-differencing quiescent stars.  Comparing the right- and left-hand panels of the figure shows that ARIMA modeling removed most of the noise structure in most of the variable stars.   

A light curve does not have to display large-amplitude variability to have autocorrelated noise.  The DW statistic traces both obvious structure and autocorrelation that is not always apparent by eye. The leftmost histogram of Figure~\ref{DW.fig} shows that the vast majority of stitched light curves are inconsistent with Gaussian white noise.  The differencing operation improves the autocorrelation only for a small handful of stars.  But 47\% of all light curves have serial autocorrelation consistent with Gaussian white noise when the residuals after ARIMA modeling are considered (right panel). 

Table 3.1 summarizes the key results presented in this section. It shows an order-of-magnitude reduction in the number of highly variable stars with spread $IQR > 0.5$\%. The Durbin-Watson and Breusch-Godfrey tests show improvement in autocorrelation, though most of the model residual light curves still show some remaining autocorrelation.   The broader ARFIMA model is more successful in reducing autocorrelation stellar variations, although we do not use these residuals for our planet search. The Augmented Dickey-Fuller and KPSS tests show that virtually all of the trends have been removed in the ARIMA residuals.  This includes the quasi-periodic brightness variations from rotationally modulated starspots that are commonly present in Kepler light curves.  

The dramatic improvements in IQR, autocorrelation, and stationarity by ARIMA modeling for many Kepler stars demonstrated here indicate that the first stage of ARPS analysis is effective in greatly reducing stellar variability in Kepler stars.  Furthermore, the modeling does not introduce significant spurious structure in quiescent stars where the original light curve is close to Gaussian white noise.

\subsection{Identification of Transits with TCF and RF} \label{resultsTransits.sec}
  
In \S\ref{Eval_KOI.sec}, Transit Comb Filter (TCF) best peaks were compared to the transits of confirmed planets among the Kepler Objects of Interest when the TCF and KOI periods coincided.  But that comparison did not actually evaluate whether the signals would have been identified independently of this coincidence.  Based on autoregressive modeling to remove the stellar variability, filtering the residuals for transits using the TCF algorithm, and training a Random Forest classifier based on many features,  the classifier can now be applied to the full Kepler database both to evaluate the recovery of training sample objects and to seek new transit-like signals.   
 
\begin{figure}
\centering
  \includegraphics[width=\textwidth]{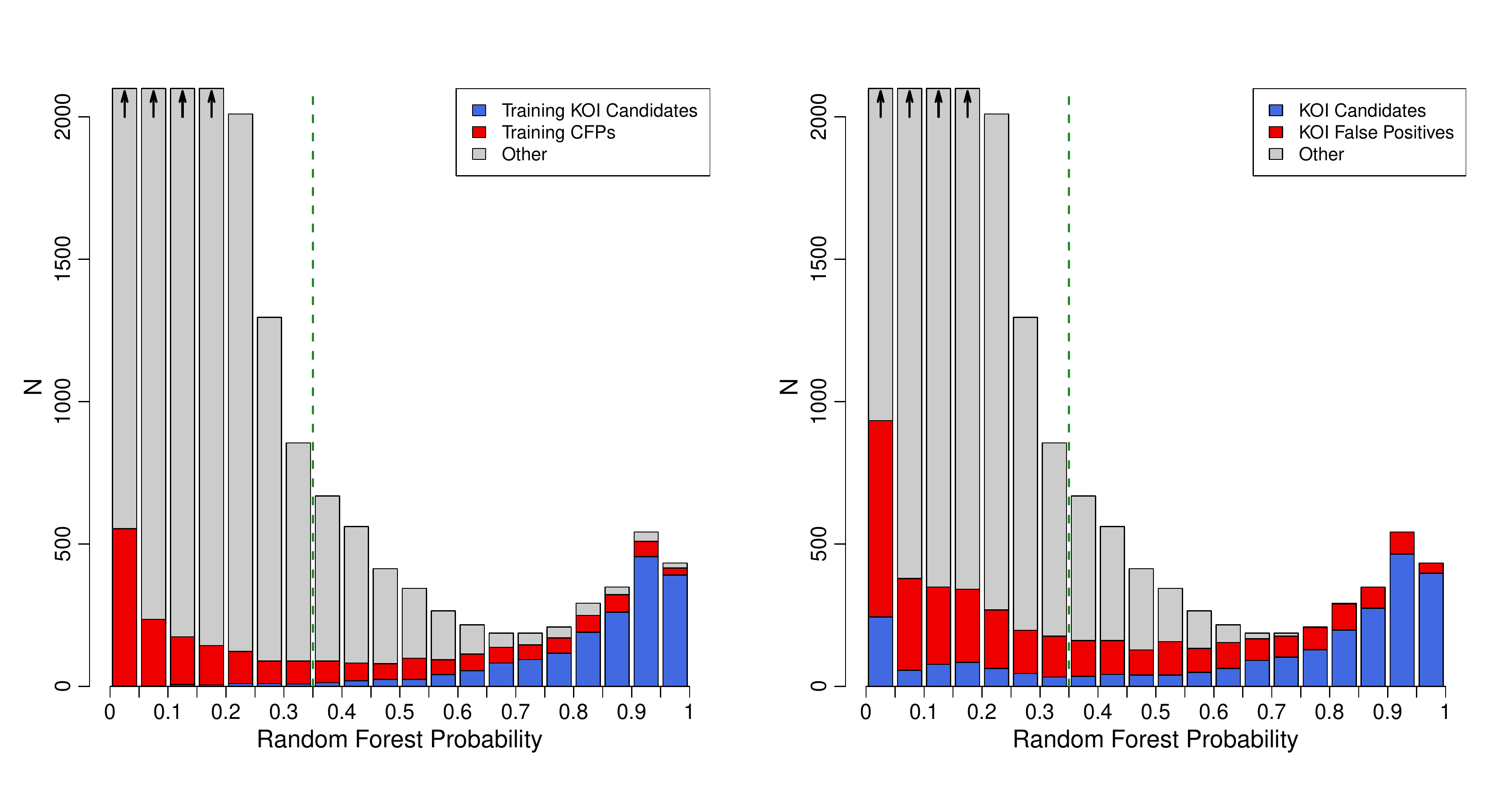} 
\caption{Histogram of Random Forest probabilities for 156,717 Kepler light curves. The left panel shows results on the data used for training. For each bar, the blue portion shows the selected KOI candidates used for training; the red portion shows Kepler Certified False Positives used for training; the gray portion shows all remaining stars. The right panel compares ARPS results with the full KOI table,: blue corresponds to all Candidate Planets (including those with periods mismatching the TCF periods) and red corresponds to all types of False Positives, while the remaining stars are gray. Dashed green line shows the selected threshold $P_{RF} = 0.35$.}
\label{RFprob_hist.fig}
\end{figure}

Figure~\ref{RFprob_hist.fig} displays histograms of the predicted RF probabilities for all 156,717 Kepler light curves analyzed.  They show the RF classifier panel for two combinations of False Positives and KOI Candidates, as well as the bulk of random stars.  As previously indicated by the ROC curve in Figure~\ref{ROC.fig} and its corresponding AUC, the model does a good job at placing the KOI candidates at the high end of the probability distribution, and is less successful with FPs. {\it We now chose a probability threshold of $P_{RF} = 0.35$ for the classifier discrimination of planetary candidates}\footnote{
As discussed in Paper I (\S4.2), the probabilities estimated by this Random Forest model are uncalibrated. A calibration may be possible using simulated planetary signals injected into real data \citep{Christiansen16}; this may be attempted in the future.}. 
This criterion is slightly more strict than the Youden index of 0.30 obtained in \S\ref{ROC.sec}. The presence of some blue objects below the threshold $P_{RF} = 0.35$ in Figure~\ref{RFprob_hist.fig} (right panel) shows that the ARPS procedure does not confirm all KOI Candidates; this was previously seen in Figure~\ref{ARPS_vs_missed_KOI.fig} and is discussed in the Appendix.  Although we could have set a more stringent threshold such as $P_{RF}=0.45$ or 0.55, we prefer to use an intermediate cutoff to ensure a promising pool of candidate new transiting systems. This discovery space is seen as the gray area above the red and blue histograms above $P_{RF}=0.35$, showing stars that have no disposition in the KOI samples.  

There are 4,668 stars in total above the $P_{RF}=0.35$ threshold shown as the dashed green line\footnote{
In relation to the Kepler pipeline, this sample selected by our RF classifier can be thought of as a data product intermediate between TCEs and KOIs.  For reference, the sample analyzed with ARPS contains 14,683 stars of 17,230 stars with TCE events.}. 
These include: 1,926 stars with KOI disposition `Planet Candidate' constituting 76\% of the full list of 2,529 KOI Candidates (blue histogam, right panel of Figure~\ref{RFprob_hist.fig}); and 1,132 stars above the threshold are KOIs with disposition `False Positive' constituting 36\% of the total sample of FP.  The potential ARPS discoveries are among the 1,610 stars that do not have KOI classifications (gray area above the threshold).  The Appendix  provides more detail on the ARPS performance on various subsets of KOIs and related subsamples of Kepler stars.  

\subsection{Vetting New Candidates} \label{vetting.sec}

Proceeding to search for new promising transits, we subject the unexamined 1,610 stars with $P_{RF}>0.35$ to a subjective vetting procedure.  We can classify light curves as False Alarms (no true periodicity present), False Positives (periodicity from non-transit causes), and likely transits.  We label this last class {\it Kepler ARPS Candidate Transits} or {\it KACTs}.  Acquiring a reliable list of KACTs is the primary scientific goal of this study.  Our subjective vetting procedure has three elements: 
\begin{description}
\item[Light curves and (P)ACF] We evaluate the early stages of ARPS analysis for FAs despite the high $P_{RF}$ value.   In a minority of stars, the noise (IQR values) and autocorrelation in the ARIMA residuals is very high, preventing meaningful searches for faint periodic planetary signals.   The original stitched light curves are also examined for sharp outliers caused by effects such as stellar flares or instrumental difficulties that did not receive data quality flags.  Outliers often have deleterious effects on the TCF periodogram at long periods (e.g. bottom panel of Figure~\ref{TCF_sample.fig}) and resulting periodogram peaks are unreliable.  Finally, some cases are found where $P_{RF}$ slightly exceeds the threshold but the TCF shows peaks consistent with noise and the folded light curve plots do not show clear planetary signals. These subjective tests for FAs eliminate many contaminants from the 1,610 stars.  
 
\item[Periodogram] We examine the significance of the main TCF periodogram peak, the presence of any harmonics, and the structure of peaks throughout the periodogram.   Cases where the periodogram is disturbed by multiply-periodic signals, as can arise in red giant stars or unusually noisy periodograms are flagged as FAs or FPs.  Also, the periodograms from ARIMA and ARFIMA residuals are compared; the presence of identical peak periods is a positive indication that a true periodicity is present.  
 
\item[Phase-folded curves] We look at whether the transit shape in the folded light curve was consistent with a planetary signal and if any secondary eclipses were present.  The absence of a box-shape signal in the folded stitched light curve (left panels of Figure~\ref{arimax.fig} and the last page of Figure~\ref{KACT_pngs.fig}) is not considered to be a problem, as the ARIMA model has not yet been applied to reduce stellar variability.  But the absence of a double-spike signal in the folded ARIMA residuals (middle panels) and of a box-shape signal in the folded ARIMAX residuals (right panels) speaks against a true transit.  Other problems with the folded light curve, such as large-scale curvature suggesting mutual illumination of two stars, periodic positive outliers, or a range in depths in the putative transit, are sometimes identified and flagged as a FP.  
\end{description}

These judgments are combined in a subjective fashion to rate the 1,610 stars that passed the $P_{RF}>0.35$ criterion.  The large majority of cases are easily rated as FAs or FPs.  We then made subjective decision to select the most promising cases, resulting in 97 KACTs presented in the next section.  To make other decisions, interested readers can easily recover the 1,610 stars by applying the threshold $P_{RF}>0.35$ to the last column of Table~\ref{ARPSfull.tbl}.  The resulting light curves, periodograms and folded light curves from the graphical outputs in the online MAST repository (\S\ref{dataprod.sec}).  
 
\section{ARPS Transit Discoveries in the Kepler Dataset} \label{KACT.sec}

\begin{figure}[ht]
\centering
  \includegraphics[width=\textwidth]{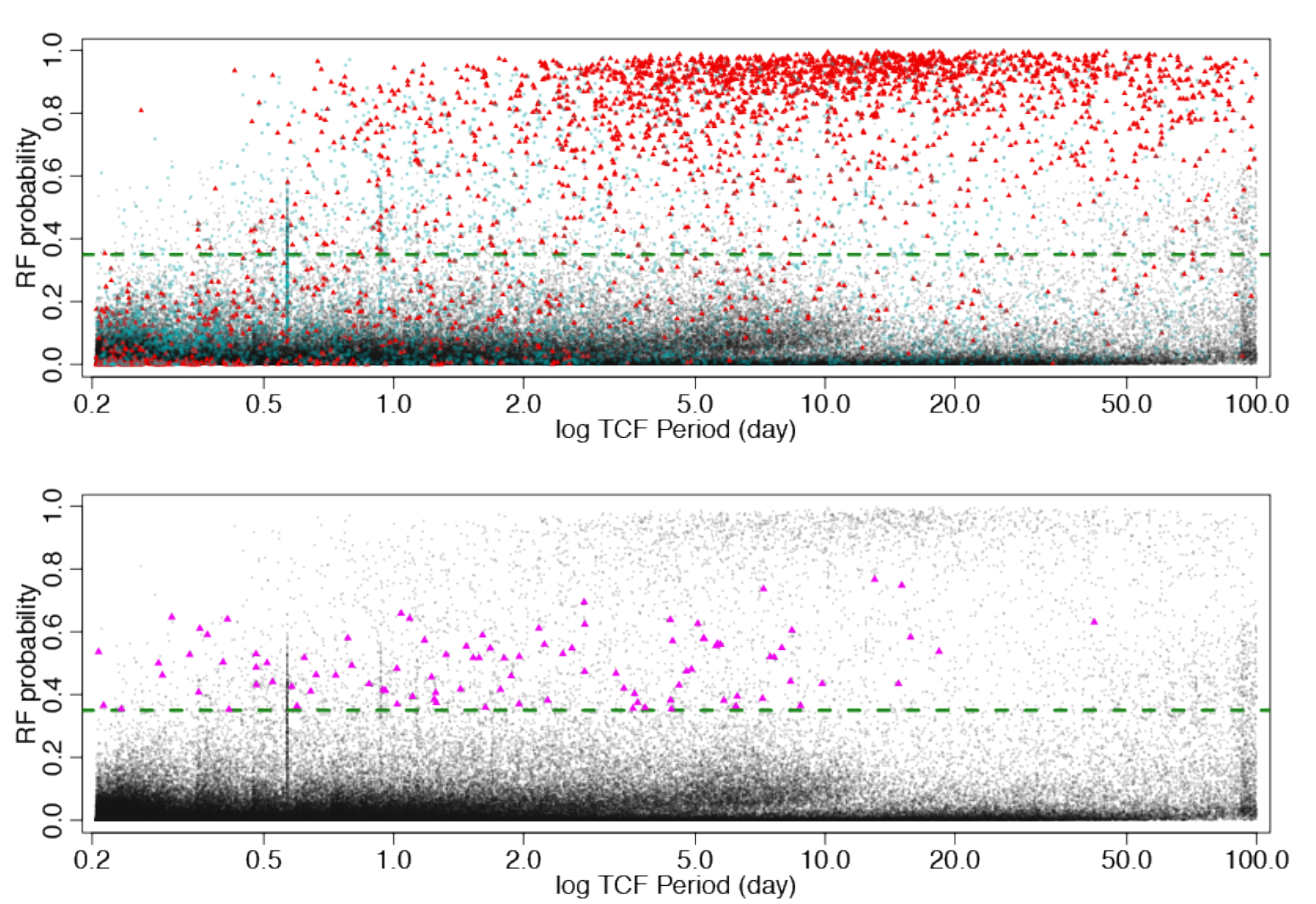} 
\caption{Random Forest probability for 156,717 Kepler stars with ARPS processing plotted as a function of best period from the TCF periodogram (gray dots). The green dashed line is the chosen threshold for acceptance as a transit-like periodic variable by ARPS. (Top) Blue triangles show Kepler Objects of Interest (KOI) Candidates and red boxes show KOI Certified False Positives used in training the Random Forest.  (Bottom) Magenta triangles show 97 Kepler Candidate Transits (KACTs) after vetting to reduce False Alarms and False Positives.}
\label{KACT.fig}
\vspace*{-0.2in}
\end{figure}

 Figure~\ref{KACT.fig} shows the Random Forest probability estimates as a function of TCF period for the full Kepler sample analyzed with ARPS. The vast majority of stars are shown as small gray dots, while stars of interest are shown as colored symbols. The upper plot shows the performance on Kepler Team KOI Candidates (blue) and Certified False Positives (red) used for RF training. The horizontal dashed line represents the selected Random Forest probability threshold of 0.35. The 1,610 symbols lying above the green line represent the object subject to subjective vetting by individual examination of  their light curves, periodogram, and phase-folded plots (\S\ref{vetting.sec}).  The stars collected into vertical stripes are duplicate periods arising mainly from overlapping point spread functions in the Kepler detector \citep{Coughlin14}.    
  
 The 97 magenta triangles in the bottom panel of Figure~\ref{KACT.fig} are the stars that lie above the RF probability threshold and survived the subjective vetting process described in \S\ref{vetting.sec}. These are the Kepler ARPS Candidate Transits or KACTs that are the principal new discoveries of this study.  Most of these stars do not show strong stellar variability in their original light curves.  Four pages of graphs for each star provide details from stages of the ARPS analysis as a Figure Set in Figure~\ref{KACT_pngs.fig}.  The graphical information for each KACT star is briefly summarized in the footnotes to Table~\ref{KACT.tbl} that are available in the electronic version of this table.  

Table~\ref{KACT.tbl} lists several properties of the KACT candidates; additional properties are available from Table~\ref{ARPSfull.tbl}.  Table~\ref{KACT.tbl} gives identifiers, stellar brightness, four quantities derived from the TCF periodogram (transit period, depth, TCF peak signal-to-noise ratio, and a code for spectral harmonics), the Random Forest probability (always above the $P_{RF}=0.35$ threshold), and a flag for UltraShort Period ($P < 1$ day). Table footnotes (available in the Machine Readable Table) summarize the light curve variability, TCF periodogram, and folded light curve shown in the Figure Set diagrams.  In addition, information from published sources on these stars is drawn from the SIMBAD (Wenger et al., 2000) and Vizier5 (Ochsenbein et al., 2000) bibliometric databases at the Centre des Donn\'ees Stellaires in Strasbourg, FR.  Recall that, from the definitions of our training and test sets for the RF classifier, a `discovery' means that it is not classified as a planetary candidate in the DR25 Golden sample; it is quite possible that the planet has been reported in some other study.  

\newpage

\startlongtable
\begin{deluxetable}{crcrrrrrrccc}
\tablecaption{Kepler ARPS Candidate Transits} 
\tablenum{5}
\tabletypesize{\tiny}
\tablewidth{0pt}
\tablehead{
 & \multicolumn{3}{c}{Kepler star} && \multicolumn{7}{c}{ARPS results}  \\ \cline{2-4} \cline{6-12}
\colhead{KACT} & \colhead{KIC} & \colhead{2MASS} & \colhead{Kepmag} && \colhead{Period} &  \colhead{Depth} & \colhead{SNR} & 
\colhead{Har} & \colhead{$P_{RF}$} & \colhead{USP}  
}
\startdata
  ~1 &  1701672 & 19045039+3716484 & 15.39 && 5.583244 & 196 & 16.1 & -1 & 0.56 & \nodata  \\ 
  ~2 &  2283362 & 19042658+3736426 & 15.77 && 0.336574 & 157 & 35.8 & 1 & 0.53 &  $\surd$  \\ 
  ~3 &  2718252 & 19301643+3759596 & 12.79 && 8.313029 & 544 & 16.9 & 2 & 0.44 &  \nodata  \\
  ~4 &  3217913 & 19042449+3819016 & 15.80 && 1.577616 & 124 & 14.2 & 2 & 0.52 &  \nodata  \\
  ~5 &  3218587 & 19052795+3823324 & 15.68 && 0.508939 & 66 & 23.4 & -1 & 0.50 &  $\surd$  \\ 
  ~6 &  3534897 & 19142179+3837528 & 14.69 && 5.634363 & 120 & 15.9 & -1 & 0.56 & \nodata  \\ 
  ~7 &  3626225 & 18592791+3843029 & 15.86 && 7.159311 & 305 & 16.2 & 2 & 0.39 &  \nodata  \\ 
  ~8 &  3643155 & 19230397+3845064 & 15.92 && 6.210501 & 255 & 15.9 & -1 & 0.36 & \nodata  \\ 
  ~9 &  3644851 & 19250163+3843455 & 14.13 && 0.523603 & 29 & 26.0 & -1 & 0.44 &  $\surd$  \\ 
 10 &  4049278 & 19172725+3908292 & 15.81 && 1.428861 & 91 & 14.9 & -1 & 0.42 &  \nodata  \\ 
 11 &  4165903 & 19341480+3915201 & 15.58 && 2.465054 & 131 & 15.6 & 2 & 0.53 &  \nodata  \\ 
 12 &  4349469 & 19063429+3927482 & 15.04 && 1.674499 & 74 & 18.1 & -1 & 0.55 &  \nodata  \\ 
 13 &  4556565 & 19204164+3937436 & 15.49 && 2.772800 & 135 & 27.8 & -1 & 0.62 & \nodata  \\ 
 14 &  4832514 & 19231946+3958458 & 15.46 && 4.382513 & 114 & 15.8 & -1 & 0.38 & \nodata  \\ 
 15 &  4851138 & 19431662+3954095 & 15.13 && 4.401514 & 122 & 14.3 & -1 & 0.35 & \nodata  \\ 
 16 &  4862360 & 19524015+3955336 & 14.98 && 2.763421 & 97 & 16.6 & 2 & 0.69 &   \nodata  \\ 
 17 &  4920605 & 19230977+4000391 & 15.30 && 1.178837 & 56 & 17.0 & -1 & 0.57 &  \nodata  \\ 
 18 &  4999297 & 19120826+4006279 & 12.09 && 5.245797 & 24 & 15.0 & 2 & 0.58 &   \nodata  \\ 
 19 &  5024556 & 19412029+4008253 & 14.79 && 0.212586 & 108 & 101.0 & 1 & 0.37 & $\surd$  \\ 
 20 &  5096682 & 19240513+4015011 & 15.38 && 0.799434 & 46 & 19.6 & 2 & 0.49 &   $\surd$  \\ 
 21 &  5130068 & 19562461+4014008 & 15.23 && 0.353229 & 58 & 30.1 & 0 & 0.41 &   $\surd$  \\ 
 22 &  5212569 & 19514560+4022324 & 14.39 && 1.528524 & 46 & 14.9 & 2 & 0.52 &   \nodata  \\ 
 23 &  5269353 & 19201879+4027188 & 15.88 && 0.479425 & 70 & 30.8 & -1 & 0.53 &  $\surd$  \\ 
 24 &  5386271 & 19510615+4032049 & 15.53 && 2.770627 & 164 & 19.2 & 2 & 0.47 &  \nodata  \\ 
 25 &  5523854 & 19145929+4043015 & 15.42 && 1.473060 & 75 & 17.7 & -1 & 0.55 &  \nodata  \\ 
 26 &  5536715 & 19310694+4047442 & 13.99 && 0.233779 & 45 & 127.0 & 1 & 0.35 &  $\surd$  \\ 
 27 &  5559584 & 19524993+4047163 & 15.13 && 0.620728 & 69 & 28.9 & 2 & 0.52 &   $\surd$  \\ 
 28 &  5600858 & 18593879+4048300 & 14.85 && 7.460664 & 164 & 15.0 & -1 & 0.52 & \nodata  \\ 
 29 &  5738241 & 19583138+4055260 & 12.48 && 1.872716 & 19 & 16.0 & 2 & 0.46 &   \nodata  \\ 
 30 &  5820733 & 19562241+4104507 & 14.84 && 5.724369 & 113 & 15.8 & -1 & 0.56 & \nodata  \\ 
 31 &  5880084 & 19320309+4107295 & 15.61 && 0.596665 & 48 & 18.2 & -1 & 0.36 &  $\surd$  \\ 
 32 &  6067645 & 19534266+4119140 & 14.74 && 0.415225 & 37 & 22.2 & -1 & 0.35 &  $\surd$  \\ 
 33 &  6104095 & 18570694+4125007 & 15.79 && 1.954359 & 128 & 17.9 & 1 & 0.52 &  \nodata  \\ 
 34 &  6128769 & 19334516+4127193 & 14.81 && 5.223006 & 88 & 13.9 & 1 & 0.58 &   \nodata  \\ 
 35 &  6143647 & 19483296+4125383 & 13.96 && 7.196896 & 95 & 14.5 & 2 & 0.74 &   \nodata  \\ 
 36 &  6382044 & 19471659+4142102 & 14.48 && 2.169515 & 60 & 15.8 & -1 & 0.61 &  \nodata  \\ 
 37 &  6544888 & 19510834+4157386 & 15.69 && 0.479849 & 85 & 38.4 & 1 & 0.49 &   $\surd$  \\ 
 38 &  6549164 & 19544365+4159364 & 14.05 && 1.251266 & 35 & 19.2 & 2 & 0.41 &   \nodata  \\ 
 39 &  6592335 & 19090291+4201563 & 15.90 && 0.412000 & 126 & 29.9 & 2 & 0.64 &  $\surd$  \\ 
 40 &  6620719 & 19430769+4201447 & 13.72 && 4.583959 & 46 & 15.5 & 0 & 0.43 &   \nodata  \\ 
 41 &  6636020 & 19560604+4203422 & 15.70 &&18.359587 & 430 & 15.8 & -1 & 0.54 & \nodata  \\ 
 42 &  6775237 & 19243571+4217431 & 14.74 && 2.589323 & 63 & 16.5 & 0 & 0.55 &   \nodata  \\ 
 43 &  6837899 & 18480859+4221344 & 14.79 &&13.025497 & 203 & 13.9 & 2 & 0.77 &  \nodata  \\ 
 44 &  6860282 & 19234998+4218386 & 14.87 && 7.936698 & 110 & 16.4 & -1 & 0.55 & \nodata  \\ 
 45 &  6880508 & 19453476+4218458 & 14.47 && 2.274646 & 79 & 20.6 & -1 & 0.38 &  \nodata  \\ 
 46 &  6963092 & 19431018+4228231 & 14.63 && 5.822575 & 84 & 14.0 & 0 & 0.38 &   \nodata  \\ 
 47 &  7014746 & 18584920+4234321 & 15.50 && 0.733749 & 59 & 16.7 & -1 & 0.46 &  $\surd$  \\ 
 48 &  7041653 & 19364121+4234085 & 15.28 && 1.040079 & 74 & 21.1 & -1 & 0.66 &  \nodata  \\ 
 49 &  7121174 & 19311355+4238202 & 15.52 && 1.105513 & 75 & 20.2 & 1 & 0.39 &   \nodata  \\ 
 50 &  7215687 & 19441629+4242014 & 15.30 && 8.767878 & 151 & 13.7 & -1 & 0.36 & \nodata  \\ 
 51 &  7282326 & 19273145+4248599 & 16.00 && 4.900259 & 190 & 14.3 & 2 & 0.48 &  \nodata  \\ 
 52 &  7284574 & 19302403+4252024 & 13.85 && 0.355377 & 44 & 29.9 & 1 & 0.61 &   $\surd$  \\ 
 53 &  7386987 & 19521066+4259063 & 15.29 && 0.402281 & 61 & 26.6 & -1 & 0.50 &  $\surd$  \\ 
 54 &  7439239 & 19202375+4303334 & 14.85 && 1.017482 & 52 & 18.0 & -1 & 0.48 &  \nodata  \\ 
 55 &  7592165 & 19035077+4315181 & 15.17 && 2.235186 & 74 & 18.8 & -1 & 0.56 &  \nodata  \\ 
 56 &  7597152 & 19123600+4317545 & 14.19 && 3.674392 & 63 & 17.4 & 0 & 0.37 &   \nodata  \\ 
 57 &  7668817 & 19052591+4323188 & 14.11 && 1.768222 & 33 & 16.1 & -1 & 0.42 &  \nodata  \\ 
 58 &  7801669 & 18500534+4335075 & 15.41 && 3.824449 & 135 & 13.3 & 2 & 0.36 &  \nodata  \\ 
 59 &  7870796 & 18495501+4336169 & 13.35 && 7.618463 & 66 & 18.0 & -1 & 0.52 &  \nodata  \\ 
 60 &  7882046 & 19134717+4340237 & 15.29 && 1.241807 & 64 & 14.9 & -1 & 0.38 &  \nodata  \\ 
 61 &  7917961 & 19581053+4339100 & 14.51 && 0.291074 & 65 & 35.4 & 1 & 0.46 &   $\surd$  \\ 
 62 &  7984390 & 19560342+4343413 & 15.79 && 0.782909 & 94 & 18.0 & 2 & 0.58 &   $\surd$  \\ 
 63 &  8128829 & 20025853+4359102 & 15.16 && 0.580214 & 62 & 18.1 & -1 & 0.43 &  $\surd$  \\ 
 64 &  8178616 & 19441508+4401077 & 13.95 && 1.018836 & 32 & 40.3 & 1 & 0.37 &   \nodata  \\ 
 65 &  8330286 & 20034696+4412196 & 13.34 && 0.956963 & 19 & 16.6 & 2 & 0.41 &   $\surd$  \\ 
 66 &  8355584 & 19105220+4419118 & 13.40 && 6.253340 & 46 & 16.7 & 1 & 0.39 &   \nodata  \\ 
 67 &  8364074 & 19250188+4419580 & 12.94 &&14.788068 & 66 & 15.6 & 2 & 0.43 &   \nodata  \\ 
 68 &  8489101 & 19180102+4433462 & 13.85 && 1.606165 & 39 & 26.3 & 2 & 0.59 &   \nodata  \\ 
 69 &  8573332 & 19461267+4437265 & 13.97 &&15.779058 & 110 & 19.9 & 2 & 0.58 &  \nodata  \\ 
 70 &  8574794 & 19475297+4436375 & 14.85 && 1.223977 & 43 & 15.5 & -1 & 0.46 &  \nodata  \\ 
 71 &  8579863 & 19525792+4438354 & 15.01 && 0.661055 & 46 & 19.4 & 0 & 0.46 &   $\surd$  \\ 
 72 &  8783707 & 20042858+4455170 & 13.95 && 1.951024 & 49 & 20.2 & 0 & 0.37 &   \nodata  \\ 
 73 &  8867434 & 18550792+4510584 & 14.12 && 5.069553 & 58 & 16.3 & -1 & 0.63 &  \nodata  \\ 
 74 &  8903917 & 19510005+4510520 & 13.87 && 0.876418 & 29 & 35.1 & 1 & 0.43 &   $\surd$  \\ 
 75 &  8956706 & 19325876+4517538 & 15.97 && 8.373183 & 415 & 44.7 & -1 & 0.60 & \nodata  \\ 
 76 &  8984572 & 20033541+4512257 & 13.72 && 3.615910 & 61 & 13.3 & -1 & 0.40 &  \nodata  \\ 
 77 &  9041683 & 19510701+4523575 & 15.38 && 1.321763 & 91 & 28.2 & 2 & 0.53 &   \nodata  \\ 
 78 &  9049660 & 19591520+4522597 & 14.75 && 4.770293 & 107 & 14.4 & -1 & 0.47 & \nodata  \\ 
 79 &  9076673 & 19042537+4524145 & 14.73 && 0.480528 & 82 & 35.7 & 2 & 0.43 &   $\surd$  \\ 
 80 &  9097652 & 19393233+4527039 & 14.39 && 9.840559 & 106 & 15.1 & 2 & 0.43 &  \nodata  \\ 
 81 &  9267794 & 19002274+4545345 & 15.52 && 3.274686 & 120 & 20.5 & 0 & 0.47 &  \nodata  \\ 
 82 &  9274173 & 19151876+4547310 & 15.00 && 4.430385 & 130 & 13.2 & -1 & 0.57 & \nodata  \\ 
 83 &  9396760 & 19134355+4555079 & 14.55 && 1.089045 & 39 & 12.6 & -1 & 0.64 &  \nodata  \\ 
 84 &  9715928 &   \nodata        & 17.56 && 0.370103 & 1250 & 31.6 & 1 & 0.59 & $\surd$  \\ 
 85 &  9791622 & 19562262+4635046 & 13.99 && 0.943949 & 65 & 94.4 & 2 & 0.42 &   $\surd$  \\ 
 86 & 10395972 & 19114454+4731120 & 15.88 && 1.255952 & 108 & 16.0 & 2 & 0.37 &  \nodata  \\ 
 87 & 10552809 & 19520658+4742178 & 14.83 && 3.584709 & 113 & 17.7 & 2 & 0.36 &  \nodata  \\ 
 88 & 10800534 & 19324624+4809312 & 15.88 && 3.416350 & 159 & 15.1 & 2 & 0.42 &  \nodata  \\ 
 89 & 11187332 & 19202414+4852350 & 15.17 && 0.305984 & 98 & 71.5 & 1 & 0.65 &   $\surd$  \\ 
 90 & 11496490 & 19024064+4926236 & 13.87 && 0.207054 & 71 & 175.3 & 1 & 0.54 &  $\surd$  \\
 91 & 11512815 & 19375412+4925566 & 14.48 && 1.631571 & 51 & 12.3 & 2 & 0.36 &   \nodata  \\ 
 92 & 11610352 & 19263403+4937144 & 15.14 && 4.377485 & 106 & 17.0 & 1 & 0.64 &  \nodata  \\ 
 93 & 11653065 & 19061659+4947435 & 14.47 && 0.284970 & 39 & 32.6 & 1 & 0.50 &   $\surd$  \\ 
 94 & 11867853 & 19352006+5009583 & 15.48 && 0.642235 & 76 & 20.5 & 2 & 0.41 &   $\surd$  \\ 
 95 & 12061096 & 19243272+5034333 & 15.61 &&15.049562 & 410 & 32.5 & -1 & 0.75 & \nodata  \\ 
 96 & 12069878 & 19423807+5031302 & 13.76 &&42.03793 & 230 & 16.3 & -1 & 0.63 &  \nodata  \\ 
 97 & 12070798 & 19440388+5030070 & 15.85 && 1.802793 & 126 & 17.8 & -1 & 0.52 & \nodata  \\ 
\enddata 
\label{KACT.tbl}
\end{deluxetable}

\begin{figure}
\centering
\includegraphics[width=0.75\textwidth]{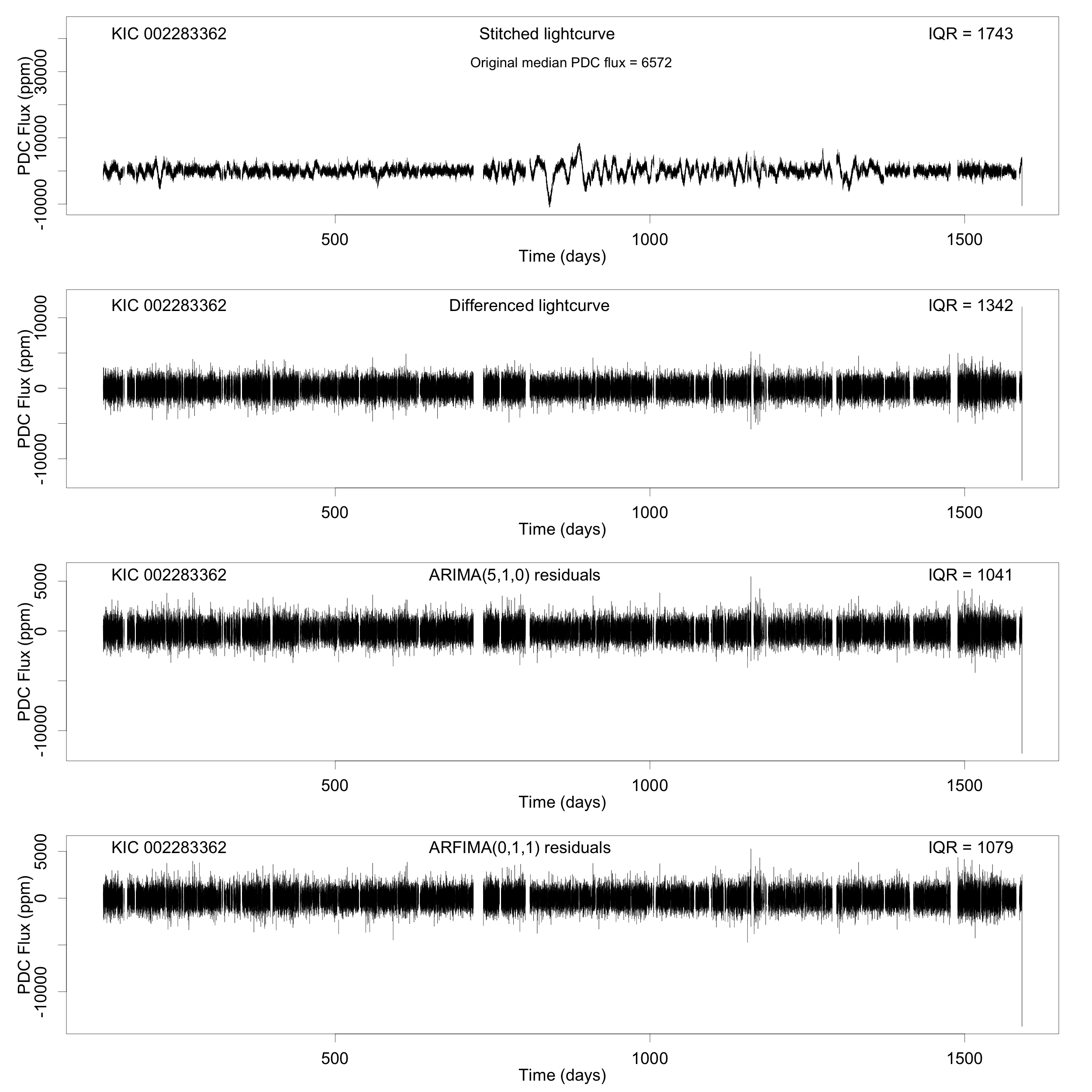}
\caption{Four pages of graphics for each of 97 Kepler ARPS Candidate Transit stars:  light curves; autocorrelation functions; TCF periodograms; and light curves folded at the `best' TCF period.  The plots for USP transit candidate KACT~2 = KIC~002283362 are shown here as an example.  The full set of 388 graphics pages appear as a Figure Set in the electronic version of this paper.}
\label{KACT_pngs.fig}
\end{figure}

\begin{figure}
\setcounter{figure}{19}
\centering
\includegraphics[width=1.0\textwidth]{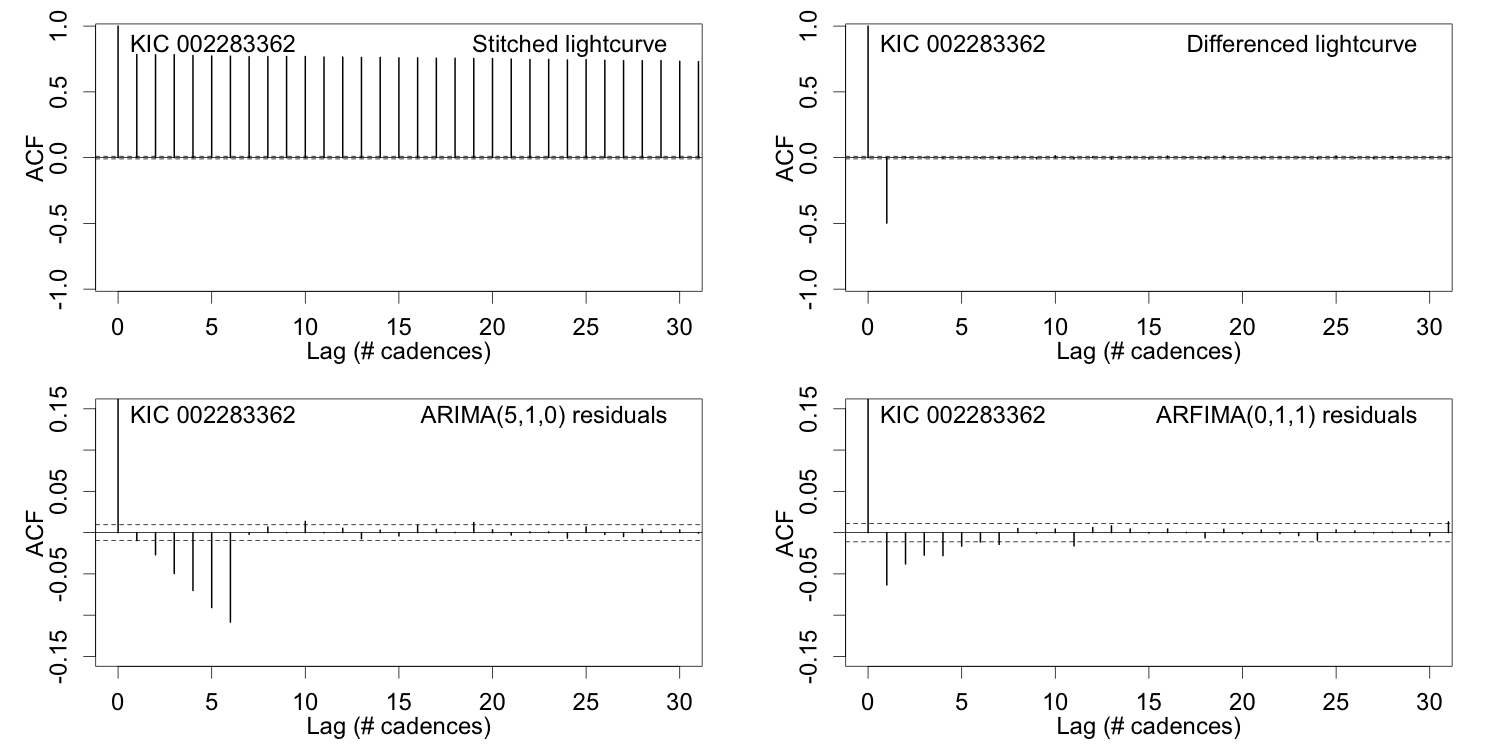}
\caption{Figure~\ref{KACT_pngs.fig} continued. }
\end{figure}

\newpage

\begin{figure}
\setcounter{figure}{19}
\centering
\includegraphics[width=0.9\textwidth]{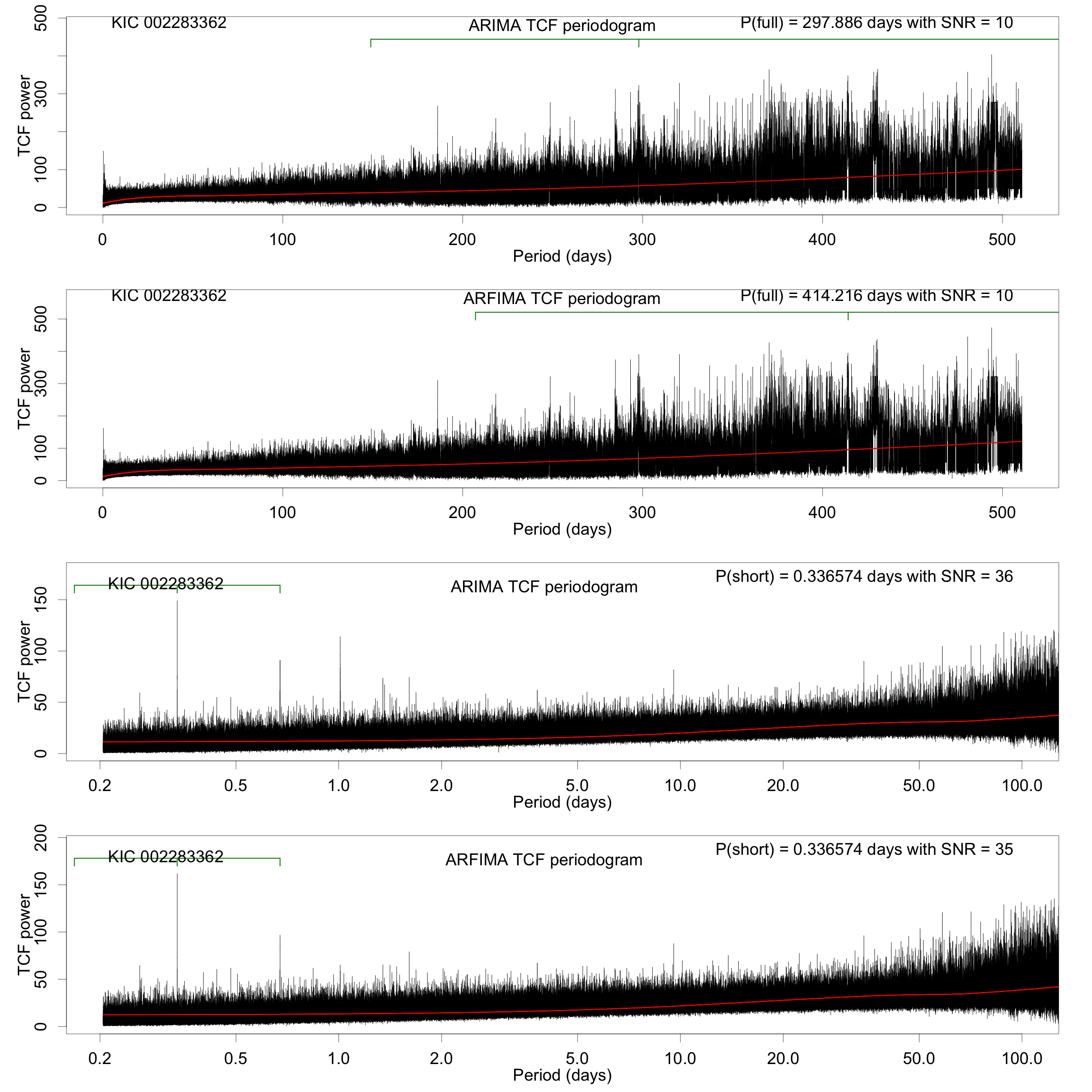}
\caption{Figure~\ref{KACT_pngs.fig} continued. }
\end{figure}

\begin{figure}
\setcounter{figure}{19}
\centering
\includegraphics[width=0.9\textwidth]{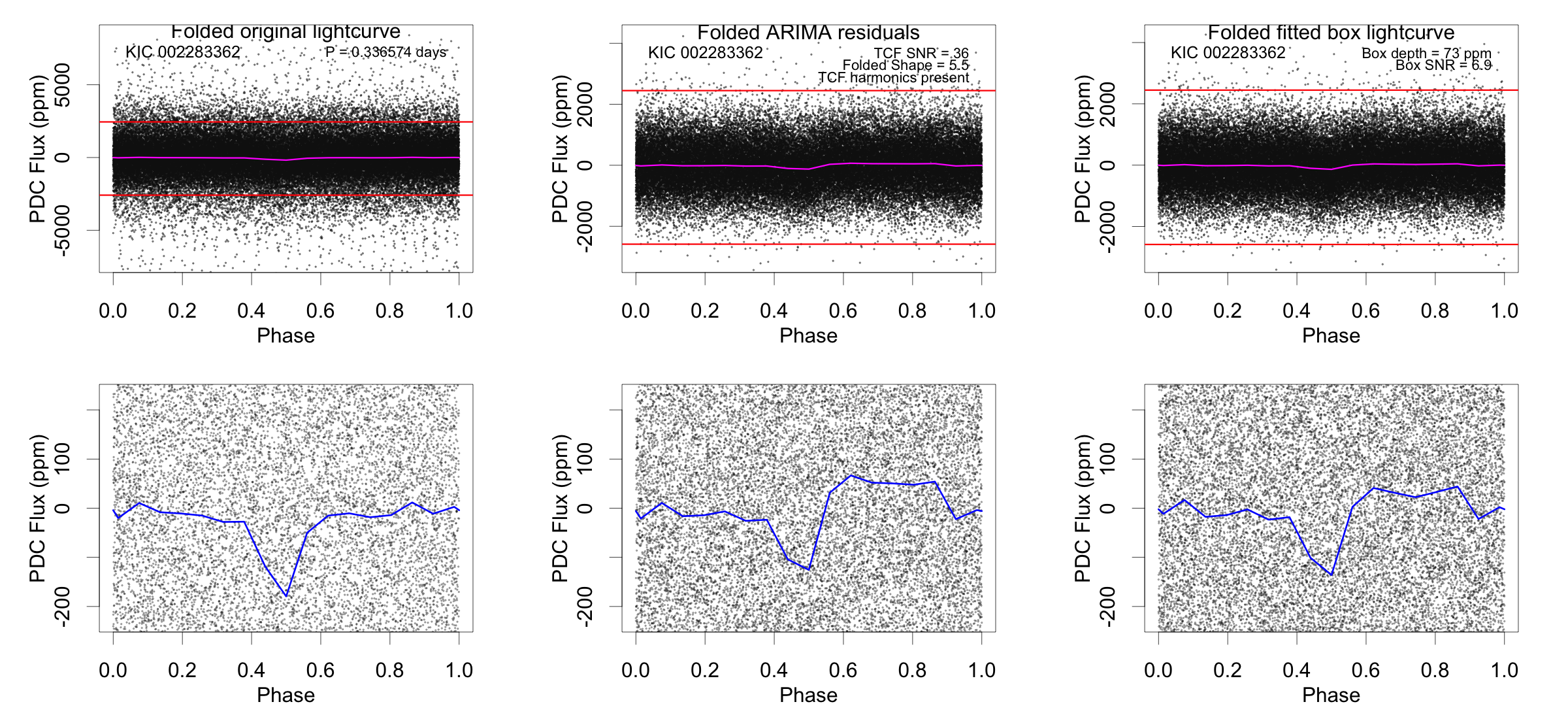}
\caption{Figure~\ref{KACT_pngs.fig} continued. }
\end{figure}

\newpage
~\\

Visual examination of the graphical output for the KACT planetary candidates in Figure~\ref{KACT_pngs.fig} rarely display the full range of planetary indicators: (i) strong peaks in the ARIMA residual TCF periodograms with clearly evident harmonics; (ii) confirmation of the TCF peak in the ARFIMA residuals; (iii) strong transit-like double-spikes in the folded ARIMA residuals; and (iv) strong transit-like box-shaped dips in the folded light curve after the ARIMAX fit is removed.  Stars with all of these clear indicators have already been identified as confirmed KOI Candidates by the Kepler Team.   Typically two or three of these four indicators are seen at weak levels.  But all received high Random Forest probabilities that reflect the collective effect of three dozen features.  The multivariate classifier can be more sensitive than the human eye in discerning subtle multivariate patterns.  

\begin{figure}
\centering
  \includegraphics[width=0.9\textwidth]{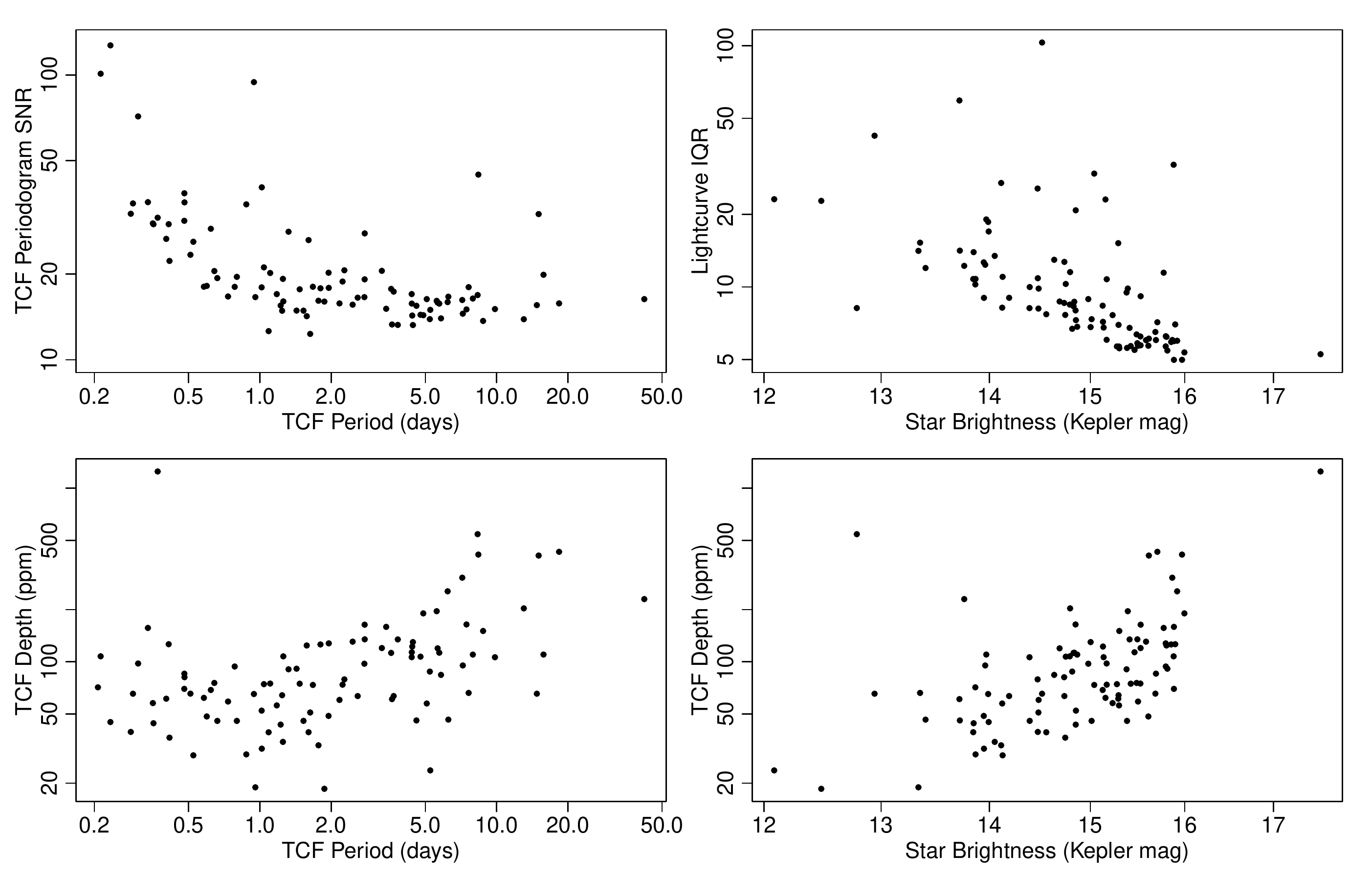} 
\caption{Properties of the 97 transit candidates discovered by the ARPS pipeline.}
\label{KACT_prop.fig}

  \includegraphics[width=0.9\textwidth]{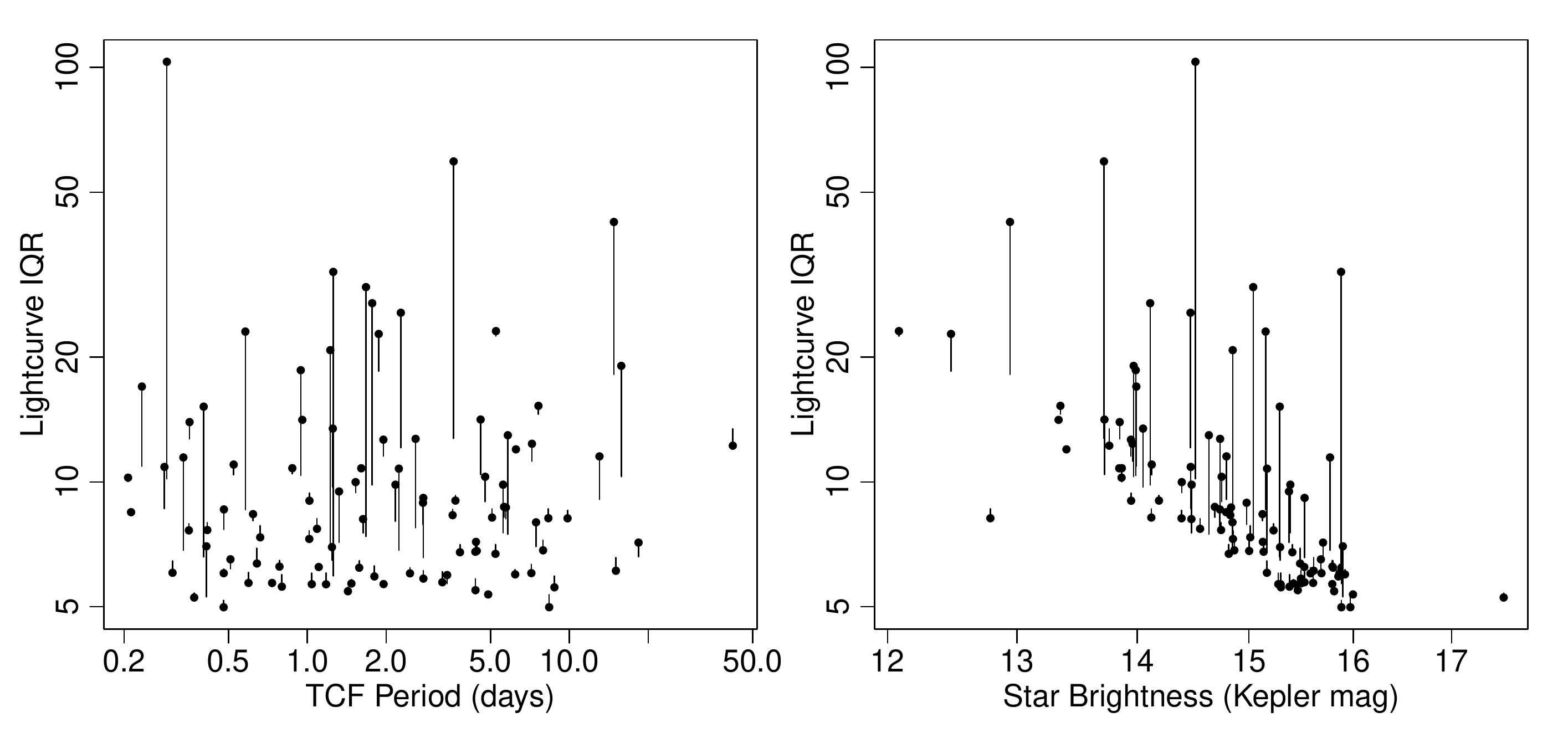} 
\caption{Effect of autoregressive modeling on the light curves of the 97 ARPS candidates. Black circles show the interquartile range of the original light curve, and the vertical lines show how much variability was reduced after ARIMA modeling.}
\label{KACT_IQR.fig}
\end{figure}

Various properties of the KACTs  are shown in Figures~\ref{KACT_prop.fig}-\ref{KACT_IQR.fig}.  The transiting exoplanet candidates have periods ranging from 0.2 to $\simeq 20$ days with host stars mostly among the fainter Kepler stars ($14-16$ mag).  Roughly 1/3 of the KACT host stars showed strong photometric variability in the original light curve for which ARIMA modeling reduced noise substantially, but ARIMA reduces autocorrelation for nearly all stars.  Perhaps the most dramatic property is the small transit depth: most are in the range $\sim 50-150$ ppm implying planetary radii only $\sim 1$\% of the stellar radius. These transit depths are typically one-tenth the amplitude of the IQR of the original lightcurves. 

The KACT discoveries, if validated, are roughly Earth-sized planets.  This result supports the completeness and reliability of the Kepler Team Planet Candidate sample at higher transit depths.  Not only does ARPS recover 97\% of confirmed KOI candidates for stronger transits (\S\ref{Eval_KOI.sec}), but ARPS does not uncover a significant population of strong transits that were missed by the Kepler Team; only 8 KACTs have estimated transit depths $>200$ ppm.  KACT ARIMAX transit depths have median $\simeq 55$~ppm, typically one-tenth of the noise level of the original and ARIMA residual light curves (stitched IQR $\simeq 570$~ppm and ARIMA residual IQR $\simeq 530$~ppm).

Especially exciting is that 29 KACTs have periods of less than one day, satisfying the designation `UltraShort Period' (USP) exoplanets. Fourteen of these have periods $P<0.5$ day, the limit of the Kepler Transit Search Procedure.  The KACT USP candidates correspond to Earth-sized planets with hot surface temperatures, probably hot enough to melt rock on the star-facing planetary surface.   Exoplanets with these characteristics are under active investigation \citep[e.g.][]{Jackson13, Steffen16, Lee17, Winn18}. 

We find that the large majority of the 97 KACTs are new to this study.  The following outlines relationships to previous work:
\begin{enumerate}

\item One of the 97 KACTs had been previously viewed as a probably planetary candidate:  KACT~89, with one of the strongest TCF signals among the KACT objects, was identified by \citet{SanchisOjeda14} as an UltraShort Period `new planet candidate' with the same $P=0.305$ day period as found with ARPS.  

\item Nine of the KACTs are previously listed in a Kepler Objects of Interest (KOI) list but did not receive a final disposition of `Candidate Planet'.  Three situations are encountered: the TCF periodogram recovers the KOI period but with high confidence it is a transit candidate (KACT 41, 64); the TCF periodogram does not confirm a signal at the KOI period (KACT 2, 70, 75); and the KOI period is a long-period harmonic the ARPS period (KACT 26, 39, 83, 90). This last situation occurs particularly when ARPS finds an UltraShort Period candidate with $P<0.5$ day, shorter than the range of periods examined by the Kepler team.  

\item In a dozen cases, KACTs are identified where the TCF periodogram unambiguously demonstrates a periodicity is present: the peak is far stronger than the noise, harmonics are clearly present, and the signal is present in both the ARIMA and ARFIMA residuals.  These are KACT 37, 52, 61, 68, 69, 74, 75, 77, 82, 85, 89, and 90. Five of these stars satisfy the criteria for the Kepler team's intermediate category called `Threshold Crossing Event' (KACT 37, 74, 75 and 90), two of which received a KOI designation.  Four of the 5 proposed TCE periods are equal to, or harmonics of, the ARPS periods. It is surprising that the Kepler Transit Search Procedure did not identify at least some of these cases.  
\end{enumerate}

\section{Discussion}  \label{Disc.sec}

\subsection{General Comments} \label{Disc_general.sec}

The ARPS transit detection procedure has both important differences and similarities to previous approaches to planetary transit detection.  The greatest difference arise from the ARIMA fits to reduce stellar variability.  ARIMA is a straight-forward maximum likelihood linear regression method with automatic model selection using the Akaike Information Criterion, in wide use for time series analysis since the 1970s under the rubric `Box-Jenkins method'.  The procedure has only one specified parameter relating to the maximum order (complexity) of the model; we choose $p+q \leq 10$ and $d=1$. Gaps in the data stream are not filled, as the algorithm accepts missing data.  In contrast, the Kepler {\it Transit Planet Search} procedure uses a complicated sequence of piecewise polynomial fitting for outlier removal, iterative Fourier whitening for reduction of quasi-periodic variability, exponential detrending at the edges of quarters, long gap filling with sigmoidal functions, and wavelet-based matched filters for identification of candidate transits \citep[][Chapter 9]{Jenkins17}.   ARIMA also differs from Gaussian Processes regression procedures for stellar variability reduction \citep[e.g.][]{Petigura13} where a global nonlinear kernel defines the autocorrelation structure but a high-dimensional fit is made locally.  

The later stages of ARPS analysis bear strong similarities to some previous treatments.  Our TCF periodogram is similar to the commonly used Box Least Squares periodogram \citep{Kovacs02} but adapted to differenced light curves.  It has different noise characteristics and different sensitivities to long-duration transits and outliers, but the TCF and BLS periodograms for a given light curve typically show analogous features (Paper I, \S3.4).  Our use of a Random Forest classifier is similar to the Kepler team's automated procedure described by \citet{McCauliff15} to replace the earlier labor intensive visual inspection of light curves and detection statistics.  And all methods end with a visual vetting procedure to treat the leakage of some non-transiting lightcurves through the earlier discrimination stages. 

We were surprised how effectively the ARPS methodology tackled the complexities of the Kepler small planet discovery effort.  Many technical issues affecting Kepler light curves treated by the Kepler Team pipeline \citep[][and \S\ref{Disc_limit.sec} below]{Jenkins17} and a great diversity of intrinsic stellar variability behaviors, are subsumed into the ARIMA modeling without special treatment.    There was little {\it a priori} reason for confidence that low-dimensional linear autoregressive models could explain these effects well.  The positive results obtained here (\S\ref{resultsAR.sec}), combined with the effectiveness of ARPS in simulations of irregularly-spaced light curves \citep{Stuhr19}, suggests that ARIMA- and ARFIMA-type parametric modeling may be a broadly useful tool for many problems in time domain astronomy \citep{Feigelson18}.  

Several aspects of the ARPS procedure for Kepler planet discovery can be discussed further:

First, ARIMA and ARFIMA modeling removed most of the trends, including quasi-periodic variations from rotationally modulated starspots.  It substantially reduced the overall noise and nearly eliminated the autocorrelation in most Kepler light curves (Figures~\ref{IQR1.fig}-\ref{DW.fig} and Table~\ref{lc_improv.tbl}).  This success is characteristic of Box-Jenkins analysis which has been used successfully to model a vast range of terrestrial applications \citep{Box15}. But perhaps there is an additional astrophysical basis for the model success: short-memory photometric variations in late-type stars may be related to solar flaring that has been interpreted as a simple autoregressive `avalanche' process \citep{Lu91, Aschwanden16}.  

Second, we were very pleased to find that the periodic signal of planetary transits were not absorbed into the ARIMA models of aperiodic behavior.  This is a risk for all methods of stellar variability reduction (wavelet transforms, Gaussian Processes regression, Independent Component Analysis, and so forth) and appeared to occur weakly in our analysis using ARFIMA modeling (Figure~\ref{AR_ARF_peak.fig}).   This achievement is easily explained: ARIMA is a maximum likelihood estimation procedure that weights all data points equally.  As the transit durations are typically a very small fraction of the orbital periods, only a small fraction of observations are affected by the planetary transit and are essentially ignored by the ARIMA model unless the transits are extraordinarily deep.  Note that in radial velocity studies, all of the observations are involved in the planetary signal, and an ARIMA model of the data cannot be applied in the simple fashion used here for transit studies.  

Third, the TCF periodogram proved to be highly effective in detecting periodicities arising from transits and other astronomical phenomena despite limitations discussed in Paper I (\S3). In particular, TCF only responds to the few observations during transitions in and out of transits, and should thus less sensitive than the traditional box-fitting procedures for long duration planets.  This is discussed in Paper I (\S3.4) and is seen in the Kepler dataset (Appendix, Figure~\ref{RFmissedFeat.fig}). Since long durations are characteristic of long period planets in circular orbits, this likely accounts for the weakening performance of ARPS planet detection as periods approach or exceed $\sim 100$ day periods (Figure~\ref{ARPS_vs_missed_KOI.fig} right panel).  For this reason, we restricted most of our study to periods shorter than 100 days.  Issues relating to long-period planets, such as discovery of Earth analogs, are thus not treated here.  TCF is also very sensitive to chance alignment of flux outliers, such as stellar flares, that occur at longer periods.  We remove these cases during the subjective vetting stage.  But despite these worries, within the constraint of $P < 100$ days, the sensitivity and reliability of TCF periodograms with respect to confirmed planetary signals was found to be excellent (\S\ref{Eval_KOI.sec}).  

Fourth, Random Forest recovery of 97\% KOI confirmed planets with matched periods is also good, although a high recovery fraction of the original training set is expected from any competent classifier. But it is satisfying that potential problems with badly imbalanced training sets, and the heterogeneity of the True Negative training set, were overcome.   The principal failure of RF was recovery of the weakest KOI transits (Figure~\ref{ARPS_vs_missed_KOI.fig} and the Appendix).  We do not know, in individual cases, whether this failure is due to insensitivity of the ARPS procedure (e.g., it is reasonable that wavelet denoising is more effective than low-dimensional ARIMA modeling some stars) or to errors (False Alarms) in the Kepler pipeline procedure.  The detailed discussion of the Kepler Team Random Forest classifier by \citet{McCauliff15} indicates that, quite naturally, true positive and false alarm cases are difficult to discriminate for putative weak transit signals.  In the latter cases,  ARPS would be correct in assessing that the unrecovered weak-KOI transits are not true planetary systems.  The same issue appears in reverse with respect to the 97 KACTs we report in Table~\ref{KACT.tbl}.  These cases either demonstrate better performance of ARIMA modeling and TCF periodograms than standard methods, or they are False Alarms or False Positives that have incorrectly passed our vetting procedure.  The solution to this important question cannot be made from statistical characteristics; additional astronomical observations and considerations are needed.   

Fifth, for badly imbalanced classes in the test set, a high recovery rate does not mean the classifier is scientifically useful if it also accepts a large False Positive population.  The right panel of Figure~\ref{ROC.fig} shows that the rejection of Certified False Positives is only moderately effective.  The table stub for Table~\ref{ARPSfull.tbl} shows two examples of successful elimination of KOIs periodic signals that are not planetary in nature.  This indicates that our multivariate machine learning approach to planet selection, including our efforts at feature engineering and treating the imbalanced training sets (Table~\ref{RFfeatures.tbl}), is effective to some degree, although not sufficiently strongly that a final subjective vetting step can be avoided.  

Sixth, the result of RF application to the full Kepler dataset, after vetting of likely False Alarms and False Positives, emerged with 97 Kepler ARPS Candidate Transits (\S\ref{KACT.sec}).  This is a reasonable number of candidates, not implausibly populous and not too small to be scientifically uninteresting.  The KACT transit depths are very small (typically $50-100$~ppm); it is therefore not surprising that examination of graphical outputs often show weak, and often visually unconvincing, signals.   We must rely on the collective power of many `features' and the effectiveness of the Random Forest algorithm to assert that they are strong candidates.  

The ARPS effort generally has strong results, with its most promising achievements in the discovery of new candidate planets with very short orbital periods and/or very small radii (Table~\ref{KACT.sec}).  But in some aspects, the ARPS methodology performed relatively poorly.  The ARPS procedure struggles with sudden flares or other instances of volatility in the light curves.  These situations can benefit from other statistical approaches, such as `change point detection' \citep{Tartakovsky14} or non-linear autoregressive models like GARCH \citep{Francq10}.  Although ARPS nominally gives two  estimates of transit depth and duration (one from the TCF peak and the other from the ARIMAX modeling), neither are trustworthy for astrophysical understanding.   Realistic transit models that include planet shape, impact parameter, stellar limb darkening, and sub-cadence structure are needed \citep{Mandel02, Hippke19}.  ARPS is designed for accurate classification and not for accurate estimation of astrophysical parameters and planetary properties.  

\subsection{Limitations Specific to the Kepler Dataset} \label{Disc_limit.sec}

{\bf Kepler operational issues}~~ The Kepler satellite is a very complicated instrument exhibiting many hardware and operational effects during its Prime Mission \citep{VanCleve16}.  These include: occasional loss of fine pointing control; satellite pointing degradation due to guide star variability; data loss due to Module 3 failure; data corruption due to the bright variable star CH Cyg; sub-pixel image motion at the edge of the field; diffuse illumination of the focal plane attributed to impact-generated debris; and pixel level flux variations due to cosmic ray hits.  Some of these hardware effects may be periodic and can produce spurious peaks in periodograms. These include: spurious spectral peaks at multiples of the Long Cadence sampling rate (566.4 $\mu$Hz = 29.42 min); a periodic 3$-$6 hour non-sinusoidal variation due to focus changes from a reaction wheel heater cycle; 3-day effect due to satellite momentum management; thermally-induced plate scale and focus changes over a 3-month quarter; and an annual variation in focus and point response function due to satellite temperature variations.  None of these effects are treated in the present study. 

{\bf PDC algorithm effects} ~~ Any algorithm that seeks to reduce instrumental or stellar variations to enhance sensitivity to planetary transits may also inadvertently incorporate the transit signal into the noise model. PDC processing is designed to treat faster variations as potential transits and slower variations as unlikely transits. With experiments based on injections of sinusoidal signals into the unprocessed light curves, the Kepler team establishes that true periodic signals will be attenuated by more than 10\% at periods longer than 10 days, more than 50\% at periods longer than 20 days, and approaching full attenuation at periods longer than 30 days \citep{Gilliland15, VanCleve16}. However, strong signals can overwhelm the processing and be detected in the PDC flux.  None of there effects are treated here. 

{\bf Quarterly analysis} ~~ In the present work, the full 4-year light curves were processed by a single ARIMA model for each star. This has the advantage of providing a large number of data points for parameter estimation, but handles all quarters in the same way. Since we sometimes observe individual quarters with very idiosyncratic behavior, it is possible that processing each quarter separately, or excluding problematic quarters, could lead to improved performance.

{\bf Noise comparison} ~~ The light curve's IQR was used to help assess how well ARIMA models capture stellar variability. This is a robust way to quantify the dispersion between all data points, but describes the data only on a global scale. To better differentiate between longer-scale variations and the noise that may obscure individual transits, it may be useful to quantify the background at a more local level. One such approach is to use Kepler's Combined Differential Photometric Precision (CDPP) which corresponds to the observed noise within a time window relevant to planet searchers \citep{Gilliland11, Christiansen12, Jenkins17}. CDPP is calculated in the Transiting Planet Search (TPS) module of the Kepler pipeline and contains estimates for different time windows per quarter for each star.  The CDPP is not used in the analysis here.  

{\bf Barycentric correction of signal arrival times} ~~ Since the formalism of the ARPS analysis requires equally-spaced time series, all of our work was done in terms of cadence. In practice, a periodic signal of interest will be subject to barycentric variations as the satellite orbits around the Sun. For the Kepler field which lies away from the ecliptic there is an offset of approximately 3 minutes at either side of the Sun \citep{Murphy13}.  This barycentric correction is not treated here. 

ARPS analysis takes cognizance of these effects only if they were removed during the Pre-Data Conditioning analysis and/or produced a non-zero quality flag in PDC fluxes. ARPS analysis differs from some other procedures by conflating instrumental, stellar and planetary transit variations into a single temporal process without attempting to separate them at an early stage. ARPS also ignores any differences in uncertainties of PDC fluxes (heteroscedasticity), treating each photometric observation equally to any other. 

A scientifically important point is that the ARPS procedure recovered less than half of the DR25 Candidate Planets with weak transit depths, $ModelSNR < 20$ (Figure~\ref{ARPS_vs_missed_KOI.fig} and the Appendix).  In some cases, this is attributable to the known reduced sensitivity of TCF compared to BLS for weak periodicities at long periods (Paper I, \S3.4), but in many cases ARPS failed to detect the signals at $3 \lesssim P \lesssim 100$ days.   Two explanations plausibly account for these discrepancies in different cases:  either the transits are False Alarms from the Kepler pipeline and do not physically exist; or the transits are real, captured by the sophisticated analysis procedure of the Kepler pipeline but not by our simpler ARPS procedure.  The present study cannot discriminate between these possibilities; additional astronomical examination is needed.
  
Finally, a quantitative analysis of ARPS's efficiency in retrieving planetary transit signals within the complex time series of realistic stellar and instrumental variations has not been attempted. Evaluating the efficiency of ARPS transit detection may be possible by applying ARPS method to artificially injected planetary signals into real light curves.  These injections would be similar to those of \citet{Christiansen16} but concentrating on the regime of KACT discoveries: small Earth-size planets orbiting at short periods. Only after this is done can reliable conclusions about the intrinsic population of exoplanets be inferred from the observed sample reported here.  This lies beyond the scope of the present study.  

\subsection{Future work} \label{Disc_future.sec}

The present effort to apply ARPS methodology to the 4-year Kepler mission dataset can be viewed as the beginning of a continuing effort to use parametric time series modeling and machine learning classification to further understanding of variability in cosmic populations:
\begin{enumerate}

\item The report here of 97 new exoplanetary candidates (\S\ref{KACT.sec}) encourages astronomical examination to confirm new transiting systems or place them in categories like False Alarms or False Positives.   This includes more precise characterization using improved estimates of stellar properties (e.g. based on Gaia mission distances), high-resolution imaging, and follow-up spectroscopy.  

\item The autoregressive modeling can be improved.  In addition to avoiding problems with model selection with more computational effort (\S\ref{armod.sec}), some variability is poorly treated by ARIMA and ARFIMA models.  Broader parametric families like GARCH can be tried.  The ARIMAX analysis in particular can be more mature with astrophysically realistic transit models and more careful computation.  

\item The multivariate classification might be improved with methods such as gradient boosted decision trees such as XGBoost or convolutional neural networks (`Deep Learning'). 

\item The ARPS procedures can be applied to other space-based transit datasets from NASA's Kepler K2 and TESS missions, and ESA's planned PLATO mission.  

\item The ARPS procedures can be applied to ground-based transit datasets.  \citet{Stuhr19} has examined the viability of converting ground-based surveys with irregular observing cadences to a regular cadence with high fractions of missing data (NA values).  These simulations, based in part on the deepesst transiting light curves from the Kepler 4-year dataset studied here, look promising for surveys with sufficiently dense cadence patterns such as the 3-telescope HAT-South network and the South Pole AST3 telescope.

\item Autoregressive modeling may be useful for a variety of other time domain problems in astronomy:  identification of weak eclipsing binaries, classification of heterogeneous ensembles of variable star light curves, automated detection of stellar flares, and study of both stellar and extragalactic accretion systems.  Some of these broader considerations are discussed by \citet{Feigelson18}. 
\end{enumerate}

\section{Conclusions} \label{Concl.sec}

We apply here the Autoregressive Planet Search (ARPS) statistical procedure, detailed in Caceres et al. (2019, Paper I), to identify new candidate transiting planets from Kepler mission light curves. It is founded on ARIMA-type parametric regression models that have a long heritage of flexibly and effectively modeling time series with complicated autocorrelated and trend behaviors at low dimensions. Best-fit ARIMA models are calculated by maximum likelihood estimation and model selection with only one parameter to limit model complexity, followed by regression diagnostics to evaluate the success of the model. Planetary transits are then sought in model residuals based on the periodogram constructed with the novel Transit Comb Filter algorithm presented in Paper I (\S3). The light curve of the model residuals folded at the period of the TCF periodogram peak is characterized, and the ARIMA model is calculated again with this periodicity as an exogenous variable.  

Using planets confidently identified by the Kepler Team as a training set, a Random Forests decision tree classifier is developed based on decision trees from a collection of several dozen features drawn from the original light curve, TCF periodogram,  ARIMAX model, and the`best' period folded light curve.  Our use of Random Forests is an example of the well-warranted proliferation of multivariate classifiers for astronomical statistical challenges.  A single RF probability threshold is applied ($P_{RF}>0.35$), and expert visual vetting is conducted to remove remaining False Alarms and False Positives.  

The effort results here in identification of 97 Kepler ARPS Candidate Transits for further study (Table~\ref{KACT.tbl}).  These are mostly (sub)Earth-sized planets orbiting close to the host star, many of them with UltraShort Periods.  Typical transit depths are $\lesssim 100$~ppm, much smaller than the noise level in the original light curves.  The ARPS analysis is reasonably computationally efficient, requiring $\sim 30$ CPU-minutes per Kepler star.

Our objective is to introduce an approach to transit detection that is complementary, with a different mathematical foundation, to transit detection procedures in common use. ARIMA-type analysis of non-planetary stellar variations are modeled using low-dimensional parametric models rather than non-parametric  approaches like wavelet decomposition or Gaussian Processes regression. We do not believe that any single method will out-perform the others in all cases. Rather, different methodologies can capture different behaviors of complex light curves. Autoregressive modeling has strengths such as unique maximum likelihood solutions, AIC-based approach to model complexity,  parameter confidence intervals, and few subjective threshold choices. The sequence of ARIMA-type modeling with TCF periodograms and Random Forest classification proves to be a particularly effective combination of methods for the specific goal of transiting planet detection.  Perhaps the most satisfying product of the ARPS analysis $-$ though scientifically less novel than new planet discovery $-$  is the widespread agreement with Kepler results, except for low-significance transits (Kepler ModelSNR$<20$).  

We can make a broad comparison of the ARPS effort with other planet searches conducted independently of the Kepler Team procedures which are outlined in \S\ref{intro.sec}.  Similar to the outcome in Table~5, \citet{Ofir13} and \citet{Huang13} produced lists of several dozen new candidate transits, primarily based on different time domain procedures for reducing stellar variability.  However, their procedures, as well as those of \citet{Jackson13} and \citet{SanchisOjeda14}, required specification of a variety of thresholds of various quantities such as median filter widths and BLS peak power.  In contrast, the initial stage of ARIMA modeling in ARPS has no free parameters, and the later stages have only one threshold to be specified ($P_{RF}=0.35$ in \S\ref{resultsTransits.sec}). ARPS and the other methods all require manual vetting of candidates surviving automated procedures.  

The results achieved so far with ARPS for planet detection are promising for application to other space-based transit surveys such as K2, TESS and Plato.  Other applications can be envisioned such as selecting eclipsing binaries, mutually illuminated binaries, and flaring stars rather than planetary signals.  We are also investigating its application to irregularly spaced time series characteristic of ground-based photometric transit surveys where the noise characteristics are dominated by atmospheric and instrumental effects, rather than by stellar variability \citep{Stuhr19}. Although ARIMA modeling is designed for evenly spaced time series, we find reasonable sensitivity to planets providing the observing cadence is sufficiently dense.  

\acknowledgements
We appreciate valuable discussions with Eric Ford, Ron Gilliland, Angie Wolfgang, and Jason Wright at Penn State.  Members of the NASA-Ames Kepler Team and an anonymous referee provided helpful advice.  EDF and GJB are affiliated with Penn State's Center for Astrostatistics. MC thanks the support from Centro de Astrofisica de Valparaiso and Centro Interdiciplinario de Estudios Atmosfericos y Astroestadistica. The ARPS project is supported at Penn State by NASA grant 80NSSC17K0122 and NSF grant AST-1614690.  

This paper includes data collected by the Kepler mission. Funding for the Kepler mission is provided by the NASA Science Mission Directorate.   Most of the data presented in this paper were obtained from the Mikulski Archive for Space Telescopes (MAST). STScI is operated by the Association of Universities for Research in Astronomy, Inc., under NASA contract NAS5-26555. Support for MAST for non-HST data is provided by the NASA Office of Space Science via grant NNX13AC07G and by other grants and contracts.  We are grateful to Scott Fleming and the MAST staff for assistance with both acquiring and depositing archive data.  Computations for this research were performed on the Pennsylvania State University's Institute for CyberScience Advanced CyberInfrastructure (ICS-ACI).

\facility{Kepler}

\appendix
\section{Further Comparison with Kepler Objects of Interest} \label{KOIcomp.sec}
\label{Comp_KOI.app}

Figure~\ref{RFprob_hist.fig} shows that the classifier has a stronger performance discriminating the cleaner samples of high-quality Kepler Team Candidates and Certified False Positives (left panel). But that performance partially degrades when considering the full list of KOIs (right panel). This Appendix examines ARPS results for the subsample of KOIs that were rejected by the Kepler Team as not satisfying strict their criteria to be confirmed exoplanetary Candidates. Disagreements between ARPS and KOI dispositions can be investigated: they can arise either because a transit claim is unreliable, or because different mathematical approaches have different sensitivities to weak transits.

\begin{figure}
\centering
  \includegraphics[width=\textwidth]{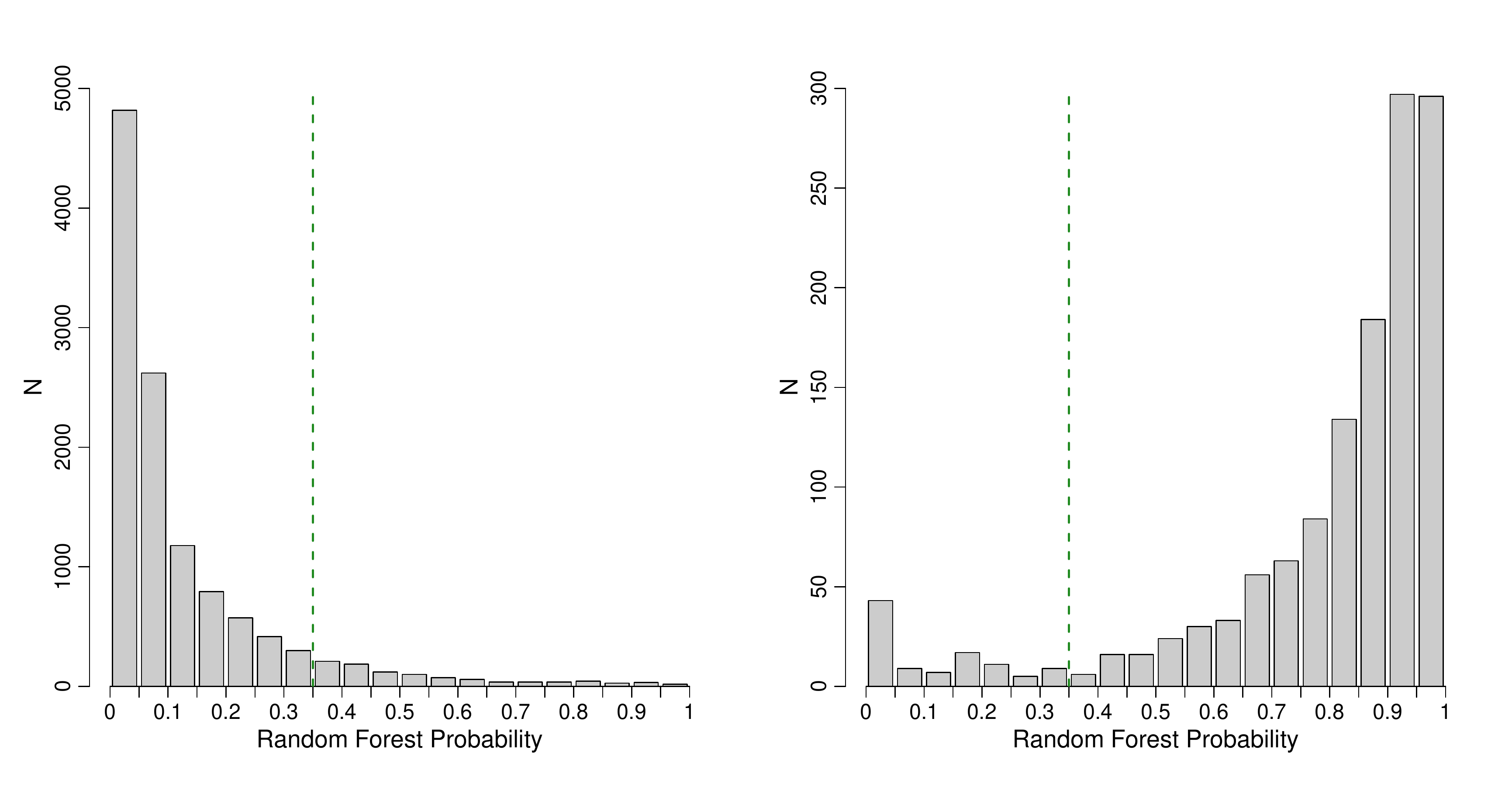} 
\caption{Histograms of Random Forest probabilities. Left panel shows 11,669 stars not included in training (corresponding to TCEs and KOIs not satisfying our training criteria). Right panel has 1,340 Kepler Objects of Interest with a disposition `CONFIRMED' on the NASA Exoplanet Archive.}
\label{RFprob_hist2.fig}
\end{figure}

Section~\ref{training.sec} discussed that 11,669 stars were left out during training of the RF classifier: stars with Threshold Crossing Events but not other characteristics, stars satisfying KOI criteria in early but not final data releases, and stars classified as False Positives for various reasons. Here, we are not referring to a hold-out set to test the model (the OOB prediction is used for this), but rather questionable candidates which we did not want to bias the model for or against being a transit.  These light curves have some type of noticeable signal since they were included in the corresponding list, yet the nature of this signal is not clear. Nevertheless, they are still run through our classifier and assigned a probability of being a transit by ARPS. These results are plotted on the left panel of Figure~\ref{RFprob_hist2.fig} showing that only 977 (8\%) are accepted by the Random Forest.  The RF probabilities here  contrast strongly with that seen for KOIs classified as confirmed Candidates (right panel).  The ARPS analysis thus concurs with the Kepler Team evaluation that most of these 11,669 stars are not convincing exoplanetary candidates.  However, the 977 stars satisfying the ARPS RF classifier threshold deserve reexamination.  

\begin{figure}
\centering
  \includegraphics[width=\textwidth]{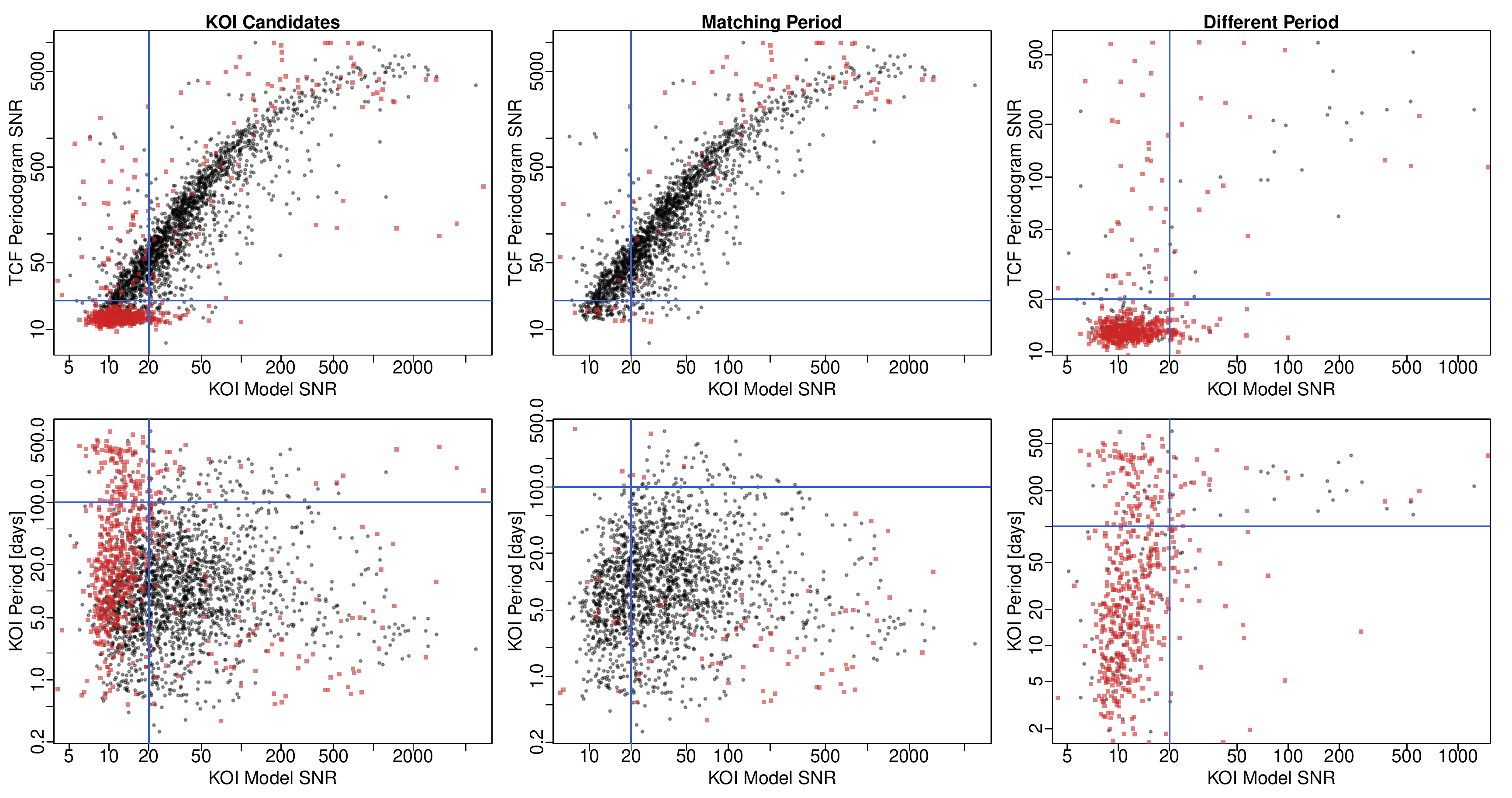} 
\caption{Comparison of Kepler Objects of Interest with Kepler Team `Candidate' disposition recovered (black circles) and missed (red squares) by the Random Forest classifier. The leftmost column is analogous to Figure~\ref{ARPS_vs_missed_KOI.fig} which showed matching TCF periods, but for Random Forest prediction. The other columns show the prediction breakdown for matching periods (center) and different periods (right). Blue lines mark boundaries for $P=100$ days and $SNR =  20$ to guide the eye.}
\label{RFmissedSNR.fig}
\end{figure}

A closer look at the KOI transit parameters can help us better understand the reasons some were missed by the RF classifier. The leftmost column of Figure~\ref{RFmissedSNR.fig} is analogous to Figure~\ref{ARPS_vs_missed_KOI.fig} but now shows whether the Random Forest correctly classified the KOI candidate rather than simply whether the periods match.  Most stars missed by the ARPS classifier (red circles) have KOI Model $SNR < 20$, TCF peak $SNR < 20$, and mismatched KOI-TCF periods. These are cases where either ARPS is not sensitive enough to capture the transit, or that the KOI signal is spurious. 

Figure~\ref{RFmissedSNR.fig} also shows some small subsets of unusual stars.  The few missed by the classifier (red symbols) that have strong TCF signals in these diagrams probably have a strong false positive signal which lead to the rejection, but further analysis might recover the KOI planet.  There are also 96 stars without matching periods passed ARPS automatic classification. An undiscovered planet might supplement the planet found by the Kepler Team in these cases.  There are also a few dozen cases where the KOI and TCF periods exactly match, but the ARPS classifier rejected the stars as an exoplanetary system.  These cases may be astronomical False Positives that were incorrectly given a Candidate disposition by the Kepler Team.   

\begin{figure}
\centering
  \includegraphics[width=\textwidth]{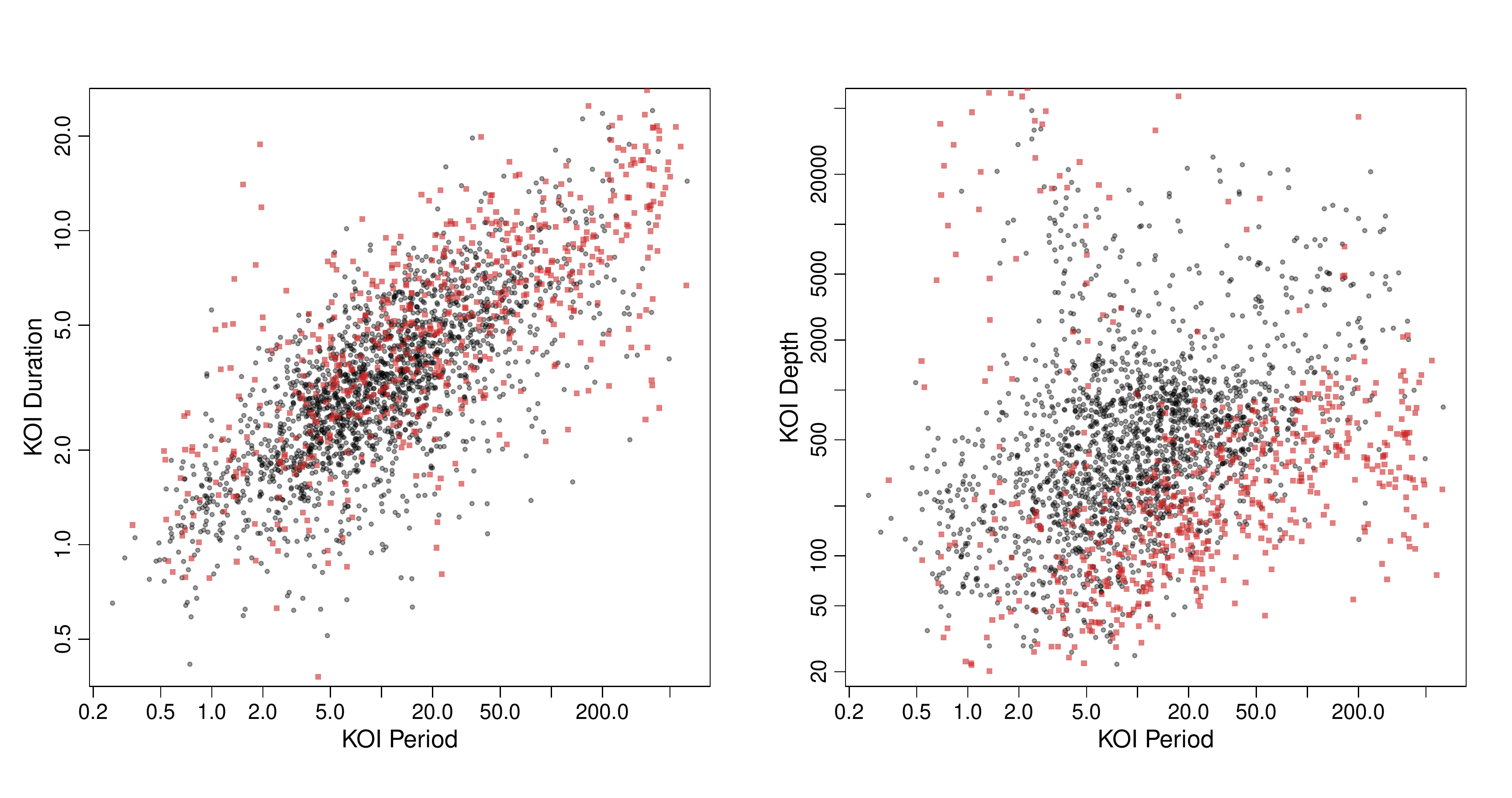} 
\caption{Examination KOI transit period, duration, and depth of KOI Candidates recovered (black circles) and missed (red squares) by the Random Forest classifier. The units of period is days, duration is hours, and depth is parts-per-million.  }
\label{RFmissedDur.fig}
\end{figure}

It is also of interest to see how well our classifier performs with respect to transit parameters like duration and depth. The effect of time series differencing on the transit signal is an ongoing worry, particularly for longer duration transits. Figure~\ref{RFmissedDur.fig} confirms that ARPS has a higher rate of recovering KOI confirmed Candidates at shorter periods ($< 20$ days) and durations ($< 5$ hours) than at longer periods, although it is not clear how much of this effect is due to differencing compared to higher periodogram noise at longer periods ($>100$ days) and durations ($>10$ hours). Most of the confirmed KOIs missed by the RF classifier lie in a specific area of a period-depth diagram, roughly below a line defined by $Depth = 50 \times P^{0.6}$ where depth is measured in parts-per-million and period is measured in days.  It is not clear whether this is due to the limits of ARPS sensitivity compared to the Kepler pipeline and/or to a prevalence of misclassified False Alarms by the Kepler Team in that region of the diagram.  A handful of KOI Candidates with short period and very high depth were rejected by ARPS, presumably because other features in the classifier pointed towards a False Positive identification. 

\begin{figure}
\centering
  \includegraphics[width=\textwidth]{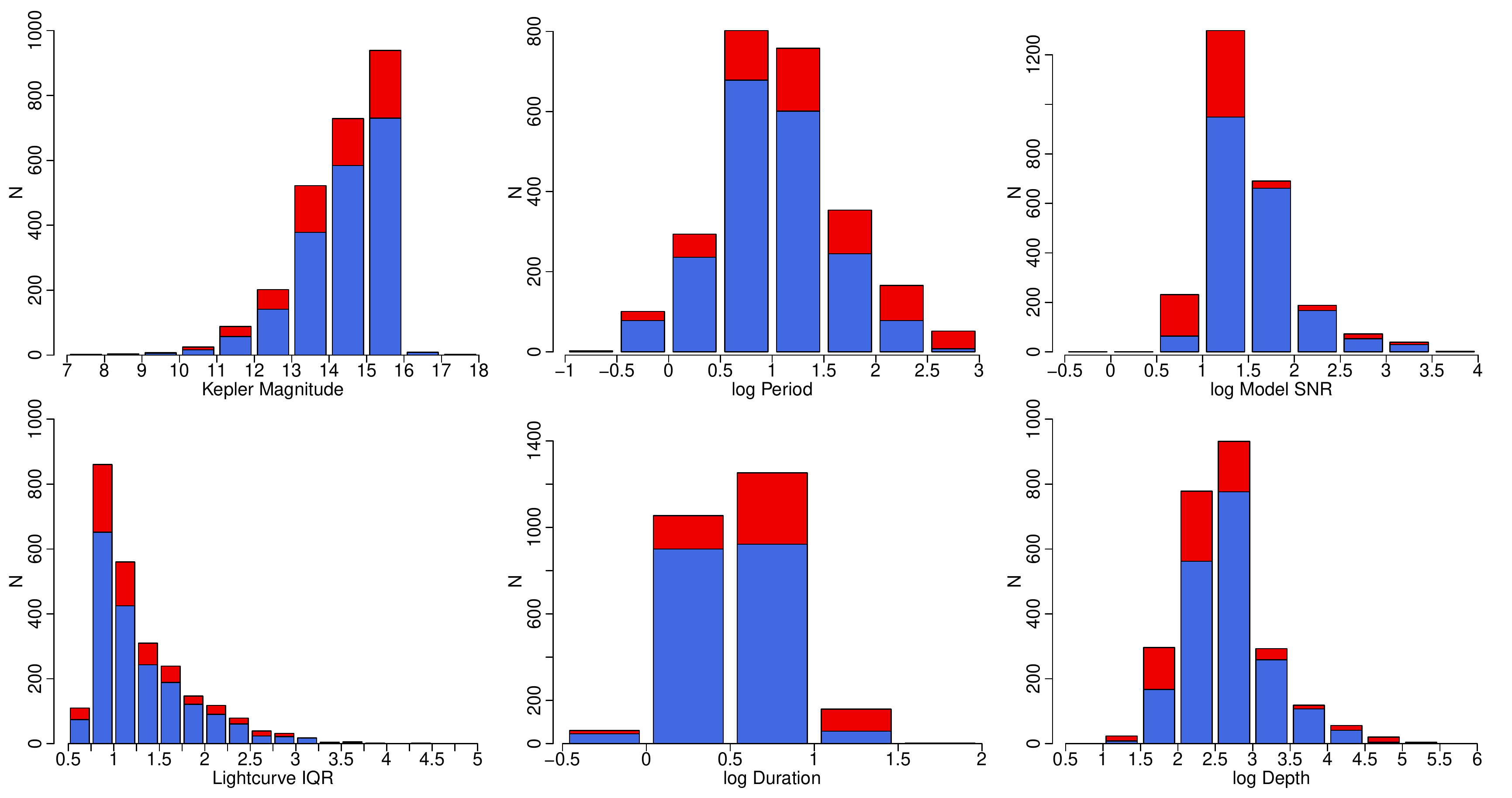} 
\caption{Histograms of light curve and KOI transit parameters of Candidates recovered (blue) and missed (red) by the Random Forest classifier.}
\label{RFmissedFeat.fig}
\end{figure}

Figure~\ref{RFmissedFeat.fig} shows histograms of several KOI transit parameters to give further insight into ARPS performance. As expected, a larger fraction of confirmed KOI candidates are recovered at higher transit depth and KOI Model SNR, and the reduced sensitivity at longer periods and duration is again seen. An interesting result is that a higher fraction of KOI candidates are missed at lower IQR.  This is reasonable since ARIMA will not be very effective for these quiescent light curves and it understandable that our analysis might not perform as well there, especially if those candidates happen to have weak signals. 

\end{document}